\documentclass[seceqn,secthm]{elsart}

\usepackage{latexsym}   
\usepackage{amssymb}
\usepackage{amsmath}
\usepackage{stmaryrd}
\usepackage{graphicx}
\usepackage{pst-all} 
\usepackage{times}

\begin{document}

\begin{frontmatter}

\title{Measurement-Only Topological Quantum Computation via Anyonic Interferometry}

\author[StationQ]{Parsa Bonderson\corauthref{cor}},
\corauth[cor]{Corresponding author.}
\ead{parsab@microsoft.com}
\author[StationQ]{Michael Freedman},
\ead{michaelf@microsoft.com}
\author[StationQ,UCSB]{Chetan Nayak}
\ead{nayak@kitp.ucsb.edu}
\address[StationQ]{Microsoft Research, Station Q, Elings Hall, University of California, Santa Barbara, CA 93106, USA}
\address[UCSB]{Department of Physics, University of California, Santa Barbara, CA 93106, USA}

\begin{abstract}
We describe measurement-only topological quantum computation using both projective and interferometrical measurement of topological charge. We demonstrate how anyonic teleportation can be achieved using ``forced measurement'' protocols for both types of measurement. Using this, it is shown how topological charge measurements can be used to generate the braiding transformations used in topological quantum computation, and hence that the physical transportation of computational anyons is unnecessary. We give a detailed discussion of the anyonics for implementation of topological quantum computation (particularly, using the measurement-only approach) in fractional quantum Hall systems.
\end{abstract}

\date{\today}

\end{frontmatter}


\section{Introduction}

Since von~Neumann's axiomatization of quantum mechanics in the 1930s~\cite{vonNeumann55}, measurement has been a kind of stepchild to unitary evolution. As the link between the quantum and classical worlds, measurement has attracted considerable skeptical scrutiny from scientists and philosophers. In the domain of quantum computation, measurement -- an act which may project out computational degrees of freedom and potentially decohere important quantum correlations -- is often dreaded and great care must be taken to avoid the occurrence of unintentional measurements (e.g. by the ``environment''). However, from a pragmatic perspective, measurement is a co-equal pillar of quantum mechanics and a tool to be exploited. In mathematics, also, unitary operators and projectors are dual. Bott periodicity states that a loop of unitaries corresponds to a projector and a loop of projectors corresponds to a unitary. In the integral quantum Hall systems, this perspective links the Chern class in the bulk to the energy current at the edge.

It is a familiar idea that measurement can \emph{stop} something from happening, e.g. the ``watched pot'' effect of the quantum Zeno paradox. It is less familiar that measurement can elicit an intended evolution of states. To get an idea for how this might work, consider adiabatic evolution of a vector $\Psi$ in a $k$-fold degenerate ground state manifold of a Hamiltonian $H$. Perturbing the Hamiltonian in time, $H(t)$, while leaving the $k$-fold degeneracy intact will evolve $\Psi$ in time\footnote{This evolution can be described by the canonical connection on the ``tautological bundle'' over the Grassmann manifold of $k$-planes.}. A discretized description of this evolution amounts to moving the $k$-dimensional ground state subspace slightly inside the entire state space (leaving $\Psi$ fixed), and then projecting $\Psi$ orthogonally back into the new ground state subspace and repeating. It well known that adiabatic evolution can effect the general unitary on the $k$-fold ground state space, so a composition of projections (``measurements'') suffices to simulate unitary evolution in this simple example. This example amounts to a variation on the ``quantum watched pot effect'' in which the ``pot'' is not holding still, but rather, it evolves in a manner dictated by how it is ``watched.''

The preceding example suggests that, under the right conditions, a quantum state can be deliberately nudged along by a sequence of measurements, as an alternative to directly constructing a unitary evolution of a state's underlying degrees of freedom. However, we are not interested in the case where we are performing an adiabatic variation of the Hamiltonian, but rather we want to employ more standard measurements, whose effects are to dramatically disturb the state with a projection, in order to generate desired unitary operators. A little more thought turns up a conundrum: How can projection, an operation which generically reduces rank, simulate an operation which has full rank such as unitary evolution? The crux of the answer (hinted at in the proceeding example) is that we do not attempt to simulate evolution of a general state via general measurements. Instead, by choosing a specific subspace of the total state space, properly chosen measurements can generate unitary evolution for states in the subspace. The archetypal example of this is quantum state teleportation, in which a properly chosen measurement preserves the rank of the quantum state, but transforms it in some particular way. In fact, teleportation is \emph{such} a good example that it turns out to be the crucial ingredient of all measurement based approaches to quantum computation. For the purposes of quantum computing, one is content as long as any desired unitary evolution on the computational subspace can be produced.

In quantum computation, the accuracy of the unitary evolution is paramount. The principle advantages of the topological approach to quantum computing are that fault tolerance is naturally provided at the hardware level by the non-local state space of non-Abelian anyons, and that the unitary operators corresponding to braid representations are essentially exact. The topological model also anticipates measuring states in the basis of topological charge for qubit readout. It makes sense to ask if topological quantum computations can be organized as a sequence of such topological charge measurements, rather than as an exercise in physically braiding anyons, which has been the hypothetical paradigm since the inception of topological quantum computation (TQC)~\cite{Kitaev03,Freedman03b}. What we have found is that performing TQC can in fact be reduced to the employment of a single computational primitive: topological charge measurement, and we call this approach ``measurement-only topological quantum computation'' (MOTQC)~\cite{Bonderson08a}.

In Ref.~\cite{Bonderson08a}, we considered (nondemolitional) topological measurements that execute precisely the von~Neumann orthogonal projections onto topological charge sectors. Using such topological charge measurements, we devised an anyonic analog of quantum state teleportation through ``forced measurement,'' a probabilistically determined adaptive sequence of measurements that assures the attainment of a desired outcome. Forced measurement teleportation is an important primitive (or rather, modular composite of primitives) for MOTQC\footnote{Readers from the computer science side of the subject who are not familiar with ``teleportation'' should not be concerned; it amounts merely to moving an anyon along a worldline which is not monotonic in time, a concept completely natural in quantum field theory.}, as we have showed that a sequence of forced measurements can be used to generate any braiding transformation used in TQC.

However, anyonic interferometry is another form of topological charge measurement, which is generally not quite the same as projective measurement for non-Abelian anyons. Anyonic interferometry is known~\cite{Bonderson07a,Bonderson07b,Bonderson07c} to amount to a combination of the usual projection onto topological charge sectors with a decohering effect between anyons inside and outside of the interferometer, that one can think of as a severing of the anyonic charge lines connecting these regions. This decoherence is a crucial difference that causes the forced projective measurement protocol of Ref.~\cite{Bonderson08a} to fail. Indeed, in the context of quantum information processing, operations associated with the term ``decoherence'' typically bear dire consequences for the encoded quantum information. Fortunately, when examined in detail, the decoherence associated with anyonic interferometry does not harm the encoded computational state information, so the situation can be salvaged. Since interferometry is a projection together with decoherence, the process studied here is more complicated than for projective measurement, and requires the use of density matrix formalism. In this paper, we will review the (projective measurement based) results of Ref.~\cite{Bonderson08a} in the density matrix formalism, providing a stepping stone in the extension to the interferometrical version. We then extend the concepts of anyonic teleportation and MOTQC to the case where interferometry measurements of topological charge are employed. This is done by appropriately modifying the forced measurement protocol with additional topological charge measurements that allow forced measurement anyonic teleportation to be achieved using interferometry. In the context of MOTQC, we require topological charge measurements of collections of at most $8$ adjacent anyons, ensuring unbounded measurement are avoided.

The extension of MOTQC to the case when topological charge is measured interferometrically is not merely an exercise in thoroughness, but rather is very pragmatically motivated by the goal of physical implementation. The most likely topological mediums in which TQC will be realized are fractional quantum Hall (FQH) systems, as these are the only physical systems that have experimentally exhibited topological order. In particular, the second Landau level FQH states at $\nu = 5/2$ and $12/5$ described by Moore--Read~\cite{Moore91}, Read--Rezayi~\cite{Read99}, and/or Bonderson--Slingerland~\cite{Bonderson07d} have quasiparticles with non-Abelian braiding statistics of either the Ising or Fibonacci type (see Appendix). The principal method of measuring topological charge in these FQH systems will undoubtedly be quasiparticle interferometry, as it has been rapidly developing on both experimental and theoretical fronts. In fact, double point-contact FQH interferometers, which could eventually be used for our desired topological charge measurements, have already been experimentally realized in the $\nu=5/2$ regime~\cite{Willett08}. In contrast, TQC braiding of computational anyons involves moving individual anyons in a highly controlled manner for which there is at present essentially no experimental basis, and theoretical proposals whose prospects are, at best, dubious. Thus, from a technological point of view, it seems that MOTQC using interferometry in FQH systems gives the clearest path to implementation of TQC. In this vein, we conclude the paper with a detailed discussion of the anyonics\footnote{The term ``anyonics'' was coined by Wilczek in reference to the science of manipulating anyons in materials and devices.} required to implement MOTQC in FQH systems, providing a concrete proposal for the actualization of TQC.

\section{Topological Charge Measurement}

Since topological charge measurement is the foundation upon which the measurement-only approach to TQC is based, we begin by describing it in detail. We use a isotopy invariant diagrammatic representation of anyonic states and operators acting on them, as described by anyon models (known in mathematical terminology as unitary braided tensor categories~\cite{Turaev94,Bakalov01}). These diagrams encode the purely topological properties of anyons, independent of any particular physical representation. We provide a terse review of useful properties of anyon models here, and recommend Refs.~\cite{Preskill-lectures,Kitaev06a,Bonderson07b,Bonderson07c} for additional details. For a few particularly relevant examples (Ising, Fibonacci, and SU$(2)_{k}$, and the non-Abelian fractional quantum Hall states related to these), see the Appendix.

An anyon model has a finite set $\mathcal{C}$ of superselection sector labels called topological or anyonic charges. These conserved charges obey a
commutative, associative fusion algebra%
\begin{equation}
a\times b=\sum\limits_{c\in \mathcal{C}}N_{ab}^{c}c
\end{equation}%
where the fusion multiplicities $N_{ab}^{c}$ are non-negative integers which indicate the number of different ways the charges $a$ and $b$ can be
combined to produce the charge $c$. There is a unique trivial ``vacuum'' charge $0 \in \mathcal{C}$ for which $N_{a0}^{c}=\delta _{ac}$, and each charge $a$ has a unique conjugate charge, or ``antiparticle,'' $\bar{a}\in \mathcal{C}$ such that $N_{ab}^{0}=\delta _{b\bar{a}}$. ($0=\bar{0}$ and $\bar{\bar{a}}=a$.) Writing the fusion multiplicities as matrices $\left[ N_{a} \right]_{b}^{c}$ with indices $b$ and $c$, we define the quantum dimension $d_{a}$ of charge $a$ as the largest eigenvalue of $N_{a}$, and the total quantum dimension as $\mathcal{D} = \sqrt{ \sum_{a} d_{a}^{2} }$. We define a topological charge $a$ to be Abelian if $d_{a}=1$ and non-Abelian if $d_{a} > 1$, which is equivalent to saying Abelian charges have unique fusion with all other charges ($\sum_{c} N_{ab}^{c} =1$ for all $b$), whereas non-Abelian charges are permitted multiple fusion channels (i.e. $\sum_{c} N_{ab}^{c} >1$ for some $b$)\footnote{Technically, this allows for a ``non-Abelian'' charge that nonetheless has only Abelian braiding, but this is not the case for the examples we consider, so we will not worry about being more careful with this distinction.}.

In the diagrammatic formalism, each segment of line is oriented (indicated with an arrow), and has a topological charge associated with it. Reversing the orientation of a line is equivalent to conjugating the charge labeling it, i.e.
\begin{equation}
\pspicture[0.4375](0.4,-0.1)(1.05,1.5)
  \small
  \psset{linewidth=0.9pt,linecolor=black,arrowscale=1.5,arrowinset=0.15}
  \psline(0.6,0.1)(0.6,1.3)
  \psline{->}(0.6,0.1)(0.6,0.9)
  \rput[bl]{0}(0.75,0.41){$a$}
  \endpspicture
=
\pspicture[0.4375](0.4,-0.1)(1.05,1.5)
  \small
  \psset{linewidth=0.9pt,linecolor=black,arrowscale=1.5,arrowinset=0.15}
  \psline(0.6,0.1)(0.6,1.3)
  \psline{-<}(0.6,0.1)(0.6,0.8)
  \rput[bl]{0}(0.75,0.41){$\bar{a}$}
  \endpspicture
.
\end{equation}
Charge lines carrying the vacuum charge $0$ can be added and removed from diagrams at will.

Each fusion product has an associated vector space $V_{ab}^{c}$ with $\dim{V_{ab}^{c}} = N_{ab}^{c}$, and its dual (splitting) space $V^{ab}_{c}$. The states in these fusion and splitting spaces are assigned to trivalent vertices with the appropriately corresponding anyonic charges:
\begin{equation}
\left( d_{c} / d_{a}d_{b} \right) ^{1/4}
\pspicture[0.6](-0.1,-0.2)(1.5,-1.2)
  \small
  \psset{linewidth=0.9pt,linecolor=black,arrowscale=1.5,arrowinset=0.15}
  \psline{-<}(0.7,0)(0.7,-0.35)
  \psline(0.7,0)(0.7,-0.55)
  \psline(0.7,-0.55) (0.25,-1)
  \psline{-<}(0.7,-0.55)(0.35,-0.9)
  \psline(0.7,-0.55) (1.15,-1)	
  \psline{-<}(0.7,-0.55)(1.05,-0.9)
  \rput[tl]{0}(0.4,0){$c$}
  \rput[br]{0}(1.4,-0.95){$b$}
  \rput[bl]{0}(0,-0.95){$a$}
 \scriptsize
  \rput[bl]{0}(0.85,-0.5){$\mu$}
  \endpspicture
=\left\langle a,b;c,\mu \right| \in
V_{ab}^{c} ,
\label{eq:bra}
\end{equation}
\begin{equation}
\left( d_{c} / d_{a}d_{b}\right) ^{1/4}
\pspicture[0.4642857](-0.1,-0.2)(1.5,1.2)
  \small
  \psset{linewidth=0.9pt,linecolor=black,arrowscale=1.5,arrowinset=0.15}
  \psline{->}(0.7,0)(0.7,0.45)
  \psline(0.7,0)(0.7,0.55)
  \psline(0.7,0.55) (0.25,1)
  \psline{->}(0.7,0.55)(0.3,0.95)
  \psline(0.7,0.55) (1.15,1)	
  \psline{->}(0.7,0.55)(1.1,0.95)
  \rput[bl]{0}(0.4,0){$c$}
  \rput[br]{0}(1.4,0.8){$b$}
  \rput[bl]{0}(0,0.8){$a$}
 \scriptsize
  \rput[bl]{0}(0.85,0.35){$\mu$}
  \endpspicture
=\left| a,b;c,\mu \right\rangle \in
V_{c}^{ab},
\label{eq:ket}
\end{equation}
where $\mu=1,\ldots ,N_{ab}^{c}$. Most anyon models of interest, for example, those described in the Appendix, have no fusion multiplicities, i.e. $N_{ab}^{c}=0,1$, in which case the vertex labels $\mu$ can be dropped (such indices can comfortably be ignored in this paper). The normalization factors $\left( d_{c}/d_{a}d_{b}\right) ^{1/4}$ are included so that diagrams are in the isotopy invariant convention throughout this paper. Isotopy invariance means that the value of a (labeled) diagram is not changed by continuous deformations, so long as open endpoints are held fixed and lines are not passed through each other or around open endpoints. Open endpoints should be thought of as ending on some boundary (e.g. a timeslice or an edge of the system) through which isotopy is not permitted. As a word of caution, we note that the diagrammatic expressions of states and operators are, by design, reminiscent of particle worldlines, but there is not a strict identification between the two. The anyonic charge lines are only a diagrammatic expression of the algebraic encoding of the topological properties of anyons, and interpreting them as worldlines is not always correct.

Diagrammatically, inner products are formed by stacking vertices so the fusing/splitting lines connect%
\begin{equation}
\label{eq:inner_product}
  \pspicture[0.45238](-0.2,-0.35)(1.2,1.75)
  \small
  \psarc[linewidth=0.9pt,linecolor=black,border=0pt] (0.8,0.7){0.4}{120}{240}
  \psarc[linewidth=0.9pt,linecolor=black,arrows=<-,arrowscale=1.4,
    arrowinset=0.15] (0.8,0.7){0.4}{165}{240}
  \psarc[linewidth=0.9pt,linecolor=black,border=0pt] (0.4,0.7){0.4}{-60}{60}
  \psarc[linewidth=0.9pt,linecolor=black,arrows=->,arrowscale=1.4,
    arrowinset=0.15] (0.4,0.7){0.4}{-60}{15}
  \psset{linewidth=0.9pt,linecolor=black,arrowscale=1.5,arrowinset=0.15}
  \psline(0.6,1.05)(0.6,1.55)
  \psline{->}(0.6,1.05)(0.6,1.45)
  \psline(0.6,-0.15)(0.6,0.35)
  \psline{->}(0.6,-0.15)(0.6,0.25)
  \rput[bl]{0}(0.07,0.55){$a$}
  \rput[bl]{0}(0.94,0.55){$b$}
  \rput[bl]{0}(0.26,1.25){$c$}
  \rput[bl]{0}(0.24,-0.05){$c'$}
 \scriptsize
  \rput[bl]{0}(0.7,1.05){$\mu$}
  \rput[bl]{0}(0.7,0.15){$\mu'$}
  \endpspicture
=\delta _{c c ^{\prime }}\delta _{\mu \mu ^{\prime }} \sqrt{\frac{d_{a}d_{b}}{d_{c}}}
  \pspicture[0.45238](0.15,-0.35)(0.8,1.75)
  \small
  \psset{linewidth=0.9pt,linecolor=black,arrowscale=1.5,arrowinset=0.15}
  \psline(0.6,-0.15)(0.6,1.55)
  \psline{->}(0.6,-0.15)(0.6,0.85)
  \rput[bl]{0}(0.75,1.25){$c$}
  \endpspicture
,
\end{equation}%
which can be applied inside more complicated diagrams, as is the case for all diagrammatic relations. This diagrammatically encodes charge conservation, and as a special case (with $c=0$) gives
\begin{equation}
\label{eq:loop=d}
  \pspicture[0.35](-0.08,0.25)(1.35,1.25)
  \small
  \psarc[linewidth=0.9pt,linecolor=black,arrows=<-,arrowscale=1.5,
    arrowinset=0.15] (0.8,0.7){0.5}{165}{363}
  \psarc[linewidth=0.9pt,linecolor=black,border=0pt]
(0.8,0.7){0.5}{0}{170}
  \rput[bl]{0}(-0.03,0.55){$a$}
 \endpspicture
=d_{a}=d_{\bar{a}}.
\end{equation}

The identity operator on a pair of anyons with charges $a$ and $b$ is written diagrammatically as
\begin{equation}
\label{eq:Id}
\mathbb{I}_{ab} =
\pspicture[0.40625](0,-0.5)(1.1,1.1)
  \small
  \psset{linewidth=0.9pt,linecolor=black,arrowscale=1.5,arrowinset=0.15}
  \psline(0.3,-0.45)(0.3,1)
  \psline{->}(0.3,-0.45)(0.3,0.50)
  \psline(0.8,-0.45)(0.8,1)
  \psline{->}(0.8,-0.45)(0.8,0.50)
  \rput[br]{0}(1.05,0.8){$b$}
  \rput[bl]{0}(0,0.8){$a$}
  \endpspicture
 = \sum\limits_{c,\mu }
\sqrt{\frac{d_{c}}{d_{a}d_{b}}} \;
 \pspicture[0.41379](-0.1,-0.45)(1.4,1)
  \small
  \psset{linewidth=0.9pt,linecolor=black,arrowscale=1.5,arrowinset=0.15}
  \psline{->}(0.7,0)(0.7,0.45)
  \psline(0.7,0)(0.7,0.55)
  \psline(0.7,0.55) (0.25,1)
  \psline{->}(0.7,0.55)(0.3,0.95)
  \psline(0.7,0.55) (1.15,1)
  \psline{->}(0.7,0.55)(1.1,0.95)
  \rput[bl]{0}(0.38,0.2){$c$}
  \rput[br]{0}(1.4,0.8){$b$}
  \rput[bl]{0}(0,0.8){$a$}
  \psline(0.7,0) (0.25,-0.45)
  \psline{-<}(0.7,0)(0.35,-0.35)
  \psline(0.7,0) (1.15,-0.45)
  \psline{-<}(0.7,0)(1.05,-0.35)
  \rput[br]{0}(1.4,-0.4){$b$}
  \rput[bl]{0}(0,-0.4){$a$}
\scriptsize
  \rput[bl]{0}(0.85,0.4){$\mu$}
  \rput[bl]{0}(0.85,-0.03){$\mu$}
  \endpspicture
\; .
\end{equation}

More complicated diagrams can be constructed by connecting lines of matching charge. The resulting vector spaces obey a notion of associativity given by unitary isomorphisms, which can be reduced using the expression of three anyon splitting/fusion spaces in terms of two anyon splitting/fusion
\begin{equation}
V_{d}^{abc}\cong \bigoplus\limits_{e}V_{e}^{ab}\otimes V_{d}^{ec}\cong
\bigoplus\limits_{f}V_{f}^{bc}\otimes V_{d}^{af},
\end{equation}
to isomorphisms called $F$-moves, which are written diagrammatically as
\begin{equation}
  \pspicture[0.444444](0,-0.45)(1.8,1.8)
  \small
  \psset{linewidth=0.9pt,linecolor=black,arrowscale=1.5,arrowinset=0.15}
  \psline(0.2,1.5)(1,0.5)
  \psline(1,0.5)(1,0)
  \psline(1.8,1.5) (1,0.5)
  \psline(0.6,1) (1,1.5)
   \psline{->}(0.6,1)(0.3,1.375)
   \psline{->}(0.6,1)(0.9,1.375)
   \psline{->}(1,0.5)(1.7,1.375)
   \psline{->}(1,0.5)(0.7,0.875)
   \psline{->}(1,0)(1,0.375)
   \rput[bl]{0}(0.05,1.6){$a$}
   \rput[bl]{0}(0.95,1.6){$b$}
   \rput[bl]{0}(1.75,1.6){${c}$}
   \rput[bl]{0}(0.5,0.5){$e$}
   \rput[bl]{0}(0.9,-0.3){$d$}
 \scriptsize
   \rput[bl]{0}(0.3,0.8){$\alpha$}
   \rput[bl]{0}(0.7,0.25){$\beta$}
  \endpspicture
= \sum_{f,\mu,\nu} \left[F_d^{abc}\right]_{(e,\alpha,\beta)(f,\mu,\nu)}
 \pspicture[0.4444444](0,-0.45)(1.8,1.8)
  \small
  \psset{linewidth=0.9pt,linecolor=black,arrowscale=1.5,arrowinset=0.15}
  \psline(0.2,1.5)(1,0.5)
  \psline(1,0.5)(1,0)
  \psline(1.8,1.5) (1,0.5)
  \psline(1.4,1) (1,1.5)
   \psline{->}(0.6,1)(0.3,1.375)
   \psline{->}(1.4,1)(1.1,1.375)
   \psline{->}(1,0.5)(1.7,1.375)
   \psline{->}(1,0.5)(1.3,0.875)
   \psline{->}(1,0)(1,0.375)
   \rput[bl]{0}(0.05,1.6){$a$}
   \rput[bl]{0}(0.95,1.6){$b$}
   \rput[bl]{0}(1.75,1.6){${c}$}
   \rput[bl]{0}(1.25,0.45){$f$}
   \rput[bl]{0}(0.9,-0.3){$d$}
 \scriptsize
   \rput[bl]{0}(1.5,0.8){$\mu$}
   \rput[bl]{0}(0.7,0.25){$\nu$}
  \endpspicture
.
\end{equation}
We write this with one line bent down as
\begin{equation}
\pspicture[0.375](0,-0.4)(1.1,1.2)
\small
  \psset{linewidth=0.9pt,linecolor=black,arrowscale=1.5,arrowinset=0.15}
  \psline(0.3,-0.45)(0.3,1)
  \psline{->}(0.3,-0.45)(0.3,-0.05)
  \psline{->}(0.3,0.5)(0.3,0.85)
  \psline(0.8,-0.45)(0.8,1)
  \psline{->}(0.8,-0.45)(0.8,-0.05)
  \psline{->}(0.8,0)(0.8,0.85)
  \psline(0.8,0.05)(0.3,0.45)
  \psline{->}(0.8,0.05)(0.45,0.33)
  \rput[bl]{0}(0.48,0.38){$e$}
  \rput[bl]{0}(0.93,0.8){$b$}
  \rput[bl]{0}(0,0.8){$a$}
  \rput[bl]{0}(0.95,-0.4){$d$}
  \rput[bl]{0}(-0.05,-0.4){$c$}
  \scriptsize
  \rput[bl]{0}(0.02,0.35){$\alpha$}
  \rput[bl]{0}(0.87,-0.1){$\beta$}
  \endpspicture
=\sum\limits_{f,\mu
,\nu }\left[ F_{cd}^{ab}\right] _{\left( e,\alpha ,\beta \right) \left(
f,\mu ,\nu \right)}
 \pspicture[0.41379](-0.1,-0.45)(1.4,1)
  \small
  \psset{linewidth=0.9pt,linecolor=black,arrowscale=1.5,arrowinset=0.15}
  \psline{->}(0.7,0)(0.7,0.45)
  \psline(0.7,0)(0.7,0.55)
  \psline(0.7,0.55) (0.25,1)
  \psline{->}(0.7,0.55)(0.3,0.95)
  \psline(0.7,0.55) (1.15,1)
  \psline{->}(0.7,0.55)(1.1,0.95)
  \rput[bl]{0}(0.38,0.1){$f$}
  \rput[br]{0}(1.4,0.8){$b$}
  \rput[bl]{0}(0,0.8){$a$}
  \psline(0.7,0) (0.25,-0.45)
  \psline{-<}(0.7,0)(0.35,-0.35)
  \psline(0.7,0) (1.15,-0.45)
  \psline{-<}(0.7,0)(1.05,-0.35)
  \rput[br]{0}(1.4,-0.4){$d$}
  \rput[bl]{0}(0,-0.4){$c$}
\scriptsize
  \rput[bl]{0}(0.85,0.4){$\mu$}
  \rput[bl]{0}(0.85,-0.03){$\nu$}
  \endpspicture
\end{equation}
which is also a unitary transformation. {}From Eq.~(\ref{eq:Id}), we immediately find that
\begin{equation}
\left[ F_{ab}^{ab}\right] _{0 \left( c,\mu ,\nu \right) }=\sqrt{\frac{d_{c}}{d_{a}d_{b}}}\;\delta _{\mu \nu }
,
\label{eq:idF}
\end{equation}
and more generally, applying Eqs.~(\ref{eq:Id}) and (\ref{eq:inner_product}) gives a relation between the two types of $F$-symbols
\begin{equation}
\left[ F_{cd}^{ab}\right] _{\left( e,\alpha ,\beta \right) \left( f,\mu
,\nu \right) } =\sqrt{\frac{d_{e}d_{f}}{d_{a}d_{d}}}\left[ F_{f}^{ceb}%
\right] _{\left( a,\alpha ,\mu \right) \left( d,\beta ,\nu \right) }^{\ast }
.
\end{equation}

The standard tensor product of operators $X$ and $Y$ acting on different sets of anyons are formed by juxtaposition in the diagrammatic representation
\begin{equation}
 \pspicture[0.46666666](-1.4,-1.5)(1.3,1.5)
  \small
  \psframe[linewidth=0.9pt,linecolor=black,border=0](-1.2,-0.5)(1.2,0.5)
  \rput[bl]{0}(-0.55,-0.15){$X \otimes Y$}
  \rput[bl]{0}(0.4,0.7){$\mathbf{\ldots}$}
  \rput[bl]{0}(0.4,-0.75){$\mathbf{\ldots}$}
  \rput[bl]{0}(-0.8,0.7){$\mathbf{\ldots}$}
  \rput[bl]{0}(-0.8,-0.75){$\mathbf{\ldots}$}
  \psset{linewidth=0.9pt,linecolor=black,arrowscale=1.5,arrowinset=0.15}
  \psline(1.0,0.5)(1.0,1)
  \psline(0.2,0.5)(0.2,1)
  \psline(1.0,-0.5)(1.0,-1)
  \psline(0.2,-0.5)(0.2,-1)
  \psline(-1.0,0.5)(-1.0,1)
  \psline(-0.2,0.5)(-0.2,1)
  \psline(-1.0,-0.5)(-1.0,-1)
  \psline(-0.2,-0.5)(-0.2,-1)
  \psline{->}(1.0,0.5)(1.0,0.9)
  \psline{->}(0.2,0.5)(0.2,0.9)
  \psline{-<}(1.0,-0.5)(1.0,-0.9)
  \psline{-<}(0.2,-0.5)(0.2,-0.9)
  \psline{->}(-1.0,0.5)(-1.0,0.9)
  \psline{->}(-0.2,0.5)(-0.2,0.9)
  \psline{-<}(-1.0,-0.5)(-1.0,-0.9)
  \psline{-<}(-0.2,-0.5)(-0.2,-0.9)
 \endpspicture
 =
  \pspicture[0.4666666666](-1,-1.5)(0.9,1.5)
  \small
  \psframe[linewidth=0.9pt,linecolor=black,border=0](-0.8,-0.5)(0.8,0.5)
  \rput[bl]{0}(-0.15,-0.12){$X$}
  \rput[bl]{0}(-0.22,0.7){$\mathbf{\ldots}$}
  \rput[bl]{0}(-0.22,-0.75){$\mathbf{\ldots}$}
  \psset{linewidth=0.9pt,linecolor=black,arrowscale=1.5,arrowinset=0.15}
  \psline(0.6,0.5)(0.6,1)
  \psline(-0.6,0.5)(-0.6,1)
  \psline(0.6,-0.5)(0.6,-1)
  \psline(-0.6,-0.5)(-0.6,-1)
  \psline{->}(0.6,0.5)(0.6,0.9)
  \psline{->}(-0.6,0.5)(-0.6,0.9)
  \psline{-<}(0.6,-0.5)(0.6,-0.9)
  \psline{-<}(-0.6,-0.5)(-0.6,-0.9)
\endpspicture
 \pspicture[0.4666666666](-1,-1.5)(0.9,1.5)
  \small
  \psframe[linewidth=0.9pt,linecolor=black,border=0](-0.8,-0.5)(0.8,0.5)
  \rput[bl]{0}(-0.15,-0.12){$Y$}
  \rput[bl]{0}(-0.22,0.7){$\mathbf{\ldots}$}
  \rput[bl]{0}(-0.22,-0.75){$\mathbf{\ldots}$}
  \psset{linewidth=0.9pt,linecolor=black,arrowscale=1.5,arrowinset=0.15}
  \psline(0.6,0.5)(0.6,1)
  \psline(-0.6,0.5)(-0.6,1)
  \psline(0.6,-0.5)(0.6,-1)
  \psline(-0.6,-0.5)(-0.6,-1)
  \psline{->}(0.6,0.5)(0.6,0.9)
  \psline{->}(-0.6,0.5)(-0.6,0.9)
  \psline{-<}(0.6,-0.5)(0.6,-0.9)
  \psline{-<}(-0.6,-0.5)(-0.6,-0.9)
 \endpspicture
.
\end{equation}

The quantum trace $\widetilde{\text{Tr}}$ of an operator $X$ is defined by closing the diagram with loops that match the outgoing lines with the
respective incoming lines at the same position%
\begin{equation}
\widetilde{\text{Tr}} \left[ X \right] =
\widetilde{\text{Tr}}
\left[
 \pspicture[0.466666666](-1,-1.5)(1,1.5)
  \small
  \psframe[linewidth=0.9pt,linecolor=black,border=0](-0.8,-0.5)(0.8,0.5)
  \rput[bl]{0}(-0.15,-0.1){$X$}
  \rput[bl]{0}(-0.22,0.7){$\mathbf{\ldots}$}
  \rput[bl]{0}(-0.22,-0.75){$\mathbf{\ldots}$}
  \psset{linewidth=0.9pt,linecolor=black,arrowscale=1.5,arrowinset=0.15}
  \psline(0.6,0.5)(0.6,1)
  \psline(-0.6,0.5)(-0.6,1)
  \psline(0.6,-0.5)(0.6,-1)
  \psline(-0.6,-0.5)(-0.6,-1)
  \psline{->}(0.6,0.5)(0.6,0.9)
  \psline{->}(-0.6,0.5)(-0.6,0.9)
  \psline{-<}(0.6,-0.5)(0.6,-0.9)
  \psline{-<}(-0.6,-0.5)(-0.6,-0.9)
  \rput[bl](-0.8,1.05){$A_1$}
  \rput[bl](0.4,1.05){$A_n$}
  \rput[tl](-0.8,-1.05){$A'_1$}
  \rput[tl](0.4,-1.05){$A'_n$}
 \endpspicture
\right]
=
 \pspicture[0.45833333](-1.1,-1.2)(2.2,1.2)
  \small
  \psframe[linewidth=0.9pt,linecolor=black,border=0](-0.8,-0.5)(0.8,0.5)
  \rput[bl]{0}(-0.15,-0.1){$X$}
  \rput[bl]{0}(-0.4,0.7){$\mathbf{\ldots}$}
  \rput[bl]{0}(-0.22,-0.75){$\mathbf{\ldots}$}
  \rput[bl]{0}(1.52,0){$\mathbf{\ldots}$}
  \psset{linewidth=0.9pt,linecolor=black,arrowscale=1.5,arrowinset=0.15}
  \psarc(1.0,0.5){0.4}{0}{180}
  \psarc(1.0,-0.5){0.4}{180}{360}
  \psarc(0,0.5){0.6}{90}{180}
  \psarc(0,-0.5){0.6}{180}{270}
  \psarc(1.5,0.5){0.6}{0}{90}
  \psarc(1.5,-0.5){0.6}{270}{360}
  \psline(1.4,-0.5)(1.4,0.5)
  \psline(0,1.1)(1.5,1.1)
  \psline(0,-1.1)(1.5,-1.1)
  \psline(2.1,-0.5)(2.1,0.5)
  \psline{->}(1.4,0.2)(1.4,-0.1)
  \psline{->}(2.1,0.2)(2.1,-0.1)
  \rput[bl](-1.07,0.6){$A_1$}
  \rput[bl](0.1,0.6){$A_n$}
 \endpspicture
,
\end{equation}%
where the capitalized topological charge labels are used to mean there can be a sum over charges on these lines. This is related to the standard trace Tr on the vector spaces (bras and kets) by
\begin{equation}
\widetilde{\text{Tr}} \left[ X \right] =\sum\limits_{c}d_{c}\text{Tr} \left[ X_{c} \right]
,
\end{equation}
where the operator $X$ is decomposed into sectors of overall topological charge $c$%
\begin{eqnarray}
&& X = \sum\limits_{c}X_{c}, \\
&& X_{c} = \Pi_{c}^{\left( 1 \ldots n \right)} \, X \, \Pi_{c}^{\left( 1 \ldots n \right)} \in V_{c}^{A_{1}\ldots A_{n}}\otimes
V_{A_{1}^{\prime }\ldots A_{n}^{\prime }}^{c}
.
\end{eqnarray}%
A diagrammatic representation of $\Pi_{c}^{\left( 1 \ldots n \right)}$, the projector of the $n$ anyons onto definite collective charge $c$, is given in Eq.~(\ref{eq:PIcn}). We note that such projectors give a partition of identity
\begin{equation}
\mathbb{I}_{a_{1} \ldots a_{n} } = \sum_{c} \Pi_{c}^{\left( 1 \ldots n \right)}
.
\end{equation}
Notice that for any operator $X$ with overall vacuum charge $X = X_{0}$, or more generally with overall charges that are Abelian $X = \sum_{c \in \left. \mathcal{C}\right|_{\text{Abelian}}} X_{c}$, the quantum trace and the standard trace coincide $\widetilde{\text{Tr}} \left[ X \right] =\text{Tr} \left[ X \right]$.
Partial quantum traces are defined by closing a subset of the charge lines with loops back on themselves, but requires specification of how these lines are closed on themselves corresponding to the physical path through which the traced out anyons are removed from the system (see Ref.~\cite{Bonderson07c} for additional discussion of this point).

The counterclockwise braiding exchange operator of two anyons is represented diagrammatically by
\begin{equation}
R_{ab}=
\pspicture[0.44](-0.1,-0.2)(1.3,1.05)
\small
  \psset{linewidth=0.9pt,linecolor=black,arrowscale=1.5,arrowinset=0.15}
  \psline(0.96,0.05)(0.2,1)
  \psline{->}(0.96,0.05)(0.28,0.9)
  \psline(0.24,0.05)(1,1)
  \psline[border=2pt]{->}(0.24,0.05)(0.92,0.9)
  \rput[bl]{0}(-0.02,0){$a$}
  \rput[br]{0}(1.2,0){$b$}
  \endpspicture
=\sum\limits_{c,\mu ,\nu }\sqrt{\frac{d_{c}}{d_{a}d_{b}}}\left[
R_{c}^{ab}\right] _{\mu \nu }
 \pspicture[0.41379](-0.1,-0.45)(1.4,1)
  \small
  \psset{linewidth=0.9pt,linecolor=black,arrowscale=1.5,arrowinset=0.15}
  \psline{->}(0.7,0)(0.7,0.45)
  \psline(0.7,0)(0.7,0.55)
  \psline(0.7,0.55) (0.25,1)
  \psline{->}(0.7,0.55)(0.3,0.95)
  \psline(0.7,0.55) (1.15,1)
  \psline{->}(0.7,0.55)(1.1,0.95)
  \rput[bl]{0}(0.38,0.2){$c$}
  \rput[br]{0}(1.4,0.8){$a$}
  \rput[bl]{0}(0,0.8){$b$}
  \psline(0.7,0) (0.25,-0.45)
  \psline{-<}(0.7,0)(0.35,-0.35)
  \psline(0.7,0) (1.15,-0.45)
  \psline{-<}(0.7,0)(1.05,-0.35)
  \rput[br]{0}(1.4,-0.4){$b$}
  \rput[bl]{0}(0,-0.4){$a$}
\scriptsize
  \rput[bl]{0}(0.85,0.4){$\nu$}
  \rput[bl]{0}(0.85,-0.03){$\mu$}
  \endpspicture
,
\end{equation}%
and similarly, the clockwise braid is
\begin{equation}
R_{ab}^{\dag}= R_{ab}^{-1}=
\pspicture[0.44](-0.1,-0.2)(1.3,1.05)
\small
  \psset{linewidth=0.9pt,linecolor=black,arrowscale=1.5,arrowinset=0.15}
  \psline{->}(0.24,0.05)(0.92,0.9)
  \psline(0.24,0.05)(1,1)
  \psline(0.96,0.05)(0.2,1)
  \psline[border=2pt]{->}(0.96,0.05)(0.28,0.9)
  \rput[bl]{0}(-0.01,0){$b$}
  \rput[bl]{0}(1.06,0){$a$}
  \endpspicture
.
\end{equation}

Some important quantities derived from braiding are the topological spin
\begin{equation}
\theta _{a}=\theta _{\bar{a}}=d_{a}^{-1}\widetilde{\text{Tr}} \left[ R_{aa} \right]
=\sum\limits_{c,\mu } \frac{d_{c}}{d_{a}}\left[ R_{c}^{aa}\right] _{\mu \mu }
= \frac{1}{d_{a}}
\pspicture[0.416666](-1.3,-0.6)(1.3,0.6)
\small
  \psset{linewidth=0.9pt,linecolor=black,arrowscale=1.5,arrowinset=0.15}
  \psarc[linewidth=0.9pt,linecolor=black] (0.7071,0.0){0.5}{-135}{135}
  \psarc[linewidth=0.9pt,linecolor=black] (-0.7071,0.0){0.5}{45}{315}
  \psline(-0.3536,0.3536)(0.3536,-0.3536)
  \psline[border=2.3pt](-0.3536,-0.3536)(0.3536,0.3536)
  \psline[border=2.3pt]{->}(-0.3536,-0.3536)(0.0,0.0)
  \rput[bl]{0}(-0.2,-0.5){$a$}
  \endpspicture
,
\end{equation}
the topological $S$-matrix
\begin{equation}
S_{ab}=\mathcal{D}^{-1} \, \widetilde{\text{Tr}}\left[ R_{ba}R_{ab}\right] =\mathcal{D}^{-1}\sum%
\limits_{c}N_{ab}^{c}\frac{\theta _{c}}{\theta _{a}\theta _{b}}d_{c}
=\frac{1}{\mathcal{D}}
\pspicture[0.363636](0.0,0.2)(2.4,1.3)
\small
  \psarc[linewidth=0.9pt,linecolor=black,arrows=<-,arrowscale=1.5,
arrowinset=0.15] (1.6,0.7){0.5}{165}{363}
  \psarc[linewidth=0.9pt,linecolor=black] (0.9,0.7){0.5}{0}{180}
  \psarc[linewidth=0.9pt,linecolor=black,border=3pt,arrows=->,arrowscale=1.5,
arrowinset=0.15] (0.9,0.7){0.5}{180}{375}
  \psarc[linewidth=0.9pt,linecolor=black,border=3pt] (1.6,0.7){0.5}{0}{160}
  \psarc[linewidth=0.9pt,linecolor=black] (1.6,0.7){0.5}{155}{170}
  \rput[bl]{0}(0.15,0.3){$a$}
  \rput[bl]{0}(2.15,0.3){$b$}
  \endpspicture
,
\end{equation}
and the monodromy scalar component
\begin{equation}
M_{ab}=\frac{\widetilde{\text{Tr}}\left[
R_{ba}R_{ab}\right]}{\widetilde{\text{Tr}} \left[ \mathbb{I}_{ab} \right]}
=\frac{1}{d_{a}d_{b}}
\pspicture[0.363636](0.0,0.2)(2.4,1.3)
\small
  \psarc[linewidth=0.9pt,linecolor=black,arrows=<-,arrowscale=1.5,
arrowinset=0.15] (1.6,0.7){0.5}{165}{363}
  \psarc[linewidth=0.9pt,linecolor=black] (0.9,0.7){0.5}{0}{180}
  \psarc[linewidth=0.9pt,linecolor=black,border=3pt,arrows=->,arrowscale=1.5,
arrowinset=0.15] (0.9,0.7){0.5}{180}{375}
  \psarc[linewidth=0.9pt,linecolor=black,border=3pt] (1.6,0.7){0.5}{0}{160}
  \psarc[linewidth=0.9pt,linecolor=black] (1.6,0.7){0.5}{155}{170}
  \rput[bl]{0}(0.15,0.3){$a$}
  \rput[bl]{0}(2.15,0.3){$b$}
  \endpspicture
=\frac{S_{ab}S_{00}}{S_{0a}S_{0b}}
\label{eq:monodromy}
.
\end{equation}

Anyonic pure states $\left| \Psi \right\rangle$ have trivial overall anyonic charge, and can have superpositions of states with different non-local topological quantum numbers, e.g. the topological charges of the different possible fusion channels of a collection of non-Abelian anyons. Localized topological charges must be definite due to superselection. For example, a general four anyon pure state would be written as
\begin{eqnarray}
\left| \Psi \right\rangle &=& \sum\limits_{c, \alpha, \beta}
\psi_{c, \alpha, \beta } \left| a_{1},a_{2};c,\alpha \right\rangle \left| c ,a_{3}; \bar{a}_{4},\beta \right\rangle \left| \bar{a}_{4},a_{4};0 \right\rangle \notag \\
&=& \sum\limits_{c, \alpha, \beta} \frac{\psi_{c, \alpha, \beta }}{\left(d_{a_{1}}d_{a_{2}}d_{a_{3}}d_{a_{4}} \right)^{1/4}}
 \pspicture[0.5](-1.5,0)(1.5,2.5)
  \small
  \psset{linewidth=0.9pt,linecolor=black,arrowscale=1.5,arrowinset=0.15}
  \psline(0.0,0.5)(1.2,2)
  \psline(0.0,0.5)(-1.2,2)
  \psline(-0.4,1)(0.4,2)
  \psline(-0.8,1.5)(-0.4,2)
    \psline{->}(-0.4,1)(0.3,1.875)
    \psline{->}(0.8,1.5)(1.1,1.875)
    \psline{->}(-0.8,1.5)(-0.5,1.875)
    \psline{->}(-0.8,1.5)(-1.1,1.875)
    \psline{->}(0,0.5)(-0.7,1.375)
  \rput[bl]{0}(-0.9,1){$c$}
  \rput[bl]{0}(-1.35,2.1){$a_{1}$}
  \rput[bl]{0}(-0.5,2.1){$a_{2}$}
  \rput[bl]{0}(0.3,2.1){$a_{3}$}
  \rput[bl]{0}(1.15,2.1){$a_{4}$}
  \scriptsize
   \rput[bl]{0}(-1.05,1.35){$\alpha$}
   \rput[bl]{0}(-0.6,0.7){$\beta$}
 \endpspicture
.
\end{eqnarray}
The normalization is such that $\left\langle \Psi | \Psi \right\rangle = \sum\limits_{c, \alpha, \beta} \left| \psi_{c, \alpha, \beta } \right|^{2} =1$.

For more general anyonic states, in particular mixed states and those with non-trivial overall topological charge, it is natural to use the density matrix formalism. It will be necessary to deal with such states whenever partial traces or superoperators (such as decoherence), which is the case when using interferometry measurements, since such effects can leave the system of interest with non-trivial overall charge. For example, a general four anyon state would be written as
\begin{equation}
\rho = \sum\limits_{\substack{ c, c^{\prime}, e, e^{\prime} f, \\ \alpha, \alpha^{\prime}, \beta, \beta^{\prime}}}
\frac{\rho_{\left(c,\alpha,e,\beta,f\right) \left(c^{\prime},\alpha^{\prime},e^{\prime},\beta^{\prime},f \right) }}{\left(d_{a_{1}} d_{a_{2}}d_{a_{3}}d_{a_{4}} d_{f} \right)^{1/2}}
 \pspicture[0.45918](-1.5,-2.4)(1.5,2.5)
  \small
  \psset{linewidth=0.9pt,linecolor=black,arrowscale=1.5,arrowinset=0.15}
  \psline(0.0,0.5)(1.2,2)
  \psline(0.0,0.5)(-1.2,2)
  \psline(-0.4,1)(0.4,2)
  \psline(-0.8,1.5)(-0.4,2)
    \psline{->}(-0.4,1)(0.3,1.875)
    \psline{->}(0.8,1.5)(1.1,1.875)
    \psline{->}(-0.8,1.5)(-0.5,1.875)
    \psline{->}(-0.8,1.5)(-1.1,1.875)
    \psline{->}(0,0.5)(-0.3,0.875)
    \psline{->}(0,0.5)(-0.7,1.375)
  \psset{linewidth=0.9pt,linecolor=black,arrowscale=1.5,arrowinset=0.15}
  \psline(0.0,-0.5)(1.2,-2)
  \psline(0.0,-0.5)(-1.2,-2)
  \psline(-0.4,-1)(0.4,-2)
  \psline(-0.8,-1.5)(-0.4,-2)
    \psline{-<}(-0.4,-1)(0.3,-1.875)
    \psline{-<}(0.8,-1.5)(1.1,-1.875)
    \psline{-<}(-0.8,-1.5)(-0.5,-1.875)
    \psline{-<}(-0.8,-1.5)(-1.1,-1.875)
    \psline{-<}(0,-0.5)(-0.3,-0.875)
    \psline{-<}(0,-0.5)(-0.7,-1.375)
  \psline(0,-0.5)(0,0.5)
  \psline{->}(0,-0.5)(0,0.1)
  \rput[bl]{0}(0.15,-0.2){$f$}
  \rput[bl]{0}(-0.9,1){$c$}
  \rput[bl]{0}(-0.5,0.5){$e$}
  \rput[bl]{0}(-1.35,2.1){$a_{1}$}
  \rput[bl]{0}(-0.5,2.1){$a_{2}$}
  \rput[bl]{0}(0.3,2.1){$a_{3}$}
  \rput[bl]{0}(1.15,2.1){$a_{4}$}
  \rput[bl]{0}(-0.95,-1.2){$c^{\prime}$}
  \rput[bl]{0}(-0.55,-0.7){$e^{\prime}$}
  \rput[bl]{0}(-1.35,-2.3){$a_{1}$}
  \rput[bl]{0}(-0.5,-2.3){$a_{2}$}
  \rput[bl]{0}(0.3,-2.3){$a_{3}$}
  \rput[bl]{0}(1.15,-2.3){$a_{4}$}
  \scriptsize
   \rput[bl]{0}(-1.05,1.35){$\alpha$}
   \rput[bl]{0}(-0.65,0.75){$\beta$}
   \rput[bl]{0}(-1.15,-1.55){$\alpha^{\prime}$}
   \rput[bl]{0}(-0.65,-1.05){$\beta^{\prime}$}
 \endpspicture
.
\label{eq:rho4}
\end{equation}
In general, when we write an anyonic state for anyons $1,\ldots,m$, we will always number the localized anyons, corresponding to the endpoints at the top and bottom of the diagrams, from left to right. For states with non-trivial overall charge, the quantum trace (and partial quantum trace) is the physical trace, so we included an extra factor $1/d_{f}$ to write the normalization condition
\begin{equation}
\widetilde{\text{Tr}}\left[ \rho \right] = \sum\limits_{c,e,f,\alpha,\beta} \rho_{\left(c,\alpha,e,\beta,f\right) \left(c,\alpha,e,\beta,f \right) } =1
.
\end{equation}
For pure states $\rho = \left| \Psi \right\rangle \left\langle \Psi \right|$, we have $\widetilde{\text{Tr}}\left[ \rho \right] = \text{Tr}\left[ \rho \right] = \left\langle \Psi |\Psi \right\rangle =1$.

  \subsection{Projective Measurement}

Given a complete set of Hermitian, orthogonal projection operators $\left\{ \Pi_{c} \right\}$ corresponding to the eigenstates of some observable, the projective measurement~\cite{vonNeumann55} of this observable for a state $\left| \Psi \right\rangle$ will have outcome $c$ with probability
\begin{equation}
\label{eq:postmeasprob}
\text{Prob}\left(c\right) = \left\langle \Psi \right| \Pi_{c} \left| \Psi \right\rangle
\end{equation}
and project (with re-normalization) the state into the corresponding subspace
\begin{equation}
\label{eq:postmeasstate}
\left| \Psi \right\rangle \mapsto \Pi_{c} \left[  \left| \Psi \right\rangle \right] = \frac{\Pi_{c} \left| \Psi \right\rangle}{\sqrt{\left\langle \Psi \right| \Pi_{c} \left| \Psi \right\rangle}}
.
\end{equation}
In an abuse of notation, we use $\Pi_{c}$ to denote both the map to the post-measurement state for outcome $c$ and the projector acting on the Hilbert space, but the meaning should be clear from context. Using the density matrix formalism to describe states, the corresponding probability and projection of the state $\rho$ for measurement outcome $c$ are
\begin{eqnarray}
\label{eq:postmeasprobrho}
&&\text{Prob}\left(c\right) = \widetilde{\text{Tr}}\left[ \Pi_{c} \rho \right] \\
\label{eq:postmeasrho}
&&\rho \mapsto \Pi_{c} \left[ \rho \right] = \frac{\Pi_{c} \rho \Pi_{c}}{\widetilde{\text{Tr}}\left[ \Pi_{c} \rho \right]}
.
\end{eqnarray}

Within the diagrammatic formalism, the projector onto definite internal vertex state $\mu$ and collective charge $c$ of two anyons $1$ and $2$ (numbered from left to right) carrying definite charges $a$ and $b$ respectively is
\begin{equation}
\Pi_{c,\mu}^{\left(12\right)} = \left| a_{1},a_{2};c, \mu \right\rangle \left\langle a_{1},a_{2};c, \mu\right|
= \sqrt{\frac{d_{c}}{d_{a_{1}}d_{a_{2}}}} \quad
 \pspicture[0.41379](0,-0.45)(1.5,1)
 \small
  \psset{linewidth=0.9pt,linecolor=black,arrowscale=1.5,arrowinset=0.15}
  \psline{->}(0.7,0)(0.7,0.45)
  \psline(0.7,0)(0.7,0.55)
  \psline(0.7,0.55) (0.25,1)
  \psline{->}(0.7,0.55)(0.3,0.95)
  \psline(0.7,0.55) (1.15,1)
  \psline{->}(0.7,0.55)(1.1,0.95)
  \rput[bl]{0}(0.38,0.2){$c$}
  \rput[br]{0}(1.55,0.75){$a_{2}$}
  \rput[bl]{0}(-0.15,0.75){$a_{1}$}
  \psline(0.7,0) (0.25,-0.45)
  \psline{-<}(0.7,0)(0.35,-0.35)
  \psline(0.7,0) (1.15,-0.45)
  \psline{-<}(0.7,0)(1.05,-0.35)
  \rput[br]{0}(1.55,-0.4){$a_{2}$}
  \rput[bl]{0}(-0.15,-0.4){$a_{1}$}
  \scriptsize
  \rput[bl]{0}(0.82,0.4){$\mu$}
  \rput[bl]{0}(0.82,0){$\mu$}
  \endpspicture
.
\end{equation}
For measurements that can distinguish the collective charge $c$, but not the different internal vertex states, the projector is given by
\begin{eqnarray}
\Pi_{c}^{\left(12\right)} &=& \left| a_{1},a_{2};c \right\rangle \left\langle a_{1},a_{2};c\right| \notag \\
&=& \sum_{\mu = 1}^{N_{ a_{1} a_{2} }^{c}} \left| a_{1},a_{2};c, \mu \right\rangle \left\langle a_{1},a_{2};c, \mu\right|
= \sum_{\mu = 1}^{N_{ a_{1} a_{2} }^{c}} \sqrt{\frac{d_{c}}{d_{a_{1}}d_{a_{2}}}} \quad
 \pspicture[0.41379](0,-0.45)(1.5,1)
 \small
  \psset{linewidth=0.9pt,linecolor=black,arrowscale=1.5,arrowinset=0.15}
  \psline{->}(0.7,0)(0.7,0.45)
  \psline(0.7,0)(0.7,0.55)
  \psline(0.7,0.55) (0.25,1)
  \psline{->}(0.7,0.55)(0.3,0.95)
  \psline(0.7,0.55) (1.15,1)
  \psline{->}(0.7,0.55)(1.1,0.95)
  \rput[bl]{0}(0.38,0.2){$c$}
  \rput[br]{0}(1.55,0.75){$a_{2}$}
  \rput[bl]{0}(-0.15,0.75){$a_{1}$}
  \psline(0.7,0) (0.25,-0.45)
  \psline{-<}(0.7,0)(0.35,-0.35)
  \psline(0.7,0) (1.15,-0.45)
  \psline{-<}(0.7,0)(1.05,-0.35)
  \rput[br]{0}(1.55,-0.4){$a_{2}$}
  \rput[bl]{0}(-0.15,-0.4){$a_{1}$}
  \scriptsize
  \rput[bl]{0}(0.82,0.4){$\mu$}
  \rput[bl]{0}(0.82,0){$\mu$}
  \endpspicture
.
\end{eqnarray}
In the rest of this paper, we will only consider anyon models with no fusion multiplicities (i.e. the fusion coefficients are either $N_{ab}^{c}=0$ or $1$, depending on whether the fusion is allow), so we can drop the internal vertex label. This subset encompasses all of the most physically relevant anyon models, and generalizing the methods and results to apply when there are fusion multiplicities is straightforward, so this is not a severe restriction.

The projector onto collective topological charge $c$ of $n$ anyons is
\begin{eqnarray}
\Pi_{c}^{\left( 1 \ldots n \right)} &=& \sum_{c_{2},\ldots,c_{n-1}}
\left| a_{1},a_{2};c_{2} \right\rangle \left| c_{2},a_{3};c_{3} \right\rangle \ldots \left| c_{n-1},a_{n};c \right\rangle \notag \\
&& \qquad \qquad \times \left\langle c_{n-1},a_{n};c\right| \ldots \left\langle c_{2},a_{3};c_{3}\right| \left\langle a_{1},a_{2};c_{2}\right| \notag \\
&=& \sum_{c_{2},\ldots,c_{n-1}} \sqrt{\frac{d_{c}}{d_{a_{1}} \ldots d_{a_{n}} }}
 \pspicture[0.4659](-0.35,-2)(2.5,2.4)
  \small
  \psset{linewidth=0.9pt,linecolor=black,arrowscale=1.5,arrowinset=0.15}
  \psline(0.0,1.75)(1,0.5)
  \psline(2.0,1.75)(1,0.5)
  \psline(0.4,1.25)(0.8,1.75)
   \psline{->}(0.4,1.25)(0.1,1.625)
   \psline{->}(0.4,1.25)(0.7,1.625)
   \psline{->}(1,0.5)(1.9,1.625)
   \psline{->}(1,0.5)(0.5,1.125)
   \rput[bl]{0}(-0.15,1.85){$a_1$}
   \rput[bl]{0}(0.75,1.85){$a_2$}
   \rput[bl]{0}(1.95,1.85){$a_n$}
\rput[bl](1.25,1.85){$\cdots$}
\rput{-45}(0.9,1.05){$\cdots$}
   \rput[bl]{0}(0.25,0.65){$c_2$}
  \psset{linewidth=0.9pt,linecolor=black,arrowscale=1.5,arrowinset=0.15}
  \psline(0.0,-1.45)(1,-0.2)
  \psline(2.0,-1.45)(1,-0.2)
  \psline(0.4,-0.95)(0.8,-1.45)
   \psline{-<}(0.4,-0.95)(0.1,-1.325)
   \psline{-<}(0.4,-0.95)(0.7,-1.325)
   \psline{-<}(1,-0.2)(1.9,-1.325)
   \psline{-<}(1,-0.2)(0.5,-0.825)
   \rput[bl]{0}(-0.15,-1.85){$a_1$}
   \rput[bl]{0}(0.75,-1.85){$a_2$}
   \rput[bl]{0}(1.95,-1.85){$a_n$}
   \rput[bl]{0}(0.25,-0.65){$c_2$}
   \rput[bl](1.25,-1.85){$\cdots$}
   \rput{45}(0.9,-0.75){$\cdots$}
  \psline(1,-0.2)(1,0.5)
  \psline{->}(1,-0.2)(1,0.3)
   \rput[bl]{0}(1.15,0.0){$c$}
  \endpspicture
.
\label{eq:PIcn}
\end{eqnarray}

When one writes operators, such as these topological charge projectors, in anyonic systems, one must also specify their topological configuration with respect to other anyons in the system, which is determined by the physical process they represent. This configuration information is explicitly contained in the diagrammatic representation of operators, but left explicit in the symbolic shorthand (e.g. $\Pi_{c}$) used to represent it. For example, when considering four anyons, we write
\begin{equation}
\label{eq:Piover}
\Pi_{c}^{\left( 14\right)} = \sqrt{ \frac{d_{c}}{d_{a_{1}} d_{a_{4}} }}
 \pspicture[0.45918](-1.5,-2.4)(1.5,2.5)
  \small
  \psset{linewidth=0.9pt,linecolor=black,arrowscale=1.5,arrowinset=0.15}
  \psline(0.0,0.5)(1.2,2)
  \psline(0.0,0.5)(-1.2,2)
  \psline(0.4,-2)(0.4,2)
  \psline(-0.4,-2)(-0.4,2)
    \psline{->}(0.4,1.5)(0.4,1.875)
    \psline{->}(0.8,1.5)(1.1,1.875)
    \psline{->}(-0.4,1.5)(-0.4,1.875)
    \psline{->}(-0.8,1.5)(-1.1,1.875)
  \psset{linewidth=0.9pt,linecolor=black,arrowscale=1.5,arrowinset=0.15}
  \psline(0.0,-0.5)(1.2,-2)
  \psline(0.0,-0.5)(-1.2,-2)
    \psline{-<}(0.8,-1.5)(1.1,-1.875)
    \psline{-<}(-0.8,-1.5)(-1.1,-1.875)
    \psline{-<}(0.4,-1.5)(0.4,-1.875)
    \psline{-<}(-0.4,-1.5)(-0.4,-1.875)
  \psline(0,-0.5)(0,0.5)
  \psline{->}(0,-0.5)(0,0.1)
  \psline[border=1.5pt](-0.4,-1.2)(-0.4,-0.8)
  \psline[border=1.5pt](-0.4,1.2)(-0.4,0.8)
  \psline[border=1.5pt](0.4,-1.2)(0.4,-0.8)
  \psline[border=1.5pt](0.4,1.2)(0.4,0.8)
  \rput[bl]{0}(0.15,-0.2){$c$}
  \rput[bl]{0}(-1.3,2.1){$a_{1}$ \,\,\, $a_{2}$ \,\,\, $a_{3}$ \,\,\, $a_{4}$ }
  \rput[bl]{0}(-1.3,-2.3){$a_{1}$ \,\,\, $a_{2}$ \,\,\, $a_{3}$ \,\,\, $a_{4}$ }
 \endpspicture
\end{equation}

\begin{figure}[t!]
\begin{center}
  \includegraphics[scale=.9]{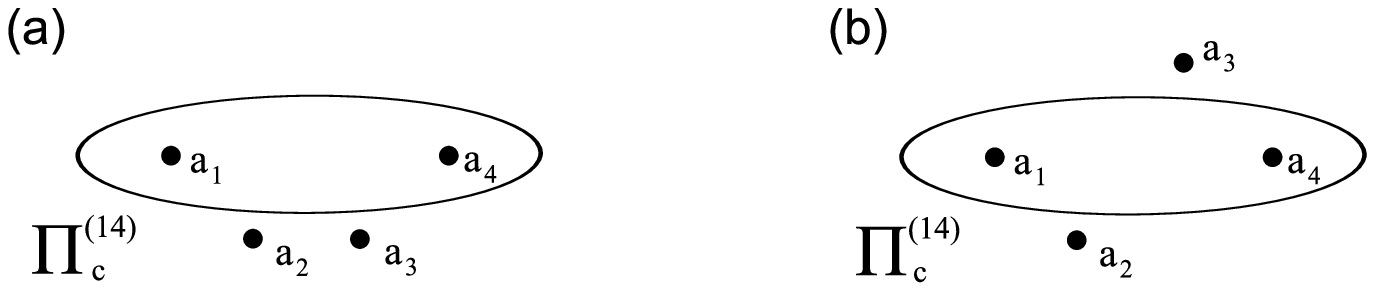}
  \caption{Topological aspects of an operator determined by the spatial configuration of a physical process are naturally encoded in their diagrammatic representation. Two topologically inequivalent spatial configuration for measuring the collective topological charge of anyons $1$ and $4$ in the presence of anyons $2$ and $3$ are shown here, with (a) corresponding to Eq.~(\ref{eq:Piover}), and (b) corresponding to Eq.~(\ref{eq:Piunder}). The ovals delineate areas inside which the collective topological charge of anyons is being measured. \newline}
  \label{fig:Pioverunder}
\end{center}
\end{figure}

to represent the projection corresponding to a charge measurement with the spatial configuration in Fig.~\ref{fig:Pioverunder}(a), whereas
\begin{equation}
\label{eq:Piunder}
\Pi_{c}^{\left( 14\right)} = \sqrt{ \frac{d_{c}}{d_{a_{1}} d_{a_{4}} }}
 \pspicture[0.45918](-1.5,-2.4)(1.5,2.5)
  \small
  \psset{linewidth=0.9pt,linecolor=black,arrowscale=1.5,arrowinset=0.15}
  \psline(0.0,0.5)(1.2,2)
  \psline(0.0,0.5)(-1.2,2)
  \psline(0.4,-2)(0.4,2)
  \psline(-0.4,-2)(-0.4,2)
    \psline{->}(0.4,1.5)(0.4,1.875)
    \psline{->}(0.8,1.5)(1.1,1.875)
    \psline{->}(-0.4,1.5)(-0.4,1.875)
    \psline{->}(-0.8,1.5)(-1.1,1.875)
  \psset{linewidth=0.9pt,linecolor=black,arrowscale=1.5,arrowinset=0.15}
  \psline(0.0,-0.5)(1.2,-2)
  \psline(0.0,-0.5)(-1.2,-2)
    \psline{-<}(0.8,-1.5)(1.1,-1.875)
    \psline{-<}(-0.8,-1.5)(-1.1,-1.875)
    \psline{-<}(0.4,-1.5)(0.4,-1.875)
    \psline{-<}(-0.4,-1.5)(-0.4,-1.875)
  \psline(0,-0.5)(0,0.5)
  \psline{->}(0,-0.5)(0,0.1)
  \psline[border=1.5pt](-0.4,-1.2)(-0.4,-0.8)
  \psline[border=1.5pt](-0.4,1.2)(-0.4,0.8)
  \psline[border=1.5pt](0.6,-1.25)(0.2,-0.75)
  \psline[border=1.5pt](0.6,1.25)(0.2,0.75)
  \rput[bl]{0}(0.15,-0.2){$c$}
  \rput[bl]{0}(-1.3,2.1){$a_{1}$ \,\,\, $a_{2}$ \,\,\, $a_{3}$ \,\,\, $a_{4}$ }
  \rput[bl]{0}(-1.3,-2.3){$a_{1}$ \,\,\, $a_{2}$ \,\,\, $a_{3}$ \,\,\, $a_{4}$ }
 \endpspicture
\end{equation}
corresponds to the spatial configuration in Fig.~\ref{fig:Pioverunder}(b). Clearly the two corresponding spatial configurations can be inferred from the diagrammatic representation of these operators. When the topological configuration is left implicit, we typically assume the obvious simplest configuration (i.e. the planar diagrams arising from direct paths), but when its important we will explicitly describe the intended spatial configuration.

Specifying a region inside which the collective topological charge is measured, such as in Fig.~\ref{fig:Pioverunder}, is like specifying a path through which the measured anyons are brought together to fuse into a single definite collective charge, and then separated and returned back to their original positions. In fact, given the ability to somehow split the anyons into their original configuration of localized anyonic charges after fusing them, this would be one possible way to perform such a charge measurement. However, there are other ways to measure the collective charge of anyons without actually fusing them, or even moving them at all, such as interferometrical measurement. Hence, the spatial configuration of a measurement gives the projection operator a path-like quality, even if the measured anyons are not actually moved. We emphasize that this means the diagrams in these projection operators should not be interpreted as the worldlines of the anyons. Rather, these diagrams only represent the fact that the anyonic state is somehow being projected into a subspace which has a definite combined topological charge. If one were to fuse anyons together, but not split them up again, this would produce a single anyon of definite topological charge $c$ with $\text{Prob}\left(c\right)$ as given in Eqs. (\ref{eq:postmeasprob},\ref{eq:postmeasprobrho}), but the projection is instead into a state space with fewer anyons.

The physical measurement in topological systems that give rise to projective topological charge measurements of this sort potentially include: Wilson loop measurements (enclosing the measured quasiparticles) in lattice models, energy splitting measurements in fractional quantum Hall (FQH) and possibly other systems, and (the asymptotic limit of) interferometry measurements when the collective charge measurement outcome $c$ is an Abelian charge (see Section~\ref{sec:Interferometry_Measurement} for more on this). While motion (or something related to it) of quasiparticles may still occur in such measurements, it is something manifestly different from the braiding of the computational anyons. In particular, while Wilson loop measurements can be related to moving quasiparticles, does not move the quasiparticles enclosed inside the loop; the motion for energy splitting measurements simply brings the quasiparticles into closer proximity and then returns them to their original location (without braiding of any sort); and for interferometry, there is of course a beam of moving probe quasiparticles interfering around the measured quasiparticles, but the measured quasiparticles are not moved at all.

When performing topological charge measurements, one must be careful to avoid carrying them out in a manner or configuration that results in undesired effects on the anyonic charge correlations of the system, such as the introduction of unintentional charge entanglement or decoherence of charge entanglement that encodes relevant information.

\subsection{Interferometry Measurement}
\label{sec:Interferometry_Measurement}

In contrast with projective measurement, interferometrical measurement of topological charge is not quite as simple and requires a density matrix formulation. We will assume the anyonic interferometer is of one of the two forms discussed in Ref.~\cite{Bonderson07b,Bonderson07c}, i.e. either an idealized Mach--Zehnder or Fabry--P\'{e}rot type. The Fabry--P\'{e}rot type interferometer has been experimentally realized in FQH systems using the double-point contact geometry in the weak tunneling regime~\cite{Camino05a,Camino07a,Willett08}. Though a form of Mach--Zehnder interferometer has been experimentally realized in FQH systems~\cite{Ji03}, unfortunately the construction of such interferometers in FQH systems unavoidably requires one of the detectors and drains to be situated inside the central interferometry region. This results in the accumulation of probe anyons which changes the topological charge contained inside this region, rendering it incapable of measuring a target charge, and hence useless for topological quantum computation (in any form).

To describe the effects of interferometry, we split the anyons of the system into three categories: the target anyons $1,\ldots,n$, located inside the interferometry region (in Ref.~\cite{Bonderson07b,Bonderson07c} these are collectively written as $A$); the (non-probe) anyons $n+1,\ldots,m$, located outside the interferometry region (in Ref.~\cite{Bonderson07b,Bonderson07c} these are collectively written as $C$); and $N$ probe anyons, which we will not number (in Ref.~\cite{Bonderson07b,Bonderson07c} each of these is written as $B_{j}$). We assume the probe anyons initially have no entanglement with the anyons $1,\ldots,m$, nor with each other. (This means that each probe anyon really has at least one other anyon with which it is initially entangled, giving a fourth category of anyons that we neglect because they will not interact with anything and are simply traced out.)

The initial state of each probe anyon is described by a density matrix, which can effectively be written as
\begin{equation}
\rho^{B} = \sum_{b} \frac{ p_{b} }{ d_{b} } \left| b \right\rangle \left\langle b \right|
= \sum_{b} \frac{ p_{b} }{ d_{b} }
\pspicture[0.47826](0,-1.15)(0.9,1.15)
  \small
  \psset{linewidth=0.9pt,linecolor=black,arrowscale=1.5,arrowinset=0.15}
  \psline(0.35,-0.775)(0.35,0.775)
  \psline{->}(0.35,-0.3)(0.35,0.2)
  \rput[bl]{0}(0.5,-0.1){$b$}
\endpspicture
,
\end{equation}
where $p_{b}$ is the probability that the probe has topological charge $b$.

The collective state of anyons $1,\ldots,m$ is described by the density matrix $\rho^{A}$, which can be written (using $F$-moves if necessary) as a linear combination of the basis elements
\begin{eqnarray}
\label{eq:rhoAbasis}
&&\left| a_{1},a_{2};g_{2} \right\rangle \left| g_{2},a_{3};g_{3} \right\rangle \ldots \left| g_{n-1},a_{n};a_{\text{in}} \right\rangle \notag \\
&& \quad \left| a_{n+1},a_{n+2};h_{2} \right\rangle \left| h_{2},a_{n+3};h_{3} \right\rangle \ldots \left| h_{m-n-1},a_{m};a_{\text{ex}} \right\rangle \left| a_{\text{in}},a_{\text{ex}};f \right\rangle \notag \\
&& \qquad \qquad \times \left\langle a_{\text{in}}^{\prime},a_{\text{ex}}^{\prime};f\right| \left\langle g_{n-1}^{\prime},a_{n}^{\prime};a_{\text{in}}\right| \ldots \left\langle g_{2}^{\prime},a_{3}^{\prime};g_{3}^{\prime}\right| \left\langle a_{1}^{\prime},a_{2}^{\prime};g_{2}^{\prime}\right| \notag \\
&& \qquad \qquad \qquad \left\langle h_{m-n-1}^{\prime},a_{m};a_{\text{ex}}\right| \ldots \left\langle h_{2}^{\prime},a_{n+3};h_{3}^{\prime}\right| \left\langle a_{n+1}^{\prime},a_{n+2}^{\prime};h_{2}^{\prime}\right| \notag \\
&& \qquad \qquad = \sqrt{\frac{d_{f}}{d_{a_{1}} \ldots d_{a_{m}} }}
 \pspicture[0.45918](-1.5,-2.4)(1.5,2.5)
  \small
  \psset{linewidth=0.9pt,linecolor=black,arrowscale=1.5,arrowinset=0.15}
  \psline(0.0,0.5)(1.2,2)
  \psline(0.0,0.5)(-1.2,2)
  \psline(0.8,1.5)(0.4,2)
  \psline(-0.8,1.5)(-0.4,2)
    \psline{->}(0.8,1.5)(0.5,1.875)
    \psline{->}(0.8,1.5)(1.1,1.875)
    \psline{->}(-0.8,1.5)(-0.5,1.875)
    \psline{->}(-0.8,1.5)(-1.1,1.875)
    \psline{->}(0,0.5)(0.5,1.125)
    \psline{->}(0,0.5)(-0.5,1.125)
  \psset{linewidth=0.9pt,linecolor=black,arrowscale=1.5,arrowinset=0.15}
  \psline(0.0,-0.5)(1.2,-2)
  \psline(0.0,-0.5)(-1.2,-2)
  \psline(0.8,-1.5)(0.4,-2)
  \psline(-0.8,-1.5)(-0.4,-2)
    \psline{-<}(0.8,-1.5)(0.5,-1.875)
    \psline{-<}(0.8,-1.5)(1.1,-1.875)
    \psline{-<}(-0.8,-1.5)(-0.5,-1.875)
    \psline{-<}(-0.8,-1.5)(-1.1,-1.875)
    \psline{-<}(0,-0.5)(0.5,-1.125)
    \psline{-<}(0,-0.5)(-0.5,-1.125)
  \psline(0,-0.5)(0,0.5)
  \psline{->}(0,-0.5)(0,0.1)
  \rput[bl]{0}(0.15,-0.2){$f$}
  \rput[bl]{0}(0.5,0.7){$a_{\text{ex}}$}
  \rput[bl]{0}(-0.95,0.7){$a_{\text{in}}$}
  \rput[bl]{0}(0.5,-0.95){$a_{\text{ex}}^{\prime}$}
  \rput[bl]{0}(-0.95,-0.95){$a_{\text{in}}^{\prime}$}
  \scriptsize
  \rput[bl]{0}(-1.35,2.1){$a_{1}$ $\ldots$ $a_{n}$ \,\, $a_{n+1}$ $\ldots$ $a_{m}$ }
  \rput[bl]{0}(-1.35,-2.3){$a_{1}$ $\ldots$ $a_{n}$ \,\, $a_{n+1}$ $\ldots$ $a_{m}$ }
 \endpspicture
.
\end{eqnarray}
In these basis elements, $a_{\text{in}}$ is the collective charge in the ket of anyons $1,\ldots,n$, which are in the interior region of the interferometer, and $a_{\text{ex}}$ is the collective charge in the ket of anyons $n+1,\ldots,m$, which are in the exterior region of the interferometer. The charges in the respective bras are denoted with primes (since they can take different values), with the except of the charges $a_{j}$ localized on the individual anyons, which must have definite values. We emphasize that the density matrix $\rho^{A}$ can have anyons $1,\ldots,m$ in an overall collective charge $f$, which represents ``hidden'' anyons that are not being kept track of with which these have anyonic charge entanglement. This means superpositions of $f$ can only be incoherent (i.e. classical probabilities). We also note that if this overall charge is trivial $f=0$, then $a_{\text{ex}} = \bar{a}_{\text{in}}$ and $a_{\text{ex}}^{\prime} = \bar{a}^{\prime}_{\text{in}}$.

To obtain the transformation of the density matrix $\rho^{A}$ after a single probe measurement, one performs the following series of operations: tensor the density matrix of the probe with $\rho^{A}$, apply the unitary evolution operator $U$ corresponding to the probe passing through the interferometer, apply the projection operator (and re-normalize) corresponding to which outcome was observed by the detectors, and finally trace out the probe (representing the fact that the probe is ``discarded'' after the measurement and no longer kept track of). For the measurement in which the probe is observed at detector $s$, this is described by
\begin{equation}
\rho^{A} \mapsto \frac{\widetilde{\text{Tr}}_{B}\left[ \Pi_{s} U \left( \rho^{A} \otimes \rho^{B} \right) U^{\dagger} \Pi_{s}  \right] }{\widetilde{\text{Tr}}\left[ \Pi_{s} U \left( \rho^{A} \otimes \rho^{B} \right) U^{\dagger} \Pi_{s}  \right] }
.
\end{equation}
This gives a POVM measurement of $\rho^{A}$ which has the effect of partially projecting the collective charge of anyons $1,\ldots,n$ and partially decohering their anyonic charge entanglement with anyons $n+1,\ldots,m$. It may be represented by multiplying the basis elements in Eq.~(\ref{eq:rhoAbasis}) by appropriate factors (see Refs.~\cite{Bonderson07b,Bonderson07c} for details). We are not so concerned with the details of the transformation of $\rho^{A}$ that results from sending a single probe through the interferometer, but rather are particularly interested in the effect on the state $\rho^{A}$ after many probes have passed through the interferometer. In the asymptotic limit of sending $N \rightarrow \infty$ probe anyons through the interferometer, the effect on $\rho^{A}$ may effectively be treated as the simultaneous projection of the collective charge of anyons $1,\ldots,n$ and the decoherence of their anyonic charge entanglement with anyons $n+1,\ldots,m$.

To be more precise, we focus on the part of Eq.~(\ref{eq:rhoAbasis}) that encodes the anyonic entanglement between the collective charge of anyons $1,\ldots,n$ and the collective charge of anyons $n+1,\ldots,m$, and apply an $F$-move to it
\begin{eqnarray}
&&\left| a_{\text{in}},a_{\text{ex}};f \right\rangle \left\langle a_{\text{in}}^{\prime},a_{\text{ex}}^{\prime};f\right| = \frac{d_{f}^{1/2}} {\left( d_{a_{\text{in}}} d_{a_{\text{ex}}} d_{a_{\text{in}}^{\prime}} d_{a_{\text{ex}}^{\prime}} \right)^{1/4} }
 \pspicture[0.4375](-0.8,-1.3)(0.6,1.1)
  \small
  \psset{linewidth=0.9pt,linecolor=black,arrowscale=1.5,arrowinset=0.15}
  \psline(0.0,-0.25)(0.0,0.25)
  \psline(0.0,0.25)(0.4,0.75)
  \psline(0.0,0.25)(-0.4,0.75)
  \psline(0.0,-0.25)(0.4,-0.75)
  \psline(0.0,-0.25)(-0.4,-0.75)
    \psline{->}(0.0,-0.25)(0.0,0.125)
    \psline{->}(0.0,0.25)(0.3,0.625)
    \psline{->}(0.0,0.25)(-0.3,0.625)
    \psline{-<}(0.0,-0.25)(0.3,-0.625)
    \psline{-<}(0.0,-0.25)(-0.3,-0.625)
  \rput[bl]{0}(0.15,-0.2){$f$}
  \rput[bl]{0}(0.3,0.85){$a_{\text{ex}}$}
  \rput[bl]{0}(-0.6,0.85){$a_{\text{in}}$}
  \rput[bl]{0}(0.3,-1.25){$a_{\text{ex}}^{\prime}$}
  \rput[bl]{0}(-0.6,-1.25){$a_{\text{in}}^{\prime}$}
 \endpspicture
\\
&& \qquad \qquad \qquad = \sum_{e} \frac{d_{f}^{1/2}} {\left( d_{a_{\text{in}}} d_{a_{\text{ex}}} d_{a_{\text{in}}^{\prime}} d_{a_{\text{ex}}^{\prime}} \right)^{1/4} } \left[ \left( F_{a_{\text{in}}^{\prime} a_{\text{ex}}^{\prime}}^{a_{\text{in}}a_{\text{ex}}} \right)^{-1} \right]_{fe}
 \pspicture[0.4375](-0.8,-1.3)(0.6,1.1)
  \small
  \psset{linewidth=0.9pt,linecolor=black,arrowscale=1.5,arrowinset=0.15}
  \psline(-0.4,-0.75)(-0.4,0.75)
  \psline(0.4,-0.75)(0.4,0.75)
  \psline(0.4,-0.25)(-0.4,0.25)
    \psline{->}(-0.4,-0.75)(-0.4,-0.3)
    \psline{->}(-0.4,-0.75)(-0.4,0.6)
    \psline{->}(0.4,-0.75)(0.4,-0.3)
    \psline{->}(0.4,-0.75)(0.4,0.6)
    \psline{->}(0.4,-0.25)(-0.1,0.0625)
  \rput[bl]{0}(-0.1,0.2){$e$}
  \rput[bl]{0}(0.3,0.85){$a_{\text{ex}}$}
  \rput[bl]{0}(-0.6,0.85){$a_{\text{in}}$}
  \rput[bl]{0}(0.3,-1.25){$a_{\text{ex}}^{\prime}$}
  \rput[bl]{0}(-0.6,-1.25){$a_{\text{in}}^{\prime}$}
 \endpspicture
.
\end{eqnarray}
Multiple anyons are allowed superpositions of their collective charge, so the density matrix will generally be a linear combination of elements like these. Written in this form, the anyonic charge entanglement between the anyons $1,\ldots,n$ in the interior of the interferometer and anyons $n+1,\ldots,m$ in the exterior of the interferometer is reduced to a particularly clean description in terms of the single charge $e$ line. These clusters of anyons are said to have no anyonic charge entanglement between them if $e=0$ is the only charge that has non-zero density matrix coefficients.

One now must determine what charges the probe is capable of distinguishing by monodromy (the full $2\pi$ winding of one anyon around another). For this, we consider the expectation value
\begin{equation}
M_{aB}= \sum_{b} p_{b} M_{ab}
\end{equation}
of the scalar component of the monodromy of the charges $a$ and $b$
\begin{equation}
M_{ab}= \frac{\widetilde{\text{Tr}} \left[ R_{ba}R_{ab}\right]}{\widetilde{\text{Tr}} \mathbb{I}_{ab} } = \frac{S_{ab} S_{00}}{S_{0a}S_{0b}}.
\end{equation}
We use this to sort the anyonic charges into classes that the probe anyons can distinguish via monodromy, i.e. all charges $a$ with the same $M_{aB}$ as a particular $a_{\kappa}$ compose the class
\begin{equation}
\mathcal{C}_{\kappa} = \left\{ a \in \mathcal{C} :  M_{aB}= M_{a_{\kappa} B} \right\}
.
\end{equation}
We define the projection operator onto the charge class $\mathcal{C}_{\kappa}$ as
\begin{equation}
\Pi_{\mathcal{C}_{\kappa}} = \sum_{c \in \mathcal{C}_{\kappa}}\Pi_{c}
.
\end{equation}

The effect of interferometry measurement in the asymptotic limit will observe the charge class $\mathcal{C}_{\kappa}$ with probability
\begin{equation}
\text{Prob}_{A}\left( \kappa \right) = \widetilde{\text{Tr}} \left[ \Pi_{\mathcal{C}_{\kappa}}^{\left(1 \ldots n \right)} \rho^{A} \right]
,
\end{equation}
for which it will transform the state $\rho^{A}$ in a way that may be described by applying the following transformation to the basis elements:
\begin{equation}
\Delta_{B;\kappa}^{\left(1 \ldots n \right)} :
 \pspicture[0.4375](-0.8,-1.3)(0.6,1.1)
  \small
  \psset{linewidth=0.9pt,linecolor=black,arrowscale=1.5,arrowinset=0.15}
  \psline(-0.4,-0.75)(-0.4,0.75)
  \psline(0.4,-0.75)(0.4,0.75)
  \psline(0.4,-0.25)(-0.4,0.25)
    \psline{->}(-0.4,-0.75)(-0.4,-0.3)
    \psline{->}(-0.4,-0.75)(-0.4,0.6)
    \psline{->}(0.4,-0.75)(0.4,-0.3)
    \psline{->}(0.4,-0.75)(0.4,0.6)
    \psline{->}(0.4,-0.25)(-0.1,0.0625)
  \rput[bl]{0}(-0.1,0.2){$e$}
  \rput[bl]{0}(0.3,0.85){$a_{\text{ex}}$}
  \rput[bl]{0}(-0.6,0.85){$a_{\text{in}}$}
  \rput[bl]{0}(0.3,-1.25){$a_{\text{ex}}^{\prime}$}
  \rput[bl]{0}(-0.6,-1.25){$a_{\text{in}}^{\prime}$}
 \endpspicture
\mapsto \Delta_{a_{\text{in}} a_{\text{in}}^{\prime} e,B}\left(\kappa\right)
 \pspicture[0.4375](-0.8,-1.3)(0.6,1.1)
  \small
  \psset{linewidth=0.9pt,linecolor=black,arrowscale=1.5,arrowinset=0.15}
  \psline(-0.4,-0.75)(-0.4,0.75)
  \psline(0.4,-0.75)(0.4,0.75)
  \psline(0.4,-0.25)(-0.4,0.25)
    \psline{->}(-0.4,-0.75)(-0.4,-0.3)
    \psline{->}(-0.4,-0.75)(-0.4,0.6)
    \psline{->}(0.4,-0.75)(0.4,-0.3)
    \psline{->}(0.4,-0.75)(0.4,0.6)
    \psline{->}(0.4,-0.25)(-0.1,0.0625)
  \rput[bl]{0}(-0.1,0.2){$e$}
  \rput[bl]{0}(0.3,0.85){$a_{\text{ex}}$}
  \rput[bl]{0}(-0.6,0.85){$a_{\text{in}}$}
  \rput[bl]{0}(0.3,-1.25){$a_{\text{ex}}^{\prime}$}
  \rput[bl]{0}(-0.6,-1.25){$a_{\text{in}}^{\prime}$}
 \endpspicture
,
\end{equation}
where
\begin{equation}
\Delta_{a_{\text{in}} a_{\text{in}}^{\prime} e,B}\left(\kappa\right) = \left\{
\begin{array}{ll}
\frac{1}{\text{Prob}_{A}\left( \kappa \right) } \quad & \text{if } a_{\text{in}},a_{\text{in}}^{\prime }\in C_{\kappa } \text{ and
} M_{eB}=1\\
0 \quad & \text{otherwise}%
\end{array}%
\right. .
\end{equation}

To elucidate this result and highlight how it differs from projective measurement, we can split the effect into the two parts: projective measurement and decoherence of anyonic charge entanglement. A projective measurement of the collective charge of the anyons $1,\ldots,n$ that observes the charge class $\mathcal{C}_{\kappa}$ [i.e. application of Eqs.~(\ref{eq:postmeasprobrho},\ref{eq:postmeasrho}) using $\Pi_{\mathcal{C}_{\kappa}}^{\left( 1 \ldots n \right)}$] transforms $\rho^{A}$ in a way that may be described by
\begin{equation}
\Pi_{\mathcal{C}_{\kappa}}^{\left(1 \ldots n \right)} :
 \pspicture[0.4375](-0.8,-1.3)(0.6,1.1)
  \small
  \psset{linewidth=0.9pt,linecolor=black,arrowscale=1.5,arrowinset=0.15}
  \psline(-0.4,-0.75)(-0.4,0.75)
  \psline(0.4,-0.75)(0.4,0.75)
  \psline(0.4,-0.25)(-0.4,0.25)
    \psline{->}(-0.4,-0.75)(-0.4,-0.3)
    \psline{->}(-0.4,-0.75)(-0.4,0.6)
    \psline{->}(0.4,-0.75)(0.4,-0.3)
    \psline{->}(0.4,-0.75)(0.4,0.6)
    \psline{->}(0.4,-0.25)(-0.1,0.0625)
  \rput[bl]{0}(-0.1,0.2){$e$}
  \rput[bl]{0}(0.3,0.85){$a_{\text{ex}}$}
  \rput[bl]{0}(-0.6,0.85){$a_{\text{in}}$}
  \rput[bl]{0}(0.3,-1.25){$a_{\text{ex}}^{\prime}$}
  \rput[bl]{0}(-0.6,-1.25){$a_{\text{in}}^{\prime}$}
 \endpspicture
\mapsto \delta_{a_{\text{in}} a_{\text{in}}^{\prime}}\left(\kappa\right)
 \pspicture[0.4375](-0.8,-1.3)(0.6,1.1)
  \small
  \psset{linewidth=0.9pt,linecolor=black,arrowscale=1.5,arrowinset=0.15}
  \psline(-0.4,-0.75)(-0.4,0.75)
  \psline(0.4,-0.75)(0.4,0.75)
  \psline(0.4,-0.25)(-0.4,0.25)
    \psline{->}(-0.4,-0.75)(-0.4,-0.3)
    \psline{->}(-0.4,-0.75)(-0.4,0.6)
    \psline{->}(0.4,-0.75)(0.4,-0.3)
    \psline{->}(0.4,-0.75)(0.4,0.6)
    \psline{->}(0.4,-0.25)(-0.1,0.0625)
  \rput[bl]{0}(-0.1,0.2){$e$}
  \rput[bl]{0}(0.3,0.85){$a_{\text{ex}}$}
  \rput[bl]{0}(-0.6,0.85){$a_{\text{in}}$}
  \rput[bl]{0}(0.3,-1.25){$a_{\text{ex}}^{\prime}$}
  \rput[bl]{0}(-0.6,-1.25){$a_{\text{in}}^{\prime}$}
 \endpspicture
,
\end{equation}
where
\begin{equation}
\delta_{a_{\text{in}} a_{\text{in}}^{\prime} }\left(\kappa\right) = \left\{
\begin{array}{ll}
\frac{1}{\text{Prob}_{A}\left( \kappa \right) } \quad & \text{if } a_{\text{in}},a_{\text{in}}^{\prime }\in C_{\kappa }\\
0 \quad & \text{otherwise}%
\end{array}%
\right. .
\end{equation}
The decoherence of the anyonic charge entanglement between the collective charge of anyons $1,\ldots,n$ and that of anyons $n+1,\ldots,m$ is described by the superoperator $D_{B}^{\left(1 \ldots n\right)}$. The transformation $\rho^{A} \mapsto D_{B}^{\left(1 \ldots n\right)} \left( \rho^{A} \right)$ may be described by the basis element transformation
\begin{equation}
D_{B}^{\left(1 \ldots n \right)} :
 \pspicture[0.4375](-0.8,-1.3)(0.6,1.1)
  \small
  \psset{linewidth=0.9pt,linecolor=black,arrowscale=1.5,arrowinset=0.15}
  \psline(-0.4,-0.75)(-0.4,0.75)
  \psline(0.4,-0.75)(0.4,0.75)
  \psline(0.4,-0.25)(-0.4,0.25)
    \psline{->}(-0.4,-0.75)(-0.4,-0.3)
    \psline{->}(-0.4,-0.75)(-0.4,0.6)
    \psline{->}(0.4,-0.75)(0.4,-0.3)
    \psline{->}(0.4,-0.75)(0.4,0.6)
    \psline{->}(0.4,-0.25)(-0.1,0.0625)
  \rput[bl]{0}(-0.1,0.2){$e$}
  \rput[bl]{0}(0.3,0.85){$a_{\text{ex}}$}
  \rput[bl]{0}(-0.6,0.85){$a_{\text{in}}$}
  \rput[bl]{0}(0.3,-1.25){$a_{\text{ex}}^{\prime}$}
  \rput[bl]{0}(-0.6,-1.25){$a_{\text{in}}^{\prime}$}
 \endpspicture
\mapsto D_{e,B}
 \pspicture[0.4375](-0.8,-1.3)(0.6,1.1)
  \small
  \psset{linewidth=0.9pt,linecolor=black,arrowscale=1.5,arrowinset=0.15}
  \psline(-0.4,-0.75)(-0.4,0.75)
  \psline(0.4,-0.75)(0.4,0.75)
  \psline(0.4,-0.25)(-0.4,0.25)
    \psline{->}(-0.4,-0.75)(-0.4,-0.3)
    \psline{->}(-0.4,-0.75)(-0.4,0.6)
    \psline{->}(0.4,-0.75)(0.4,-0.3)
    \psline{->}(0.4,-0.75)(0.4,0.6)
    \psline{->}(0.4,-0.25)(-0.1,0.0625)
  \rput[bl]{0}(-0.1,0.2){$e$}
  \rput[bl]{0}(0.3,0.85){$a_{\text{ex}}$}
  \rput[bl]{0}(-0.6,0.85){$a_{\text{in}}$}
  \rput[bl]{0}(0.3,-1.25){$a_{\text{ex}}^{\prime}$}
  \rput[bl]{0}(-0.6,-1.25){$a_{\text{in}}^{\prime}$}
 \endpspicture
,
\end{equation}
where
\begin{equation}
D_{e,B} = \left\{
\begin{array}{ll}
1 \quad & \text{if } M_{eB}=1\\
0 \quad & \text{otherwise}%
\end{array}%
\right. .
\end{equation}
This decoherence, uncovered in Ref.~\cite{Bonderson07a}, results from probe anyons passing between the anyon clusters in the interior and exterior of the interferometer, and exhibits an effect specific to non-Abelian anyons. In particular, while one expects off-diagonal $a_{\text{in}}, a_{\text{in}}^{\prime}$ terms of the density matrix to decohere when the probe can distinguish between $a_{\text{in}}$ and $a_{\text{in}}^{\prime}$ (a familiar effect in quantum mechanics), there are also terms that will decohere even when the probe cannot distinguish between $a_{\text{in}}$ and $a_{\text{in}}^{\prime}$ (potentially even for $a_{\text{in}}=a_{\text{in}}^{\prime}$). Specifically, this decoherence occurs for terms corresponding to $e$-channels that the probe can ``see'' (i.e. $M_{eB} \neq 1$), which can exist for $a_{\text{in}}$ and $a_{\text{in}}^{\prime}$ that the probe cannot distinguish only for non-Abelian anyons, because they have multiple fusion channels. The decoherence superoperator $D_{B}^{\left(1 \ldots n\right)}$ obviously has no effect on a state if all the $e$ that occur in this state are such that $M_{eB}=1$. In particular, under such conditions, an interferometry measurement (in the asymptotic limit) is a projective measurement. This is always the case when the charge measurement result is an Abelian charge. We also emphasize that this decoherence (and thus interferometry measurements) can change the overall collective charge of the anyons $1,\ldots,m$. In other words, by applying an $F$-move to return to the original basis, one finds that it is not generally true that
\begin{equation}
\sum_{e} \left[ \left( F_{a_{\text{in}}^{\prime} a_{\text{ex}}^{\prime}}^{a_{\text{in}}a_{\text{ex}}} \right)^{-1} \right]_{fe} D_{e,B} \left[ F_{a_{\text{in}}^{\prime} a_{\text{ex}}^{\prime}}^{a_{\text{in}}a_{\text{ex}}} \right]_{ef^{\prime}} = \delta_{f f^{\prime}}
.
\end{equation}

Combining these, we find that an interferometry measurement of anyons $1,\ldots,n$ using $N \rightarrow \infty$ probes $B$ resulting in the outcome $\mathcal{C}_{\kappa}$ is given by
\begin{eqnarray}
&& \text{Prob}_{A}\left( \kappa \right) = \widetilde{\text{Tr}} \left[ \Pi_{\mathcal{C}_{\kappa}}^{\left(1 \ldots n \right)} \rho^{A} \right] \\
&& \rho^{A} \mapsto \Delta_{B;\kappa}^{\left(1 \ldots n\right)} \left[ \rho^{A} \right] = D_{B}^{\left(1 \ldots n\right)} \circ \Pi_{\mathcal{C}_{\kappa}}^{\left(1 \ldots n\right)}  \left[ \rho^{A} \right]
.
\end{eqnarray}
Note that the transformations $\Pi_{\mathcal{C}_{\kappa}}$ and $D_{B}$ commute with each other, so the order of their composition is unimportant. As mentioned above, if $M_{eB}=1$ for all $e$ that occur for $a_{\text{in}}, a_{\text{in}}^{\prime} \in \mathcal{C}_{\kappa}$ with non-zero density matrix coefficients, then the action of $D_{B}$ is trivial and interferometry measurement is simply a projective measurement, and, in particular, this is the case when the measurement outcome is an Abelian charge.

It is often the case that the probe anyons used for interferometry are such that all charges are distinguishable by the probe, i.e. making all charge classes $\mathcal{C}_{\kappa}$ singletons, and only vacuum has trivial monodromy with the probe, i.e. $M_{aB}=1$ iff $a=0$. This is the case for all examples considered in this paper, so we restrict our attention to this case from now on. When this is the case, the interferometry measurement transformation $\Delta_{B,c}^{\left( 1 \ldots n \right)}$ simplifies so that the projection is onto a definite charge ($\Pi_{c}^{\left(1 \ldots n\right)}$) and anyonic charge entanglement between anyons $1,\ldots,n$ and anyons $n+1, \ldots, m$ is completely decohered ($D_{e,B} = \delta_{0e}$), i.e. for a charge measurement outcome $c$ this is represented on the basis elements by
\begin{equation}
\label{eq:simppostintmeas}
\Delta_{B;c}^{\left(1 \ldots n\right)} :
 \pspicture[0.4375](-0.8,-1.3)(0.6,1.1)
  \small
  \psset{linewidth=0.9pt,linecolor=black,arrowscale=1.5,arrowinset=0.15}
  \psline(-0.4,-0.75)(-0.4,0.75)
  \psline(0.4,-0.75)(0.4,0.75)
  \psline(0.4,-0.25)(-0.4,0.25)
    \psline{->}(-0.4,-0.75)(-0.4,-0.3)
    \psline{->}(-0.4,-0.75)(-0.4,0.6)
    \psline{->}(0.4,-0.75)(0.4,-0.3)
    \psline{->}(0.4,-0.75)(0.4,0.6)
    \psline{->}(0.4,-0.25)(-0.1,0.0625)
  \rput[bl]{0}(-0.1,0.2){$e$}
  \rput[bl]{0}(0.3,0.85){$a_{\text{ex}}$}
  \rput[bl]{0}(-0.6,0.85){$a_{\text{in}}$}
  \rput[bl]{0}(0.3,-1.25){$a_{\text{ex}}^{\prime}$}
  \rput[bl]{0}(-0.6,-1.25){$a_{\text{in}}^{\prime}$}
 \endpspicture
\mapsto \frac{\delta_{a_{\text{in}}c} \delta_{a_{\text{in}}^{\prime} c} \delta_{a_{\text{ex}} a_{\text{ex}}^{\prime}} \delta_{0e} }{\text{Prob}_{A} \left(c\right) }
 \pspicture[0.4375](-0.8,-1.3)(0.6,1.1)
  \small
  \psset{linewidth=0.9pt,linecolor=black,arrowscale=1.5,arrowinset=0.15}
  \psline(-0.4,-0.75)(-0.4,0.75)
  \psline(0.4,-0.75)(0.4,0.75)
    \psline{->}(-0.4,-0.75)(-0.4,0.2)
    \psline{->}(0.4,-0.75)(0.4,0.2)
  \rput[bl]{0}(0.3,0.85){$a_{\text{ex}}$}
  \rput[bl]{0}(-0.55,0.9){$c$}
  \rput[bl]{0}(0.3,-1.2){$a_{\text{ex}}$}
  \rput[bl]{0}(-0.55,-1.15){$c$}
 \endpspicture
.
\end{equation}
This form makes it particularly easy to apply the interferometry measurement to a state, as one only needs to re-write the diagrams (using $F$-moves) describing the state to take the form of the left-hand side of this equation and then apply this simple transformation.

A point that is glossed over in the above discussion is that there is a phase parameter $\beta$ in the interference term of the outcome probabilities that is an experimental variable. For example, the tunneling probability for a double-point contact interferometer in the weak tunneling limit is
\begin{equation}
\label{eq:pint}
p^{\shortleftarrow}_{a_{\text{in}}} \simeq \left| t_{1} \right|^{2} + \left| t_{2} \right|^{2} + 2\left| t_{1} t_{2} \right| \text{Re} \left\{ M_{a_{\text{in}}B} e^{i \beta} \right\}
\end{equation}
where $t_{1}$ and $t_{2}$ are the tunneling amplitudes of the two point contacts. The interferometer can only distinguish between target charges $a_{\text{in}}$ and $a_{\text{in}}^{\prime}$ if they give different values of this tunneling probability $p^{\shortleftarrow}_{a_{\text{in}}} \neq p^{\shortleftarrow}_{a_{\text{in}}^{\prime}}$, which leads to out definition of $\mathcal{C}_{\kappa}$. There are two values (mod $2 \pi$) of $\beta$ which give $p^{\shortleftarrow}_{a_{\text{in}}} = p^{\shortleftarrow}_{a_{\text{in}}^{\prime}}$ when $M_{a_{\text{in}}B} \neq M_{a_{\text{in}}^{\prime}B}$, but these are non-generic experimental variable values that require the ability to tune $\beta$ with infinite precision, which is, of course, physically impossible. While we do not need to worry that these non-generic points will make charges indistinguishable, it is important to stay as far away from them as possible in order to increase the distinguishability of the measured charge, which decreases the number probes needed for a given level of confidence in the measurement outcome. Specifically, to distinguish between target charges $a_{\text{in}}$ and $a_{\text{in}}^{\prime}$ with a confidence level $1-\alpha$, one needs
\begin{equation}
N \gtrsim 2 \left[ \text{erfc}^{-1} \left( \alpha \right) \right]^{2} \left[ \frac{ \sqrt{ p^{\shortleftarrow}_{a_{\text{in}}} \left( 1-p^{\shortleftarrow}_{a_{\text{in}}}\right) } + \sqrt{p^{\shortleftarrow}_{a_{\text{in}}^{\prime}}\left( 1-p^{\shortleftarrow}_{a_{\text{in}}^{\prime}} \right) } }{\Delta p} \right] ^{2},
\end{equation}
where
\begin{equation}
\Delta p = \left| p^{\shortleftarrow}_{a_{\text{in}}}- p^{\shortleftarrow}_{a_{\text{in}}^{\prime}} \right|
.
\end{equation}
In order to minimize the necessary duration of a measurement, one should tune $\beta$ so as to maximize $\Delta p$. For the double-point contact interferometer in the weak tunneling limit, this takes the maximal value
\begin{equation}
\Delta p_{\text{max}} = 2 \left| t_{1} t_{2} \right| \Delta M
,
\end{equation}
where $\Delta M = \left| M_{a_{\text{in}}B} - M_{a_{\text{in}}^{\prime}B} \right|$ generally determines how distinguishable $a_{\text{in}}$ and $a_{\text{in}}^{\prime}$ are by $B$ probes. For such an appropriately tuned interferometer with $\left| t_{1} \right| \sim \left| t_{2} \right| \sim t <1/4$, we have the estimate
\begin{equation}
\label{eq:N2pointest}
N \gtrsim 8 \left[\frac{ \text{erfc}^{-1} \left( \alpha \right) }{ t \Delta M}\right]^{2}
.
\end{equation}

\section{Quantum State Teleportation by ``Forced Measurement''}
\label{sec:Teleportation}

Quantum entanglement is the primary source of philosophical quandaries concerning the foundations of quantum physics~\cite{Einstein35,Bell64}, and yet it is also the primary resource of quantum information science. One of the abecedarian examples of entanglement's use as a resource is quantum state teleportation~\cite{Bennett93}, which enjoys the validation of having been experimentally realized~\cite{Bouwmeester97}. A novel use of teleportation is as a means of incorporating the entanglement needed to perform a quantum computation. This is the concept underlying measurement-based approaches to conventional quantum computation~\cite{Gottesman99,Raussendorf01,Nielsen03,Aliferis04}. Not surprisingly, our measurement-only approach to topological quantum computation utilizes an anyonic version of quantum state teleportation to incorporate the necessary entanglement. This is done by using teleportation to generate the braiding transformations that in turn comprise the computational gates. In order to extend the concept of teleportation to anyonic states, we introduce and employ a protocol that we call ``forced measurement.'' The idea is to perform a quantum state teleportation, without the ability to apply unitary operators to qubits, simply by performing a series of measurements until the desired outcome is achieved. Forced measurement is a probabilistically determined adaptive series of measurements in which the measurements to be carried out are predetermined, but the number of times that they need to be carried out is probabilistically determined by the first attainment of the desired measurement outcome. In this section, we will first show how such a forced measurement protocol would be implemented for conventional teleportation; then we apply it to anyons, reviewing its implementation for projective measurement in the density matrix formalism, and finally describing the modified protocol for implementing anyonic teleportation using interferometry measurements.

  \subsection{Conventional Qubits (Spin-$1/2$ Systems)}

We begin by recalling the usual procedure for quantum teleportation of conventional qubits. Let us write the standard Bell states
\begin{eqnarray}
\left| \Phi^{\pm} \right\rangle &=& \frac{1}{\sqrt{2}} \left( \left| \uparrow \uparrow \right\rangle \pm \left| \downarrow \downarrow \right\rangle \right) \\
\left| \Psi^{\pm} \right\rangle &=& \frac{1}{\sqrt{2}} \left( \left| \uparrow \downarrow \right\rangle \pm \left| \downarrow \uparrow \right\rangle \right)
\end{eqnarray}
in the following way
\begin{eqnarray}
\left| \Phi_{0} \right\rangle &=& \frac{1}{\sqrt{2}} \left( \left| \uparrow \downarrow \right\rangle - \left| \downarrow \uparrow \right\rangle \right) \\
\left| \Phi_{\mu} \right\rangle &=& \mathbb{I} \otimes \sigma_{\mu} \left| \Phi_{0} \right\rangle
\end{eqnarray}
where $\sigma_{0} = \mathbb{I}$ and $\sigma_{j}$ are the Pauli matrices. The (normalized) state to be teleported is
\begin{equation}
\left| \psi \right\rangle = \psi_{\uparrow} \left| \uparrow \right\rangle + \psi_{\downarrow} \left| \downarrow \right\rangle
.
\end{equation}
The entanglement resource for the teleportation will be a Bell state $\left| \Phi_{0} \right\rangle$, which we create and then place its two spins at locations $1$ and $2$. The spin system encoding the state $\psi$ is at location $3$, which is in close enough proximity to $2$ so that the collective state of spins $2$ and $3$ can be measured. This gives the tensor product of states
\begin{equation}
\left| \Phi_{0} \right\rangle_{12} \left| \psi \right\rangle_{3} =
\frac{1}{2} \sum_{\mu=0,\ldots,3} \sigma^{\left( 1 \right)}_{\mu} \left| \psi \right\rangle_{1} \left| \Phi_{\mu} \right\rangle_{23}
\end{equation}
upon which we perform an orthogonal projective measurement in the Bell basis on spins $2$ and $3$ (i.e. using $\Pi^{\left( 23 \right)}_{\mu} = \left| \Phi_{\mu} \right\rangle_{23}\left\langle \Phi_{\mu} \right|_{23} $), so that the state transforms as
\begin{equation}
\left| \Phi_{0} \right\rangle_{12} \left| \psi \right\rangle_{3} \mapsto
\Pi_{\mu}^{\left( 23 \right)} \left[ \left| \Phi_{0} \right\rangle_{12} \left| \psi \right\rangle_{3} \right] =
\sigma^{\left( 1\right)}_{\mu} \left| \psi \right\rangle_{1} \left| \Phi_{\mu} \right\rangle_{23}
\end{equation}
with probability $\text{Prob}\left( \mu \right) = 1/4$. The usual prescription for teleportation at this point would send the measurement outcome $\mu$ (two classical bits of information) from someone named ``Bob'' at position $2$ to someone named ``Alice'' at position $1$, who then applies $\sigma_{\mu}$ to the spin there to recover the state $\left| \psi \right\rangle_{1}$ (assuming that someone named ``Charlie'' or ``Eve'' has not decided to complicate the situation).

There is however an alternate way to obtain the state at position $1$ without applying a recovery Pauli transformation. Notice that in addition to teleporting the state $\psi$ to position $1$, the entanglement resource Bell state has also been teleported to positions $2$ and $3$. As long as the measurement method is non-demolitional, the entanglement resource may be used again. If we now measure the spins at $1$ and $2$ in the Bell basis, we get outcome $\nu$ with probability $\text{Prob} \left(\nu \right)=1/4$, and the state transforms as
\begin{equation}
\sigma^{\left( 1\right)}_{\mu} \left| \psi \right\rangle_{1} \left| \Phi_{\mu} \right\rangle_{23} \mapsto
\Pi_{\nu}^{\left( 12 \right)} \left[ \sigma^{\left( 1\right)}_{\mu} \left| \psi \right\rangle_{1} \left| \Phi_{\mu} \right\rangle_{23} \right] =
\pm \left| \Phi_{\nu} \right\rangle_{12} \sigma^{\left( 3\right)}_{\nu} \left| \psi \right\rangle_{3}
.
\end{equation}
The $\pm$ comes from the fact that $\sigma_{\mu} \sigma_{\nu} = \sigma_{\nu} \sigma_{\mu}$ if $\mu = 0$, $\nu = 0$, or $\mu = \nu$ and $\sigma_{\mu} \sigma_{\nu} = -\sigma_{\nu} \sigma_{\mu}$ otherwise. We can now try to teleport the state from position $3$ to $1$ again, obtaining outcome $\mu_{2}$ with probability $\text{Prob} \left( \mu_{2} \right)=1/4$ and the state transformation
\begin{equation}
\left| \Phi_{\nu} \right\rangle_{12} \sigma^{\left( 3\right)}_{\nu} \left| \psi \right\rangle_{3} \mapsto
\Pi_{\mu_{2}}^{\left( 23 \right)} \left[ \left| \Phi_{\nu} \right\rangle_{12} \sigma^{\left( 3\right)}_{\nu} \left| \psi \right\rangle_{3} \right] =
\pm \sigma^{\left( 1\right)}_{\mu_{2}} \left| \psi \right\rangle_{1} \left| \Phi_{\mu_{2}} \right\rangle_{23}
.
\end{equation}
This can be repeated indefinitely, or rather until the $n^{th}$ try when we get the desired measurement outcome $\mu_{n}=0$, which gives us the state $\left| \psi \right\rangle_{1}$ at position $1$ without a Pauli transformation. Each attempt has probability $\text{Prob} \left( 0 \right)=1/4$ of obtaining this desired outcome, so the average number of attempts needed to obtain the desired outcome is
\begin{equation}
\left\langle n \right\rangle = 4
\end{equation}
and the probability of needing $n>N$ attempts to obtain the desired outcome is
\begin{equation}
\text{Prob}\left(\mu_{1},\ldots,\mu_{N} \neq 0 \right) = \left( \frac{3}{4} \right)^{N} ,
\end{equation}
i.e. failure is exponentially suppressed in the number of attempts. This gives a series of measurements with the set of outcomes $M=\left\{ \mu_{n}=0,\nu_{n},\ldots,\mu_1,\nu_{1}=0 \right\}$ such that $\mu_{n} = \nu_{1} = 0$ (the $\nu_{1}=0$ initialization is included for convenience). We call such a series of measurements a ``forced measurement'' (because continue measuring until we get the desired outcome), and write its corresponding operator as
\begin{equation}
\breve{\Pi}_{M}^{\left(23\leftarrow 12 \right)} = \Pi_{\mu_{n}=0}^{\left(23 \right)} \circ \Pi_{\nu_{n}}^{\left(12 \right)} \circ \ldots \circ \Pi_{\mu_{1}}^{\left(23 \right)} \circ \Pi_{\nu_{1}=0}^{\left(12 \right)}
,
\end{equation}
so that
\begin{equation}
\breve{\Pi}_{M}^{\left(23\leftarrow 12 \right)} \left[  \left| \Phi_{0} \right\rangle_{12} \left| \psi \right\rangle_{3} \right] = \left(-1 \right)^{M} \left| \psi \right\rangle_{1} \left| \Phi_{0} \right\rangle_{23}
,
\end{equation}
where the (irrelevant) overall sign depends on the series of measurement outcomes. In density matrix notation, writing
\begin{eqnarray}
\rho &=& \left| \psi \right\rangle \left\langle \psi \right| \\
\rho^{\text{ER}} &=& \left| \Phi_{0} \right\rangle \left\langle \Phi_{0} \right|
,
\end{eqnarray}
this forced measurement teleportation takes the form
\begin{equation}
\breve{\Pi}_{M}^{\left(23\leftarrow 12 \right)} \left[ \rho^{\text{ER} \left( 12\right)} \otimes \rho^{\left( 3\right)} \right] = \rho^{\left( 1\right)} \otimes \rho^{\text{ER} \left( 23\right)}
.
\end{equation}

In the context of conventional qubits, the ``forced measurement'' prescription is obviously a much less efficient alternative to the usual recovery procedure (which is presumably why it is never considered), as it requires multiple uses of a quantum channel in order to shuttle spin system $2$ back and forth between spins $1$ and $3$, rather than two uses of a classical channel. However, it serves the purpose of a conceptual template for anyonic teleportation, where one does not have an operation equivalent to application of a Pauli matrix to use for recovery of the state.

  \subsection{Anyons}

In anyonic systems, it is important to use the forced measurement approach to teleportation, because undesired measurement outcomes have the effect of introducing non-trivial anyonic charge entanglement between the state system and other anyons. Unlike the Spin-$1/2$ case, correcting for undesired anyonic charge entanglement is not as simple as applying a local operator to one of the anyons, so there is no longer a more efficient alternative to forced measurement. In the case of projective topological charge measurement, the unwanted entanglement introduced by measurement is with the entanglement resource pair. This case is very similar to that of conventional qubits, and essentially the same protocol is used for forced measurement teleportation. In the case of interferometry topological charge measurement, the unwanted entanglement manifests itself in the introduction of a non-trivial overall topological charge of system receiving the teleported state (the entanglement resource pair is automatically unentangled from the teleported state). This is a result of the decoherence aspect of interferometry measurement discussed in Section~\ref{sec:Interferometry_Measurement} that results from entanglement with probe anyons that are taken to ``infinity'' and traced out. This difference for interferometry measurements requires an additional topological charge measurement to determine whether the desired outcome is achieved, and thus results in a slightly modified forced measurement protocol for teleportation.

    \subsubsection{Using Projective Measurements}

In Ref.~\cite{Bonderson08a}, we introduced the projective measurement forced measurement procedure using the state vector formalism. Here we will review those results using the density matrix formalism in order to provide a more uniform treatment of the two cases and allow a more direct comparison with the interferometry forced measurement.

We consider the state encoded in the non-local internal degrees of freedom of some anyons given by the density matrix $\rho$. Since we are presently only interested in manipulating one particular anyon which has definite charge $a$, we represent it as
\begin{equation}
\rho \left(a, \ldots \right) =
 \pspicture[0.5](-.5,-1.3)(.5,1.1)
  \small
  \psframe[linewidth=0.9pt,linecolor=black,border=0](-0.35,-0.25)(0.35,0.25)
  \rput[bl]{0}(-0.1,-0.15){$\rho$}
  \psset{linewidth=0.9pt,linecolor=black,arrowscale=1.5,arrowinset=0.15}
  \psline(0,0.25)(0,0.75)
  \psline(0,-0.25)(0,-0.75)
  \psline{->}(0,0.25)(0,0.65)
  \psline{-<}(0,-0.25)(0,-0.65)
  \rput[bl](-0.1,0.85){$a$}
  \rput[bl](-0.1,-1.05){$a$}
 \endpspicture
,
\end{equation}
where ``$\ldots$'' represents all the other anyons that comprise the state $\rho$, which we leave implicit (i.e. there should really be additional anyonic charge lines emanating from the box). Normalization factors giving $\widetilde{\text{Tr}} \left[ \rho \right] =1$ are absorbed into the box.

In order to teleport the state information encoded by an anyon of definite charge $a$ to another anyon of definite charge $a$, we introduce a particle-antiparticle pair produced from vacuum, given by the state
\begin{equation}
\label{eq:aabarpair}
\rho^{\text{ER}} = \left| a,\bar{a};0 \right\rangle \left\langle a, \bar{a} ;0\right| = \frac{1} {d_{a} }
 \pspicture[0.4375](-0.8,-1.3)(0.6,1.1)
  \small
  \psset{linewidth=0.9pt,linecolor=black,arrowscale=1.5,arrowinset=0.15}
  \psline(0.0,0.25)(0.4,0.75)
  \psline(0.0,0.25)(-0.4,0.75)
  \psline(0.0,-0.25)(0.4,-0.75)
  \psline(0.0,-0.25)(-0.4,-0.75)
    \psline{->}(0.0,0.25)(0.3,0.625)
    \psline{->}(0.0,0.25)(-0.3,0.625)
    \psline{-<}(0.0,-0.25)(0.3,-0.625)
    \psline{-<}(0.0,-0.25)(-0.3,-0.625)
  \rput[bl]{0}(0.3,0.85){$\bar{a}$}
  \rput[bl]{0}(-0.55,0.85){$a$}
  \rput[bl]{0}(0.3,-1.05){$\bar{a}$}
  \rput[bl]{0}(-0.55,-1.05){$a$}
 \endpspicture
.
\end{equation}
The state in Eq.~(\ref{eq:aabarpair}) has maximal anyonic entanglement between its two anyons, and is the analog of the maximally entangled Bell states typically used as the entanglement resource in quantum state teleportation of conventional qubits.

We now tensor these together
\begin{equation}
\rho^{\text{ER} \left(12\right)} \otimes \rho^{\left(3 \ldots \right)} \left(a, \ldots \right)
= \frac{1} {d_{a}}
 \pspicture[0.5](-2,-1.3)(.5,1.1)
   \small
  \psset{linewidth=0.9pt,linecolor=black,arrowscale=1.5,arrowinset=0.15}
  \psline(-1.2,0.25)(-0.8,0.75)
  \psline(-1.2,0.25)(-1.6,0.75)
  \psline(-1.2,-0.25)(-0.8,-0.75)
  \psline(-1.2,-0.25)(-1.6,-0.75)
    \psline{->}(-1.2,0.25)(-0.9,0.625)
    \psline{->}(-1.2,0.25)(-1.5,0.625)
    \psline{-<}(-1.2,-0.25)(-0.9,-0.625)
    \psline{-<}(-1.2,-0.25)(-1.5,-0.625)
  \rput[bl]{0}(-0.9,0.85){$\bar{a}$}
  \rput[bl]{0}(-1.75,0.85){$a$}
  \rput[bl]{0}(-0.9,-1.05){$\bar{a}$}
  \rput[bl]{0}(-1.75,-1.05){$a$}
  \small
  \psframe[linewidth=0.9pt,linecolor=black,border=0](-0.35,-0.25)(0.35,0.25)
  \rput[bl]{0}(-0.1,-0.15){$\rho$}
  \psset{linewidth=0.9pt,linecolor=black,arrowscale=1.5,arrowinset=0.15}
  \psline(0,0.25)(0,0.75)
  \psline(0,-0.25)(0,-0.75)
  \psline{->}(0,0.25)(0,0.65)
  \psline{-<}(0,-0.25)(0,-0.65)
  \rput[bl](-0.1,0.85){$a$}
  \rput[bl](-0.1,-1.05){$a$}
 \endpspicture
\end{equation}
to obtain the combined state on which we will perform measurements. What we would like to do is perform a measurement of the collective charge of anyons $2$ and $3$ for which the result is vacuum charge $0$.

Introducing the notation
\begin{equation}
\rho_{e}^{\text{ER} \left( 12 \right)} \rho_{e}^{\left(3 \ldots \right)} \left(a, \ldots \right) =
\frac{1}{d_{a}}
 \pspicture[0.5](-1,-2.3)(1.2,2.1)
  \small
  \psframe[linewidth=0.9pt,linecolor=black,border=0](-0.35,-0.25)(0.35,0.25)
  \rput[bl]{0}(-0.1,-0.15){$\rho$}
  \psset{linewidth=0.9pt,linecolor=black,arrowscale=1.5,arrowinset=0.15}
  \psline(0,0.25)(0,0.75)
  \psline(0,-0.25)(0,-0.75)
  \psline{->}(0,0.25)(0,0.65)
  \psline{-<}(0,-0.25)(0,-0.65)
  \rput[bl](0.15,0.45){$a$}
  \rput[bl](0.15,-0.65){$a$}
  \psset{linewidth=0.9pt,linecolor=black,arrowscale=1.5,arrowinset=0.15}
  \psline(-0.8,1.75)(0,0.75)
  \psline(0.8,1.75) (0,0.75)
  \psline(-0.4,1.25) (0,1.75)
  \psline(-0.8,-1.75)(0,-0.75)
  \psline(0.8,-1.75) (0,-0.75)
  \psline(-0.4,-1.25) (0,-1.75)
   \psline{->}(-0.4,1.25)(-0.7,1.625)
   \psline{->}(-0.4,1.25)(-0.1,1.625)
   \psline{->}(0,0.75)(0.7,1.625)
   \psline{->}(0,0.75)(-0.3,1.125)
   \psline{-<}(-0.4,-1.25)(-0.7,-1.625)
   \psline{-<}(-0.4,-1.25)(-0.1,-1.625)
   \psline{-<}(0,-0.75)(0.7,-1.625)
   \psline{-<}(0,-0.75)(-0.3,-1.125)
   \rput[bl]{0}(-0.95,1.85){$a$}
   \rput[bl]{0}(-0.05,1.85){$\bar{a}$}
   \rput[bl]{0}(0.75,1.85){${a}$}
   \rput[bl]{0}(-0.5,0.75){$e$}
   \rput[bl]{0}(-0.95,-2.05){$a$}
   \rput[bl]{0}(-0.05,-2.05){$\bar{a}$}
   \rput[bl]{0}(0.75,-2.05){${a}$}
   \rput[bl]{0}(-0.5,-0.95){$e$}
 \endpspicture
,
\end{equation}
it is clear that $\rho_{0}^{\text{ER} \left( 12 \right)} \rho_{0}^{\left(3 \ldots \right)} \left(a, \ldots \right) = \rho^{\text{ER} \left( 12 \right)} \otimes \rho^{\left(3 \ldots \right)} \left(a, \ldots \right)$. Applying a projective topological charge measurement of anyons $2$ and $3$ to the state $\rho_{e}^{\text{ER} \left( 12 \right)} \rho_{e}^{\left(3 \ldots \right)} \left(a, \ldots \right)$, we get the measurement outcome $f$ with probability and post-measurement state
\begin{eqnarray}
&& \text{Prob} \left( f \right) =  \left| \left[F_{a}^{a\bar{a}a}\right]_{ef} \right|^{2}  \\
&& \Pi_{f}^{\left(23\right)} \left[ \rho_{e}^{\text{ER} \left( 12 \right)} \rho_{e}^{\left(3 \ldots \right)} \left(a, \ldots \right) \right] =
\rho_{f}^{\text{ER} \left( 32 \right)} \rho_{f}^{\left(1 \ldots \right)} \left(a, \ldots \right)
\end{eqnarray}
To see this, we use the unitary $F$-move that transforms between the two fusion bases of the three anyons
\begin{equation}
  \pspicture[0.454545](0,-0.4)(1.8,1.8)
  \small
  \psset{linewidth=0.9pt,linecolor=black,arrowscale=1.5,arrowinset=0.15}
  \psline(1,0.5)(1,0)
  \psline(0.2,1.5)(1,0.5)
  \psline(1.8,1.5) (1,0.5)
  \psline(0.6,1) (1,1.5)
   \psline{->}(0.6,1)(0.3,1.375)
   \psline{->}(0.6,1)(0.9,1.375)
   \psline{->}(1,0.5)(1.7,1.375)
   \psline{->}(1,0.5)(0.7,0.875)
   \psline{->}(1,0)(1,0.375)
   \rput[bl]{0}(0.05,1.6){$a$}
   \rput[bl]{0}(0.95,1.6){$\bar{a}$}
   \rput[bl]{0}(1.75,1.6){${a}$}
   \rput[bl]{0}(0.5,0.5){$e$}
   \rput[bl]{0}(.95,-0.3){$a$}
  \endpspicture
= \sum_{f} \left[F_{a}^{a\bar{a}a}\right]_{ef}
 \pspicture[0.454545](0,-0.4)(1.8,1.8)
  \small
  \psset{linewidth=0.9pt,linecolor=black,arrowscale=1.5,arrowinset=0.15}
  \psline(1,0.5)(1,0)
  \psline(0.2,1.5)(1,0.5)
  \psline(1.8,1.5) (1,0.5)
  \psline(1.4,1) (1,1.5)
   \psline{->}(0.6,1)(0.3,1.375)
   \psline{->}(1.4,1)(1.1,1.375)
   \psline{->}(1,0.5)(1.7,1.375)
   \psline{->}(1,0.5)(1.3,0.875)
   \psline{->}(1,0)(1,0.375)
   \rput[bl]{0}(0.05,1.6){$a$}
   \rput[bl]{0}(0.95,1.6){$\bar{a}$}
   \rput[bl]{0}(1.75,1.6){${a}$}
   \rput[bl]{0}(1.25,0.45){$f$}
   \rput[bl]{0}(.95,-0.3){$a$}
  \endpspicture
\end{equation}
(which should make clear where the probability $\text{Prob} \left(f \right)$ comes from) and apply the projective measurement
\begin{eqnarray}
&& \Pi_{f}^{\left(23\right)} \left[ \rho_{e}^{\text{ER} \left( 12 \right)} \rho_{e}^{\left(3 \ldots \right)} \left(a, \ldots \right) \right] =
\frac{\Pi_{f}^{\left(23\right)} \rho_{e}^{\text{ER} \left( 12 \right)} \rho_{e}^{\left(3 \ldots \right)} \left(a, \ldots \right) \Pi_{f}^{\left(23\right)} }{ \text{Prob} \left(f \right) } \notag \\
&& \qquad = \frac{d_{f}}{ \text{Prob} \left( f \right) d_{a}^{3} }
 \pspicture[0.5](-1.1,-3.8)(1.2,3.6)
  \small
  \psframe[linewidth=0.9pt,linecolor=black,border=0](-0.35,-0.25)(0.35,0.25)
  \rput[bl]{0}(-0.1,-0.15){$\rho$}
  \psset{linewidth=0.9pt,linecolor=black,arrowscale=1.5,arrowinset=0.15}
  \psline(0,0.25)(0,0.75)
  \psline(0,-0.25)(0,-0.75)
  \psline{->}(0,0.25)(0,0.65)
  \psline{-<}(0,-0.25)(0,-0.65)
  \rput[bl](0.15,0.45){$a$}
  \rput[bl](0.15,-0.65){$a$}
  \psset{linewidth=0.9pt,linecolor=black,arrowscale=1.5,arrowinset=0.15}
  \psline(-0.8,1.75)(0,0.75)
  \psline(0.8,1.75) (0,0.75)
  \psline(-0.4,1.25) (0,1.75)
  \psline(-0.8,-1.75)(0,-0.75)
  \psline(0.8,-1.75) (0,-0.75)
  \psline(-0.4,-1.25) (0,-1.75)
  \psline(-0.8,1.75)(-0.8,3.25)
  \psline(-0.8,-1.75)(-0.8,-3.25)
  \psline(0.8,1.75)(0.4,2.25)
  \psline(0,1.75)(0.4,2.25)
  \psline(0.4,2.25)(0.4,2.75)
  \psline(0.4,2.75)(0.8,3.25)
  \psline(0.4,2.75)(0,3.25)
  \psline(0.8,-1.75)(0.4,-2.25)
  \psline(0,-1.75)(0.4,-2.25)
  \psline(0.4,-2.25)(0.4,-2.75)
  \psline(0.4,-2.75)(0.8,-3.25)
  \psline(0.4,-2.75)(0,-3.25)
   \psline{->}(-0.4,1.25)(-0.7,1.625)
   \psline{->}(-0.4,1.25)(-0.1,1.625)
   \psline{->}(0,0.75)(0.7,1.625)
   \psline{->}(0,0.75)(-0.3,1.125)
   \psline{-<}(-0.4,-1.25)(-0.7,-1.625)
   \psline{-<}(-0.4,-1.25)(-0.1,-1.625)
   \psline{-<}(0,-0.75)(0.7,-1.625)
   \psline{-<}(0,-0.75)(-0.3,-1.125)
   \psline{->}(0.4,2.25)(0.4,2.625)
   \psline{->}(0.4,2.75)(0.7,3.125)
   \psline{->}(0.4,2.75)(0.1,3.125)
   \psline{-<}(0.4,-2.25)(0.4,-2.625)
   \psline{-<}(0.4,-2.75)(0.7,-3.125)
   \psline{-<}(0.4,-2.75)(0.1,-3.125)
   \rput[bl]{0}(-0.95,3.35){$a$}
   \rput[bl]{0}(-0.05,3.35){$\bar{a}$}
   \rput[bl]{0}(0.75,3.35){${a}$}
   \rput[bl]{0}(-0.05,1.3){$\bar{a}$}
   \rput[bl]{0}(0.75,1.3){${a}$}
   \rput[bl]{0}(0.55,2.3){${f}$}
   \rput[bl]{0}(-0.5,0.75){$e$}
   \rput[bl]{0}(-0.95,-3.55){$a$}
   \rput[bl]{0}(-0.05,-3.55){$\bar{a}$}
   \rput[bl]{0}(0.75,-3.55){${a}$}
   \rput[bl]{0}(-0.05,-1.5){$\bar{a}$}
   \rput[bl]{0}(0.75,-1.5){${a}$}
   \rput[bl]{0}(0.55,-2.7){${f}$}
   \rput[bl]{0}(-0.5,-0.95){$e$}
 \endpspicture
\quad =
\frac{1}{d_{a}}
 \pspicture[0.5](-1.1,-2.3)(1.2,2.1)
  \small
  \psframe[linewidth=0.9pt,linecolor=black,border=0](-0.35,-0.25)(0.35,0.25)
  \rput[bl]{0}(-0.1,-0.15){$\rho$}
  \psset{linewidth=0.9pt,linecolor=black,arrowscale=1.5,arrowinset=0.15}
  \psline(0,0.25)(0,0.75)
  \psline(0,-0.25)(0,-0.75)
  \psline{->}(0,0.25)(0,0.65)
  \psline{-<}(0,-0.25)(0,-0.65)
  \rput[bl](-0.35,0.45){$a$}
  \rput[bl](-0.35,-0.65){$a$}
  \psset{linewidth=0.9pt,linecolor=black,arrowscale=1.5,arrowinset=0.15}
  \psline(-0.8,1.75)(0,0.75)
  \psline(0.8,1.75) (0,0.75)
  \psline(0.4,1.25) (0,1.75)
  \psline(-0.8,-1.75)(0,-0.75)
  \psline(0.8,-1.75) (0,-0.75)
  \psline(0.4,-1.25) (0,-1.75)
   \psline{->}(-0.4,1.25)(-0.7,1.625)
   \psline{->}(0.4,1.25)(0.1,1.625)
   \psline{->}(0,0.75)(0.7,1.625)
   \psline{->}(0,0.75)(0.3,1.125)
   \psline{-<}(-0.4,-1.25)(-0.7,-1.625)
   \psline{-<}(0.4,-1.25)(0.1,-1.625)
   \psline{-<}(0,-0.75)(0.7,-1.625)
   \psline{-<}(0,-0.75)(0.3,-1.125)
   \rput[bl]{0}(-0.95,1.85){$a$}
   \rput[bl]{0}(-0.05,1.85){$\bar{a}$}
   \rput[bl]{0}(0.75,1.85){${a}$}
   \rput[bl]{0}(0.35,0.65){$f$}
   \rput[bl]{0}(-0.95,-2.05){$a$}
   \rput[bl]{0}(-0.05,-2.05){$\bar{a}$}
   \rput[bl]{0}(0.75,-2.05){${a}$}
   \rput[bl]{0}(0.35,-1.05){$f$}
 \endpspicture
.
\end{eqnarray}

If we obtain an undesired measurement outcome $f \neq 0$, we can ``undo'' the measurement by subsequently performing a measurement of anyons $1$ and $2$, as long as the measurement processes are non-demolitional. If this measurement has outcome $e_{2}$, the combined system is put in the post-measurement state $\rho_{e_{2}}^{\text{ER} \left( 12 \right)} \rho_{e_{2}}^{\left(3 \ldots \right)} \left(a, \ldots \right)$. Now we can perform a measurement of anyons $2$ and $3$ again, with an entirely new chance of obtaining the desired outcome. This procedure may be repeated until we obtain the desired measurement outcome, obtaining a string of measurement outcomes $M=\left\{ e_{1},f_{1},\ldots,e_{n},f_{n} \right\}$ (including the initialization $e_{1}$ for convenience), where $e_{1}=f_{n}=0$ and $f_{j} \neq 0$ for $j<n$. The probabilities of measurement outcomes $e_{j}$ and $f_{j}$ are respectively
\begin{eqnarray}
\text{Prob}\left(e_{j} \right) &=& \left| \left[ F^{a \bar{a} a}_{a} \right]_{e_{j} f_{j-1}} \right|^{2} \\
\text{Prob}\left(f_{j} \right) &=& \left| \left[ F^{a \bar{a} a}_{a} \right]_{e_{j} f_{j}} \right|^{2}
.
\end{eqnarray}
The probability of obtaining the desired outcome $f=0$ at the $j^{th}$ measurement attempt in this procedure is
\begin{equation}
\text{Prob}\left(f_{j}=0 \right) = \left| \left[ F^{a \bar{a} a}_{a} \right]_{e_{j}0} \right|^{2} =N_{a \bar{a}}^{e_{j}} \frac{ d_{e_{j}} }{ d_{a}^{2} }
.
\end{equation}
This probability has the non-zero lower bound
\begin{equation}
\text{Prob}\left(f_{j}=0 \right) \geq d_{a}^{-2},
\end{equation}
since $d_{x} \geq 1$ for any $x$. The average number of attempts until a desired outcome is achieved in a forced measurement is thus
\begin{equation}
\left\langle n \right\rangle \leq d_{a}^{2} ,
\end{equation}
and the probability of needing $n>N$ attempts to obtain the desired outcome is
\begin{equation}
\text{Prob}\left(f_{1},\ldots,f_{N} \neq 0 \right)\leq \left( 1-d_{a}^{-2} \right)^{N} ,
\end{equation}
i.e. failure is exponentially suppressed in the number of attempts.

Thus, the forced projective measurement given by the probabilistically determined adaptive series of measurements
\begin{equation}
\breve{\Pi}_{M}^{\left(32\leftarrow 12 \right)} = \Pi_{f_{n}=0}^{\left(23 \right)} \circ \Pi_{e_{n}}^{\left(12 \right)} \circ
\ldots \circ \Pi_{f_{1}}^{\left(23 \right)} \circ \Pi_{e_{1}=0}^{\left(12 \right)}
\end{equation}
enables use to perform anyonic state teleportation
\begin{equation}
\breve{\Pi}_{M}^{\left(32\leftarrow 12 \right)} \left[ \rho^{\text{ER} \left( 12 \right)} \otimes \rho^{\left(3 \ldots \right)} \left(a, \ldots \right)\right] = \rho^{\text{ER} \left( 32 \right)} \otimes \rho^{\left(1 \ldots \right)} \left(a, \ldots \right)
\end{equation}
using projective measurements. The notation $32\leftarrow 12$ means the entanglement resource was originally encoded in anyons $1$ and $2$, but after the forced measurement is encoded in anyons $3$ and $2$. One can also infer from this that the role of anyon $3$ in encoding the state $\rho$ has now been transferred to anyon $1$. We emphasize that while it is important to perform all the $\Pi^{\left(12\right)}$ measurements in order to teleport the state information, the actual outcomes $e_{j}$ of these measurements are unimportant.

    \subsubsection{Using Interferometry Measurements}
      \label{sec:Interferometry_Teleportation}

In contrast with projective measurement, interferometrical measurement of topological charge is not quite as simple and requires a density matrix formulation. We will assume the asymptotic limit of $N \rightarrow \infty$ probe measurements in which interferometry may effectively be treated as a projective measurement of the target anyons' collective charge, together with decoherence of anyonic charge entanglement between the target anyons and those exterior to the interferometry region (see Section~\ref{sec:Interferometry_Measurement}). We will also assume that the probe anyons can distinguish between all topological charges and only has trivial monodromy with the vacuum charge, allowing us to use Eq.~(\ref{eq:simppostintmeas}) for interferometry measurements, which can be described as the combination of projection onto a definite charge ($\Pi_{c}$) and complete anyonic charge decoherence between the interior and exterior of the interferometry region ($D_{e,B} = \delta_{0e}$). Since we otherwise do not care about the specific details of the probe anyons, we will drop the label $B$ and simply denote interferometry measurements using $\Delta_{c} = D \circ \Pi_{c}$.

We again consider the state of some anyons given by the density matrix $\rho$, and are presently only interested in manipulating one particular anyon which has definite charge $a$, so we represent it as
\begin{equation}
\rho \left(a, \ldots \right) =
 \pspicture[0.48](-.5,-1.25)(.5,1.25)
  \small
  \psframe[linewidth=0.9pt,linecolor=black,border=0](-0.35,-0.25)(0.35,0.25)
  \rput[bl]{0}(-0.1,-0.15){$\rho$}
  \psset{linewidth=0.9pt,linecolor=black,arrowscale=1.5,arrowinset=0.15}
  \psline(0,0.25)(0,1)
  \psline(0,-0.25)(0,-1)
  \psline{->}(0,0.25)(0,0.8)
  \psline{-<}(0,-0.25)(0,-0.8)
  \rput[bl](-0.1,1.1){$a$}
  \rput[bl](-0.1,-1.3){$a$}
 \endpspicture
,
\end{equation}
where ``$\ldots$'' represents all the other anyons that comprise the state $\rho$, which we leave implicit (i.e. there should really be additional anyonic charge lines emanating from the box). We will need to assume that $\rho$ has trivial overall charge $\rho = \rho_{0}$ in order to be able to implement our forced measurement procedure. This is not a severe restriction, as one can always achieve such a condition by keeping track of additional anyons. Furthermore, the anyonic states used to encode topological qubits will have trivial overall charge, so this restriction automatically includes those of interest to us for TQC. One might worry that such a restriction is pointless, since we are going to perform interferometry measurements, and we know that interferometry can change the overall charge. In fact, it is exactly this effect that we correct for by using forced measurement. We will see that the interferometry measurements employed are such that they change the overall charge and introduce anyonic charge entanglement in a very particular way that be represented by the attachment of a charge line connecting the $a$ lines on either end of the density matrix. Thus, we define the density matrix that can occur at intermediate steps in the forced measurement
\begin{equation}
\rho_{x} \left(a, \ldots \right) = \frac{1}{\sqrt{d_{x}}}
 \pspicture[0.4666666](-1.3,-1.5)(.5,1.5)
  \small
  \psframe[linewidth=0.9pt,linecolor=black,border=0](-0.35,-0.25)(0.35,0.25)
  \rput[bl]{0}(-0.2,-0.15){$\rho_{0}$}
  \psset{linewidth=0.9pt,linecolor=black,arrowscale=1.5,arrowinset=0.15}
  \psline(0,0.25)(0,1.25)
  \psline(0,-0.25)(0,-1.25)
  \psline(-0.75,0.25)(0,0.75)
  \psline(-0.75,-0.25)(0,-0.75)
  \psline(-0.75,-0.25)(-0.75,0.25)
  \psline{->}(0,0.25)(0,0.65)
  \psline{-<}(0,-0.25)(0,-0.6)
  \psline{->}(0,0.75)(0,1.15)
  \psline{-<}(0,-0.75)(0,-1.15)
  \psline{->}(-0.75,-0.25)(-0.75,0.15)
  \rput[bl](-0.1,1.35){$a$}
  \rput[bl](-0.1,-1.55){$a$}
  \rput[bl](0.15,0.5){$a$}
  \rput[bl](0.15,-0.7){$a$}
  \rput[bl](-1.1,-0.1){$x$}
 \endpspicture
,
\end{equation}
where $x=0$ obviously gives the original state $\rho_{0}$, and the normalization factor is included so that $\widetilde{\text{Tr}} \left[ \rho_{x} \right] = \widetilde{\text{Tr}} \left[ \rho_{0} \right] =1 $. We note that measuring the collective charge of all the anyons that comprise this state (i.e. measuring the overall charge of $\rho_{x}$) gives the measurement outcome $x$. Hence, the restriction that the original state $\rho_{0}$ have trivial overall charge makes it possible to correctly identify the charge $x$ by performing a measurement\footnote{If the original density matrix had overall charge that was non-Abelian, this measurement could not unambiguously identify the charge $x$. One would still be able to properly identify the charge $x$ using such a measurement if this condition was relaxed to allow the initial state to have a known, definite Abelian overall charge, however such a generalization is unnecessary for our purposes.}. We will later see that it is essential in our forced measurement procedure to be able to properly identify the charge $x$, hence the reason for restricting to such initial $\rho_{0}$.

The entanglement resource for interferometry measurement based teleportation is
\begin{equation}
\rho_{e}^{\text{ER}} = \frac{1}{d_{e}} \left| a,\bar{a};e \right\rangle \left\langle a, \bar{a} ;e\right| = \frac{1} {d_{a} \sqrt{d_{e}}}
 \pspicture[0.4375](-0.8,-1.3)(0.6,1.1)
  \small
  \psset{linewidth=0.9pt,linecolor=black,arrowscale=1.5,arrowinset=0.15}
  \psline(0.0,-0.25)(0.0,0.25)
  \psline(0.0,0.25)(0.4,0.75)
  \psline(0.0,0.25)(-0.4,0.75)
  \psline(0.0,-0.25)(0.4,-0.75)
  \psline(0.0,-0.25)(-0.4,-0.75)
    \psline{->}(0.0,-0.25)(0.0,0.125)
    \psline{->}(0.0,0.25)(0.3,0.625)
    \psline{->}(0.0,0.25)(-0.3,0.625)
    \psline{-<}(0.0,-0.25)(0.3,-0.625)
    \psline{-<}(0.0,-0.25)(-0.3,-0.625)
  \rput[bl]{0}(0.15,-0.1){$e$}
  \rput[bl]{0}(0.3,0.85){$\bar{a}$}
  \rput[bl]{0}(-0.55,0.85){$a$}
  \rput[bl]{0}(0.3,-1.05){$\bar{a}$}
  \rput[bl]{0}(-0.55,-1.05){$a$}
 \endpspicture
.
\end{equation}
We do not require $e=0$ here because it is not necessary for the procedure to work (nor does it significantly increase the probability of obtaining a desired measurement result). Furthermore, $e$ will generally change in each step, so if we wanted to require $e=0$, we would have to perform additional steps to re-initialize the entanglement resource, making the procedure less efficient.

We now tensor these together
\begin{equation}
\rho_{e}^{\text{ER} \left(12\right)} \otimes \rho_{x}^{\left(3 \ldots \right)} \left(a, \ldots \right)
= \frac{1} {d_{a} \sqrt{d_{e}d_{x}}}
 \pspicture[0.466666](-2,-1.5)(.5,1.5)
   \small
  \psset{linewidth=0.9pt,linecolor=black,arrowscale=1.5,arrowinset=0.15}
  \psline(-1.2,-0.75)(-1.2,0.75)
  \psline(-1.2,0.75)(-0.8,1.25)
  \psline(-1.2,0.75)(-1.6,1.25)
  \psline(-1.2,-0.75)(-0.8,-1.25)
  \psline(-1.2,-0.75)(-1.6,-1.25)
    \psline{->}(-1.2,-0.25)(-1.2,0.125)
    \psline{->}(-1.2,0.75)(-0.9,1.125)
    \psline{->}(-1.2,0.75)(-1.5,1.125)
    \psline{-<}(-1.2,-0.75)(-0.9,-1.125)
    \psline{-<}(-1.2,-0.75)(-1.5,-1.125)
  \rput[bl]{0}(-1.55,-0.1){$e$}
  \rput[bl]{0}(-0.9,1.35){$\bar{a}$}
  \rput[bl]{0}(-1.75,1.35){$a$}
  \rput[bl]{0}(-0.9,-1.55){$\bar{a}$}
  \rput[bl]{0}(-1.75,-1.55){$a$}
  \small
  \psframe[linewidth=0.9pt,linecolor=black,border=0](-0.35,-0.25)(0.35,0.25)
  \rput[bl]{0}(-0.2,-0.15){$\rho_{0}$}
  \psset{linewidth=0.9pt,linecolor=black,arrowscale=1.5,arrowinset=0.15}
  \psline(0,0.25)(0,1.25)
  \psline(0,-0.25)(0,-1.25)
  \psline(-0.75,0.25)(0,0.75)
  \psline(-0.75,-0.25)(0,-0.75)
  \psline(-0.75,-0.25)(-0.75,0.25)
  \psline{->}(0,0.25)(0,0.65)
  \psline{-<}(0,-0.25)(0,-0.6)
  \psline{->}(0,0.75)(0,1.15)
  \psline{-<}(0,-0.75)(0,-1.15)
  \psline{->}(-0.75,-0.25)(-0.75,0.15)
  \rput[bl](-0.1,1.35){$a$}
  \rput[bl](-0.1,-1.55){$a$}
  \rput[bl](0.15,0.5){$a$}
  \rput[bl](0.15,-0.7){$a$}
  \rput[bl](-1.05,-0.1){$x$}
 \endpspicture
.
\end{equation}
Performing an interferometry measurement on anyons $2$ and $3$ will have the charge measurement outcome $f$ with probability and post-measurement state given by
\begin{eqnarray}
&&\text{Prob}\left( f \right) =  N_{a \bar{a}}^{f} \frac{d_{f}}{d_{a}^{2}} \\
&& \Delta_{f}^{\left(23 \right)} \left[ \rho_{e}^{\text{ER} \left(12\right)} \otimes \rho_{x}^{\left(3 \ldots \right)} \left(a, \ldots \right)  \right]
= \sum_{y} \text{Prob}\left( y | f \right) \rho_{f}^{\text{ER} \left(32 \right)} \otimes \rho_{y}^{\left(1 \ldots \right)} \left(a, \ldots \right)
\label{eq:intteleportDelta}
\end{eqnarray}
where
\begin{equation}
\text{Prob}\left( y | f \right) = \sum_{q} \left| \left[F_{a}^{e \bar{f} a} \right]_{qa} \right|^{2} \left| \left[F_{a}^{qxa} \right]_{ya} \right|^{2}.
\end{equation}
The probability $\text{Prob}\left( f \right)$ should be obvious, since anyons $2$ and $3$ are initially unentangled (no charge lines connect them) and thus have randomly determined collective charge (i.e. probabilities weighted by their quantum dimension). To demonstrate the result for the post-measurement state, we break the interferometry measurement into the two parts (projection and decoherence) $\Delta_{f}^{\left(23 \right)} = D^{\left(23 \right)} \circ \Pi_{f}^{\left(23 \right)}$
\begin{eqnarray}
&& \!\!\!\!\!\!\!\!\!\!\!\! \Pi_{f}^{\left(23 \right)} \left[ \rho_{e}^{\text{ER} \left(12\right)} \otimes \rho_{x}^{\left(3 \ldots \right)} \left(a, \ldots \right)  \right] = \frac{\Pi_{f}^{\left(23 \right)} \rho_{e}^{\text{ER} \left(12\right)} \otimes \rho_{x}^{\left(3 \ldots \right)} \left(a, \ldots \right) \Pi_{f}^{\left(23 \right)}}{\text{Prob} \left(f\right)} \notag \\
&& \!\!\!\!\!\!\!\! =\frac{1} {d_{a} \sqrt{d_{e}d_{x}}}
 \pspicture[0.483333](-2,-3)(.5,3)
   \small
  \psset{linewidth=0.9pt,linecolor=black,arrowscale=1.5,arrowinset=0.15}
  \psline(-1.2,-0.75)(-1.2,0.75)
  \psline(-1.2,0.75)(-0.8,1.25)
  \psline(-1.2,0.75)(-1.6,1.25)
  \psline(-1.2,-0.75)(-0.8,-1.25)
  \psline(-1.2,-0.75)(-1.6,-1.25)
    \psline{->}(-1.2,-0.25)(-1.2,0.125)
    \psline{->}(-1.2,0.75)(-0.9,1.125)
    \psline{->}(-1.2,0.75)(-1.5,1.125)
    \psline{-<}(-1.2,-0.75)(-0.9,-1.125)
    \psline{-<}(-1.2,-0.75)(-1.5,-1.125)
  \rput[bl]{0}(-1.55,-0.1){$e$}
  \rput[bl]{0}(-0.9,2.85){$\bar{a}$}
  \rput[bl]{0}(-1.75,2.85){$a$}
  \rput[bl]{0}(-0.9,-3.05){$\bar{a}$}
  \rput[bl]{0}(-1.75,-3.05){$a$}
  \small
  \psframe[linewidth=0.9pt,linecolor=black,border=0](-0.35,-0.25)(0.35,0.25)
  \rput[bl]{0}(-0.2,-0.15){$\rho_{0}$}
  \psset{linewidth=0.9pt,linecolor=black,arrowscale=1.5,arrowinset=0.15}
  \psline(0,0.25)(0,1.25)
  \psline(0,-0.25)(0,-1.25)
  \psline(-0.75,0.25)(0,0.75)
  \psline(-0.75,-0.25)(0,-0.75)
  \psline(-0.75,-0.25)(-0.75,0.25)
  \psline{->}(0,0.25)(0,0.65)
  \psline{-<}(0,-0.25)(0,-0.6)
  \psline{->}(0,0.75)(0,1.15)
  \psline{-<}(0,-0.75)(0,-1.15)
  \psline{->}(-0.75,-0.25)(-0.75,0.15)
  \rput[bl](-0.1,2.85){$a$}
  \rput[bl](-0.1,-3.05){$a$}
  \rput[bl](0.15,0.5){$a$}
  \rput[bl](0.15,1){$a$}
  \rput[bl](-0.75,1){$\bar{a}$}
  \rput[bl](0.15,-1.2){$a$}
  \rput[bl](-0.75,-1.2){$\bar{a}$}
  \rput[bl](0.15,-0.7){$a$}
  \rput[bl](-1.05,-0.1){$x$}
  \psline(-1.6,1.25)(-1.6,2.75)
  \psline(-1.6,-1.25)(-1.6,-2.75)
  \psline(-0.8,1.25)(-0.4,1.75)
  \psline(-0.8,-1.25)(-0.4,-1.75)
  \psline(0,1.25)(-0.4,1.75)
  \psline(0,-1.25)(-0.4,-1.75)
  \psline(-0.4,1.75)(-0.4,2.25)
  \psline(-0.4,-1.75)(-0.4,-2.25)
  \psline(-0.4,2.25)(-0.8,2.75)
  \psline(-0.4,2.25)(0,2.75)
  \psline(-0.4,-2.25)(-0.8,-2.75)
  \psline(-0.4,-2.25)(0,-2.75)
  \psline{->}(-0.4,1.75)(-0.4,2.125)
  \psline{-<}(-0.4,-1.75)(-0.4,-2.125)
  \psline{->}(-0.4,2.25)(-0.1,2.625)
  \psline{->}(-0.4,2.25)(-0.7,2.625)
  \psline{-<}(-0.4,-2.25)(-0.1,-2.625)
  \psline{-<}(-0.4,-2.25)(-0.7,-2.625)
  \rput[bl]{0}(-0.3,1.8){$f$}
  \rput[bl]{0}(-0.3,-2.2){$f$}
 \endpspicture
=\sum_{w} \frac{\left[\left( F_{\bar{f} f}^{\bar{f} f} \right)^{-1} \right]_{0w}} {d_{a} \sqrt{d_{e}d_{x}}}
 \pspicture[0.49](-2,-5)(2.5,5)
   \small
  \psset{linewidth=0.9pt,linecolor=black,arrowscale=1.5,arrowinset=0.15}
  \psline(-1.2,-0.75)(-1.2,0.75)
  \psline(-1.2,0.75)(-0.8,1.25)
  \psline(-1.2,0.75)(-1.6,1.25)
  \psline(-1.2,-0.75)(-0.8,-1.25)
  \psline(-1.2,-0.75)(-1.6,-1.25)
    \psline{->}(-1.2,-0.25)(-1.2,0.125)
    \psline{->}(-1.2,0.75)(-0.9,1.125)
    \psline{->}(-1.2,0.75)(-1.5,1.125)
    \psline{-<}(-1.2,-0.75)(-0.9,-1.125)
    \psline{-<}(-1.2,-0.75)(-1.5,-1.125)
  \rput[bl]{0}(-1.55,-0.1){$e$}
  \rput[bl]{0}(-0.9,4.35){$\bar{a}$}
  \rput[bl]{0}(-1.75,4.35){$a$}
  \rput[bl]{0}(-0.9,-4.55){$\bar{a}$}
  \rput[bl]{0}(-1.75,-4.55){$a$}
  \small
  \psframe[linewidth=0.9pt,linecolor=black,border=0](-0.35,-0.25)(0.35,0.25)
  \rput[bl]{0}(-0.2,-0.15){$\rho_{0}$}
  \psset{linewidth=0.9pt,linecolor=black,arrowscale=1.5,arrowinset=0.15}
  \psline(0,0.25)(0,1.25)
  \psline(0,-0.25)(0,-1.25)
  \psline(-0.75,0.25)(0,0.75)
  \psline(-0.75,-0.25)(0,-0.75)
  \psline(-0.75,-0.25)(-0.75,0.25)
  \psline{->}(0,0.25)(0,0.65)
  \psline{-<}(0,-0.25)(0,-0.6)
  \psline{->}(0,0.75)(0,1.15)
  \psline{-<}(0,-0.75)(0,-1.15)
  \psline{->}(-0.75,-0.25)(-0.75,0.15)
  \rput[bl](-0.1,4.35){$a$}
  \rput[bl](-0.1,-4.55){$a$}
  \rput[bl](0.15,1){$a$}
  \rput[bl](-0.75,1){$\bar{a}$}
  \rput[bl](0.15,-1.2){$a$}
  \rput[bl](-0.75,-1.2){$\bar{a}$}
  \rput[bl](0.15,0.5){$a$}
  \rput[bl](0.15,-0.7){$a$}
  \rput[bl](-1.05,-0.1){$x$}
  \psline(-1.6,1.25)(-1.6,4.25)
  \psline(-1.6,-1.25)(-1.6,-4.25)
  \psline(-0.8,1.25)(-0.4,1.75)
  \psline(-0.8,-1.25)(-0.4,-1.75)
  \psline(0,1.25)(-0.4,1.75)
  \psline(0,-1.25)(-0.4,-1.75)
  \psline(-0.4,3.75)(-0.8,4.25)
  \psline(-0.4,3.75)(0,4.25)
  \psline(-0.4,-3.75)(-0.8,-4.25)
  \psline(-0.4,-3.75)(0,-4.25)
  \psline(-0.4,1.75)(0,2.25)
  \psline(-0.4,-1.75)(-0,-2.25)
  \psline(0,2.25)(0.8,1.25)
  \psline(0,-2.25)(0.8,-1.25)
  \psline(-0.4,3.75)(1.6,1.25)
  \psline(-0.4,-3.75)(1.6,-1.25)
  \psline(1.6,-1.25)(1.6,1.25)
  \psline(0.8,-1.25)(0.8,1.25)
  \psline(1.6,-0.25)(0.8,0.25)
  \psline{->}(-0.4,3.75)(-0.1,4.125)
  \psline{->}(-0.4,3.75)(-0.7,4.125)
  \psline{-<}(-0.4,-3.75)(-0.1,-4.125)
  \psline{-<}(-0.4,-3.75)(-0.7,-4.125)
  \psline{->}(1.6,-1.25)(1.6,1)
  \psline{->}(0.8,-1.25)(0.8,1)
  \psline{->}(1.6,-1.25)(1.6,-0.75)
  \psline{->}(0.8,-1.25)(0.8,-0.75)
  \psline{->}(1.6,-0.25)(1.1,0.0625)
  \rput[bl]{0}(1.75,0.7){$f$}
  \rput[bl]{0}(1.75,-1.1){$f$}
  \rput[bl]{0}(0.95,0.7){$\bar{f}$}
  \rput[bl]{0}(0.95,-1.1){$\bar{f}$}
  \rput[bl]{0}(1.1,0.2){$w$}
 \endpspicture
\qquad
\label{eq:intteleportPi}
\end{eqnarray}
where the $F$-move applied in the last step is in preparation for applying the decoherence operator $D^{\left( 23\right)}$, which then simply projects onto $w=0$. Applying $D^{\left( 23\right)}$ to the result of Eq.~(\ref{eq:intteleportPi}) gives the post-interferometry measurement state (using isotopy and braidings whose effects cancel)
\begin{eqnarray}
&& \!\!\!\!\!\!\!\!\!\!\!\! \Delta_{f}^{\left(23 \right)} \left[ \rho_{e}^{\text{ER} \left(12\right)} \otimes \rho_{x}^{\left(3 \ldots \right)} \left(a, \ldots \right)  \right]
= D^{\left(23 \right)} \circ \Pi_{f}^{\left(23 \right)} \left[ \rho_{e}^{\text{ER} \left(12\right)} \otimes \rho_{x}^{\left(3 \ldots \right)} \left(a, \ldots \right)  \right] \notag \\
&& \!\!\!\!\!\!\!\! = \frac{1} {d_{a} d_{f} \sqrt{ d_{e}d_{x} }}
 \pspicture[0.490909](-2.2,-2.8)(1.7,2.7)
   \small
  \psset{linewidth=0.9pt,linecolor=black,arrowscale=1.5,arrowinset=0.15}
  \psline(1.2,-1.75)(1.2,1.75)
  \psline(1.2,1.75)(0.8,2.25)
  \psline(1.2,1.75)(1.6,2.25)
  \psline(1.2,-1.75)(0.8,-2.25)
  \psline(1.2,-1.75)(1.6,-2.25)
    \psline{->}(1.2,-1.25)(1.2,0.125)
    \psline{->}(1.2,1.75)(0.9,2.125)
    \psline{->}(1.2,1.75)(1.5,2.125)
    \psline{-<}(1.2,-1.75)(0.9,-2.125)
    \psline{-<}(1.2,-1.75)(1.5,-2.125)
  \rput[bl]{0}(1.35,-0.15){$f$}
  \rput[bl]{0}(0.7,2.35){$\bar{a}$}
  \rput[bl]{0}(1.55,2.35){$a$}
  \rput[bl]{0}(0.7,-2.55){$\bar{a}$}
  \rput[bl]{0}(1.55,-2.55){$a$}
  \small
  \psframe[linewidth=0.9pt,linecolor=black,border=0](-0.35,-0.25)(0.35,0.25)
  \rput[bl]{0}(-0.2,-0.15){$\rho_{0}$}
  \psset{linewidth=0.9pt,linecolor=black,arrowscale=1.5,arrowinset=0.15}
  \psline(0,0.25)(0,2.25)
  \psline(0,-0.25)(0,-2.25)
  \psline(-0.75,0.25)(0,0.75)
  \psline(-0.75,-0.25)(0,-0.75)
  \psline(-0.75,-0.25)(-0.75,0.25)
  \psline(-1.25,0.25)(0,1.25)
  \psline(-1.25,-0.25)(0,-1.25)
  \psline(-1.25,-0.25)(-1.25,0.25)
  \psline(-1.75,0.25)(0,1.75)
  \psline(-1.75,-0.25)(0,-1.75)
  \psline(-1.75,-0.25)(-1.75,0.25)
  \psline{->}(0,0.25)(0,0.65)
  \psline{->}(0,0.25)(0,1.15)
  \psline{->}(0,0.25)(0,1.65)
  \psline{-<}(0,-0.25)(0,-0.6)
  \psline{-<}(0,-0.25)(0,-1.1)
  \psline{-<}(0,-0.25)(0,-1.6)
  \psline{->}(0,0.75)(0,2.15)
  \psline{-<}(0,-0.75)(0,-2.15)
  \psline{->}(-0.75,-0.25)(-0.75,0.15)
  \psline{->}(-1.25,-0.25)(-1.25,0.15)
  \psline{->}(-1.75,-0.25)(-1.75,0.15)
  \rput[bl](-0.1,2.35){$a$}
  \rput[bl](-0.1,-2.55){$a$}
  \rput[bl](0.15,0.5){$a$}
  \rput[bl](0.15,1){$a$}
  \rput[bl](0.15,1.5){$a$}
  \rput[bl](0.15,-0.7){$a$}
  \rput[bl](0.15,-1.2){$a$}
  \rput[bl](0.15,-1.7){$a$}
  \rput[bl](-1.05,-0.1){$x$}
  \rput[bl](-1.55,-0.15){$\bar{f}$}
  \rput[bl](-2.05,-0.1){$e$}
 \endpspicture
=\sum_{y}  \frac{\text{Prob}\left( y |f \right)} {d_{a} \sqrt{d_{f}d_{y}}}
 \pspicture[0.46666](-1.2,-1.5)(1.5,1.5)
   \small
  \psset{linewidth=0.9pt,linecolor=black,arrowscale=1.5,arrowinset=0.15}
  \psline(1.2,-0.75)(1.2,0.75)
  \psline(1.2,0.75)(0.8,1.25)
  \psline(1.2,0.75)(1.6,1.25)
  \psline(1.2,-0.75)(0.8,-1.25)
  \psline(1.2,-0.75)(1.6,-1.25)
    \psline{->}(1.2,-0.25)(1.2,0.125)
    \psline{->}(1.2,0.75)(0.9,1.125)
    \psline{->}(1.2,0.75)(1.5,1.125)
    \psline{-<}(1.2,-0.75)(0.9,-1.125)
    \psline{-<}(1.2,-0.75)(1.5,-1.125)
  \rput[bl]{0}(1.35,-0.15){$f$}
  \rput[bl]{0}(0.7,1.35){$\bar{a}$}
  \rput[bl]{0}(1.55,1.35){$a$}
  \rput[bl]{0}(0.7,-1.55){$\bar{a}$}
  \rput[bl]{0}(1.55,-1.55){$a$}
  \small
  \psframe[linewidth=0.9pt,linecolor=black,border=0](-0.35,-0.25)(0.35,0.25)
  \rput[bl]{0}(-0.2,-0.15){$\rho_{0}$}
  \psset{linewidth=0.9pt,linecolor=black,arrowscale=1.5,arrowinset=0.15}
  \psline(0,0.25)(0,1.25)
  \psline(0,-0.25)(0,-1.25)
  \psline(-0.75,0.25)(0,0.75)
  \psline(-0.75,-0.25)(0,-0.75)
  \psline(-0.75,-0.25)(-0.75,0.25)
  \psline{->}(0,0.25)(0,0.65)
  \psline{-<}(0,-0.25)(0,-0.6)
  \psline{->}(0,0.75)(0,1.15)
  \psline{-<}(0,-0.75)(0,-1.15)
  \psline{->}(-0.75,-0.25)(-0.75,0.15)
  \rput[bl](-0.1,1.35){$a$}
  \rput[bl](-0.1,-1.55){$a$}
  \rput[bl](0.15,0.5){$a$}
  \rput[bl](0.15,-0.7){$a$}
  \rput[bl](-1.05,-0.1){$y$}
 \endpspicture
\qquad \qquad
\end{eqnarray}
which gives our claimed post-interferometry measurement state. In the last step, we used the diagrammatic evaluation
\begin{equation}
 \pspicture[0.490909](-1.2,-2.8)(1.2,2.7)
   \small
  \psset{linewidth=0.9pt,linecolor=black,arrowscale=1.5,arrowinset=0.15}
  \psline(0.5,-1)(0.5,1)
  \psline(0,-1.5)(0,1.5)
  \psline(-0.5,-2)(-0.5,2)
  \psline(1,0.5)(-1,2.5)
  \psline(1,-0.5)(-1,-2.5)
  \psline{->}(0.5,-1)(0.5,0.1)
  \psline{->}(0,-1.5)(0,0.1)
  \psline{->}(-0.5,-2)(-0.5,0.1)
  \psline{->}(1,0.5)(0.65,0.85)
  \psline{->}(1,0.5)(0.15,1.35)
  \psline{->}(1,0.5)(-0.35,1.85)
  \psline{->}(1,0.5)(-0.85,2.35)
  \psline{-<}(1,-0.5)(0.65,-0.85)
  \psline{-<}(1,-0.5)(0.15,-1.35)
  \psline{-<}(1,-0.5)(-0.35,-1.85)
  \psline{-<}(1,-0.5)(-0.85,-2.35)
  \rput[bl](0.15,-0.1){$x$}
  \rput[bl](-0.35,-0.15){$\bar{f}$}
  \rput[bl](-0.85,-0.1){$e$}
  \rput[bl](0.7,0.9){$a$}
  \rput[bl](0.2,1.4){$a$}
  \rput[bl](-0.3,1.9){$a$}
  \rput[bl](-0.8,2.4){$a$}
  \rput[bl](0.7,-1.1){$a$}
  \rput[bl](0.2,-1.6){$a$}
  \rput[bl](-0.3,-2.1){$a$}
  \rput[bl](-0.8,-2.6){$a$}
 \endpspicture
=\sum_{y,q} \left| \left[F_{a}^{e \bar{f} a} \right]_{qa} \right|^{2} \left| \left[F_{a}^{qxa} \right]_{ya} \right|^{2} \sqrt{ \frac{d_{e} d_{f} d_{x} }{d_{y} } }
 \pspicture[0.4857](-0.7,-1.8)(0.7,1.7)
   \small
  \psset{linewidth=0.9pt,linecolor=black,arrowscale=1.5,arrowinset=0.15}
  \psline(0,-1)(0,1)
  \psline(0.5,0.5)(-0.5,1.5)
  \psline(0.5,-0.5)(-0.5,-1.5)
  \psline{->}(0,-1)(0,0.1)
  \psline{->}(0.5,0.5)(0.15,0.85)
  \psline{->}(0.5,0.5)(-0.35,1.35)
  \psline{-<}(0.5,-0.5)(0.15,-0.85)
  \psline{-<}(0.5,-0.5)(-0.35,-1.35)
  \rput[bl](-0.35,-0.1){$y$}
  \rput[bl](0.2,0.9){$a$}
  \rput[bl](-0.3,1.4){$a$}
  \rput[bl](0.2,-1.1){$a$}
  \rput[bl](-0.3,-1.6){$a$}
 \endpspicture
.
\end{equation}

After the interferometry measurement $\Delta_{f}^{\left(23 \right)}$, the entanglement resource has moved to anyons $2$ and $3$ and has no entanglement with the anyons encoding the state $\rho_{0}$. However, we are left with incoherent superpositions of the charge $y$, and the possibility of having an undesired non-trivial charge $y \neq 0$. We now measure this charge $y$ to determine whether or not we have achieved the desired result $y=0$. This is done by performing an interferometry measurement on anyons $1,\ldots$, where ``$\ldots$'' stands for all other anyons that have non-trivial anyonic charge entanglement with anyon $1$, which of course now excludes anyons $2$ and $3$. This measurement will have outcome $z$ (given the previous measurement outcome $f$) with probability $\text{Prob}\left(z |f \right)$, and result in the post-measurement state
\begin{equation}
\Delta_{z}^{\left( 1\ldots \right)} \left[ \sum_{y} \text{Prob}\left( y | f \right) \rho_{y}^{\left(1 \ldots \right)} \left(a, \ldots \right) \right]
=\rho_{z}^{\left(1 \ldots \right)} \left(a, \ldots \right)
.
\end{equation}
Since the overall charge $y$ was already fully decohered, this interferometry measurement is simply a projective measurement of the incoherent superpositions of $y$ (i.e. this is essentially just a classical measurement).

Thus, the total probability of ending up with $\rho_{z}^{\left(1 \ldots \right)} \left(a, \ldots \right)$ after performing these two measurement ($\Delta^{\left( 23 \right)}$ and $\Delta^{\left( 1\ldots \right)}$) on the state $\rho_{e}^{\text{ER} \left(12\right)} \otimes \rho_{x}^{\left(3 \ldots \right)} \left(a, \ldots \right)$ is
\begin{eqnarray}
\text{Prob} \left( z \right) &=& \sum_{f} \text{Prob} \left( z | f \right) \text{Prob} \left( f \right) \\
&=& \sum_{f,q} N_{a \bar{a}}^{f} \frac{d_{f}}{ d_{a}^{2}} \left| \left[F_{a}^{e \bar{f} a} \right]_{qa} \right|^{2} \left| \left[F_{a}^{qxa} \right]_{za} \right|^{2}
\end{eqnarray}
and in particular, the probability of achieving the desired outcome $z=0$ from this process is
\begin{equation}
\text{Prob} \left( z=0 \right) = d_{a}^{-2}
\label{eq:probintsuccess}
.
\end{equation}
To obtain this result, we combined the relation
\begin{equation}
\left| \left[F_{a}^{e \bar{f} a} \right]_{\bar{x} a} \right| = \sqrt{\frac{d_{x}}{d_{f}}} \left| \left[F_{a}^{x e a} \right]_{fa} \right|
,
\end{equation}
which can be derived from the pentagon equations, with unitarity of the $F$-moves to perform the sum
\begin{equation}
\sum_{f} N_{a \bar{a}}^{f} \frac{d_{f}}{d_{x}} \left| \left[F_{a}^{e \bar{f} a} \right]_{\bar{x} a} \right|^{2} = \sum_{f} N_{a \bar{a}}^{f} \left| \left[F_{a}^{x e a} \right]_{f a} \right|^{2} = 1
.
\end{equation}

If the measurement results in an undesired outcome $z \neq 0$, we now undo the ``failed'' teleportation by performing an interferometry measurement $\Delta^{\left( 12 \right)}$ on anyons $1$ and $2$. For measurement outcome $e_{2}$, this gives the post-measurement state
\begin{equation}
\Delta_{e_{2}}^{\left(12 \right)} \left[ \rho_{f}^{\text{ER} \left(32 \right)} \otimes \rho_{z}^{\left(1 \ldots \right)} \left(a, \ldots \right) \right]
= \sum_{x_{2}} \text{Prob}\left( x_{2} | e_{2} \right) \rho_{e_{2}}^{\text{ER} \left(12\right)} \otimes \rho_{x_{2} }^{\left(3 \ldots \right)} \left(a, \ldots \right)
\end{equation}
One could now measure $x_{2}$, but this provides no important information and has no significant effect on the forced measurement process, so we will not do so in order to be more efficient with our measurements (especially the more difficult ones involving more than two anyons). Leaving the charge $x_{2}$ unmeasured will simply introduce a sum over the probabilities of having charge $x_{2}$ in the next step. Now we can repeat the measurements $\Delta^{\left( 23 \right)}$ and $\Delta^{\left( 1\ldots \right)}$ with a new chance of obtaining the desired outcome $z_{2}=0$. This procedure may be repeated until the desired outcome is achieved on the $n$th attempt, giving the set of measurement outcomes
\begin{equation}
M = \left\{ e_{1},f_{1},z_{1}, \ldots ,e_{n},f_{n},z_{n} \right\}
\end{equation}
where $z_{n}=0$ and $z_{j} \neq 0$ for $j<n$ (and the initialization $e_{1}$ is included for convenience). The probabilities of measurement outcomes $e_{j}$, $f_{j}$, and $z_{j}$ are respectively
\begin{eqnarray}
\text{Prob} \left( e_{j} \right) &=& N_{a \bar{a}}^{e_{j}} \frac{d_{e_{j}}}{d_{a}^{2}} \\
\text{Prob} \left( f_{j} \right) &=& N_{a \bar{a}}^{f_{j}} \frac{d_{f_{j}}}{d_{a}^{2}} \\
\text{Prob} \left( z_{j} \right) &=& \sum_{f_{j}} \text{Prob} \left( z_{j} | f_{j} \right) \text{Prob} \left( f_{j} \right) \notag \\
&=& \sum_{f_{j},q} N_{a \bar{a}}^{f_{j}} \frac{d_{f_{j}}}{ d_{a}^{2}} \left| \left[F_{a}^{e_{j} \bar{f_{j}} a} \right]_{qa} \right|^{2} \left| \left[F_{a}^{qx_{j}a} \right]_{z_{j}a} \right|^{2}
.
\end{eqnarray}
Of course, the only probability that matters for our purposes is the total probability $\text{Prob} \left( z_{j}=0 \right) = d_{a}^{-2}$ from Eq.~(\ref{eq:probintsuccess}) of achieving the desired outcome $z_{j}=0$ in each attempt, which is independent of $j$. From this, the average number of attempts $n$ until a desired $z_{n}=0$ outcome is achieved in an interferometry forced measurement is found to be
\begin{equation}
\left\langle n \right\rangle = d_{a}^{2},
\end{equation}
and the probability of needing $n>N$ attempts to obtain the desired outcome is exponentially suppressed
\begin{equation}
\text{Prob}\left(z_{1},\ldots,z_{N} \neq 0 \right) = \left( 1-d_{a}^{-2} \right)^{N}
,
\end{equation}
just as in the previous cases of forced measurement teleportation. For the non-Abelian anyon models considered in this paper
\begin{equation}
\begin{array}{ll}
\text{Prob} \left( z_{j}=0 \right) = \frac{1}{d_{a}^{2}} = \frac{1}{2} & \text{ for Ising using } a=\frac{1}{2} = \sigma \\
\text{Prob} \left( z_{j}=0 \right) = \frac{1}{d_{a}^{2}} =\phi^{-2} \approx .38 & \text{ for Fib using } a=1 = \varepsilon \\
\text{Prob} \left( z_{j}=0 \right) = \frac{1}{d_{a}^{2}} =\frac{1}{4 \cos^{2} \left( \frac{\pi}{k+2} \right)} \quad & \text{ for SU$(2)_{k}$ using } a=\frac{1}{2}
\end{array}
\end{equation}

Thus, the forced interferometry measurement given by the probabilistically determined adaptive series of interferometry measurements
\begin{equation}
\breve{\Delta}_{M}^{\left( 32 \leftarrow 12 \right)} = \Delta_{z_{n} = 0}^{\left( 1 \ldots \right)} \circ \Delta_{f_{n}}^{\left( 23 \right)} \circ \Delta_{e_{n}}^{\left( 12 \right)} \circ \ldots \circ \Delta_{z_{1}}^{\left( 1 \ldots \right)} \circ \Delta_{f_{1}}^{\left( 23 \right)} \circ \Delta_{e_{1} }^{\left( 12 \right)}
\end{equation}
enables us to perform anyonic state teleportation
\begin{equation}
\breve{\Delta}_{M}^{\left( 32 \leftarrow 12 \right)} \left[ \rho_{e_{1}}^{\text{ER} \left(12\right)} \otimes \rho_{0 }^{\left(3 \ldots \right)} \left(a, \ldots \right) \right] = \rho_{f_{n}}^{\text{ER} \left(32 \right)} \otimes \rho_{0}^{\left(1 \ldots \right)} \left(a, \ldots \right)
\end{equation}
using interferometry measurements. We emphasize that while it is important to perform all the $\Delta^{\left( 12 \right)}$ and $\Delta^{\left( 23 \right)}$ measurements in order to teleport the state information, the actual outcomes $e_{j}$ and $f_{j}$ of these measurements are unimportant.

It is sometimes possible to use knowledge of measurement outcomes previous to a $y_{j}$ measurement (usually at or near the beginning of a forced measurement) to infer (by classical information processing) that the probability of a measurement with outcome $z_{j}=0$ is either $0$ or $1$, thus making the measurement unnecessary (e.g. if any two of $e$, $x$, and $f$ are known to be Abelian). This could help with the efficiency of measurement use, but only some fraction of the time.

\section{Measurement-Generated Braiding Transformations}

As motivation for this section (and in fact for the entire paper), we consider the state of a number anyons, two of which have charge $a$. Focusing only on these two anyons, we write the state as
\begin{equation}
\left| \psi \left(a,a,\ldots \right) \right\rangle =
 \pspicture[0.35](-1,-.5)(1.1,1.5)
  \small
  \psframe[linewidth=0.9pt,linecolor=black,border=0](-0.75,-0.25)(0.75,0.25)
  \rput[bl]{0}(-0.15,-0.15){$\psi$}
  \psset{linewidth=0.9pt,linecolor=black,arrowscale=1.5,arrowinset=0.15}
  \psline(0.6,0.25)(0.6,1)
  \psline(-0.6,0.25)(-0.6,1)
  \psline{->}(0.6,0.25)(0.6,0.8)
  \psline{->}(-0.6,0.25)(-0.6,0.8)
  \rput[bl](0.5,1.1){$a$}
  \rput[bl](-0.7,1.1){$a$}
 \endpspicture
.
\end{equation}
If we introduce a charge conjugate pair
\begin{equation}
\left| a,\bar{a};0 \right\rangle = \frac{1}{\sqrt{d_{a}}}
  \pspicture[0.35294](0.4,-0.05)(1.6,0.8)
  \small
  \psset{linewidth=0.9pt,linecolor=black,arrowscale=1.5,arrowinset=0.15}
  \psline(1,0)(0.6,0.5)
  \psline(1,0)(1.4,0.5)
   \psline{->}(1,0)(0.7,0.375)
   \psline{->}(1,0)(1.3,0.375)
   \rput[bl]{0}(0.5,0.6){$a$}
   \rput[bl]{0}(1.35,0.6){$\bar{a}$}
  \endpspicture
\end{equation}
nearby (tensor these states together), then applying three projective topological charge measurements, all with outcomes $0$, we have
\begin{eqnarray}
&&\Pi^{\left(23\right)}_{0} \circ \Pi^{\left(24\right)}_{0} \circ \Pi^{\left(12\right)}_{0} \left[
\left| \bar{a},a;0 \right\rangle_{\left(23\right)} \otimes \left| \psi \left(a,a,\ldots \right) \right\rangle_{\left(14 \ldots \right)} \right] \notag \\
&& \qquad = \frac{1}{\sqrt{d_{a}}}
 \pspicture[0.4634](-1,-.6)(1.1,3.5)
  \small
  \psframe[linewidth=0.9pt,linecolor=black,border=0](-0.75,-0.25)(0.75,0.25)
  \rput[bl]{0}(-0.15,-0.15){$\psi$}
  \psset{linewidth=0.5pt,linecolor=black,arrowscale=1.0,arrowinset=0.15}
  \psline(0.6,0.25)(0.6,0.75)
  \psline(-0.6,0.25)(-0.6,0.75)
  \psline(0,0.5)(0.2,0.75)
  \psline(0,0.5)(-0.2,0.75)
  \psline(-0.6,0.75)(-0.4,1)
  \psline(-0.2,0.75)(-0.4,1)
  \psline(-0.6,1.5)(-0.4,1.25)
  \psline(-0.2,1.5)(-0.4,1.25)
  \psline(0.2,0.75)(0.2,1.5)
  \psline(0.6,0.75)(0.6,1.5)
  \psline(-0.6,1.5)(-0.6,2.25)
  \psline(-0.2,1.5)(0.2,1.75)
  \psline(0.6,1.5)(0.2,1.75)
  \psline(0.2,2)(-0.2,2.25)
  \psline(0.2,2)(0.6,2.25)
  \psline(-0.6,2.25)(-0.6,3)
  \psline(0.6,2.25)(0.6,3)
  \psline(0,2.5)(0.2,2.25)
  \psline(0,2.5)(-0.2,2.25)
  \psline(0,2.75)(0.2,3)
  \psline(0,2.75)(-0.2,3)
  \psline(0.2,1.5)(0.2,2.25)
    \psline[border=1.2pt](0.2,1.55)(0.2,2.15)
  \psline{->}(0.6,0.25)(0.6,0.6875)
  \psline{->}(-0.6,0.25)(-0.6,0.6875)
  \psline{->}(0,0.5)(0.15,0.6875)
  \psline{->}(0,0.5)(-0.15,0.6875)
  \psline{->}(0,2.75)(0.15,2.9375)
  \psline{->}(0,2.75)(-0.15,2.9375)
  \psline{->}(0.6,2.75)(0.6,2.9375)
  \psline{->}(-0.6,2.75)(-0.6,2.9375)
   \psset{linewidth=0.25pt,linecolor=black,linestyle=dashed}
  \psline(-0.8,0.75)(0.8,0.75)
  \psline(-0.8,1.5)(0.8,1.5)
  \psline(-0.8,2.25)(0.8,2.25)
  \rput[bl](0.5,3.1){$a$}
  \rput[bl](-0.7,3.1){$a$}
  \rput[bl](0.1,3.1){$a$}
  \rput[bl](-0.3,3.1){$\bar{a}$}
 \endpspicture
= \frac{\theta_{a}^{\ast}}{\sqrt{d_{a}}}
 \pspicture[0.4615](-1,-.5)(1.1,2.1)
  \small
  \psframe[linewidth=0.9pt,linecolor=black,border=0](-0.75,-0.25)(0.75,0.25)
  \rput[bl]{0}(-0.15,-0.15){$\psi$}
  \psset{linewidth=0.5pt,linecolor=black,arrowscale=1.0,arrowinset=0.15}
  \psline(0.6,0.25)(-0.6,1.75)
  \psline(0,1.5)(0.2,1.75)
  \psline(0,1.5)(-0.2,1.75)
  \psline(-0.6,0.25)(0.6,1.75)
  \psline{->}(0.6,0.25)(-0.55,1.6875)
  \psline{->}(-0.6,0.25)(0.55,1.6875)
  \psline{->}(0,1.5)(0.15,1.6875)
  \psline{->}(0,1.5)(-0.15,1.6875)
  \psline[border=1.5pt](-0.4,0.5)(0.4,1.5)
  \rput[bl](0.5,1.85){$a$}
  \rput[bl](-0.7,1.85){$a$}
  \rput[bl](0.1,1.85){$a$}
  \rput[bl](-0.3,1.85){$\bar{a}$}
 \endpspicture
\notag \\
&& \qquad = \theta_{a}^{\ast} \left| \bar{a},a;0 \right\rangle_{\left( 23 \right)} \otimes R^{\left(14\right)}_{aa} \left| \psi \left(a,a,\ldots \right) \right\rangle_{\left( 14 \ldots \right)}
\end{eqnarray}
where $\theta_{a} = e^{i 2 \pi s_{a}}$ is a phase, known as the topological spin of $a$, obtained from straightening out a counterclockwise twist in a charge $a$ line. The dashed line partitions have been inserted to help clarify the contributions to the diagram from each of the three projectors. From this we see that topological charge measurement could potentially be used to generate a braiding transformation. Specifically, performing these three measurements with vacuum $0$ as their charge measurement outcomes transforms the state $\psi$ exactly the same way it would if anyons $1$ and $4$ were exchanged by a counterclockwise half twist (i.e. braided), up to an overall phase. However, we know that measurement outcomes are probabilistic, so the only way we could achieve such a transformation via topological charge measurements is if we had some way to force each measurement to have the desired vacuum charge outcome. Of course, this is exactly what we have produced in the Section~\ref{sec:Teleportation}, where we demonstrated that one could perform teleportation through a ``forced measurement'' procedure.

Based on this observation, we can now produce braiding transformations
\begin{equation}
R_{aa}=
 \pspicture[0.4](-0.1,0)(1.4,1)
  \psset{linewidth=0.9pt,linecolor=black,arrowscale=1.5,arrowinset=0.15}
  \psline(0.96,0.05)(0.2,1)
  \psline{->}(0.96,0.05)(0.28,0.9)
  \psline(0.24,0.05)(1,1)
  \psline[border=2pt]{->}(0.24,0.05)(0.92,0.9)
  \rput[bl]{0}(-0.05,0.1){$a$}
  \rput[br]{0}(1.25,0.1){$a$}
  \endpspicture
,\qquad
R_{aa}^{-1}= R_{aa}^{\dag}=
 \pspicture[0.4](-0.1,0)(1.4,1)
  \psset{linewidth=0.9pt,linecolor=black,arrowscale=1.5,arrowinset=0.15}
  \psline{->}(0.24,0.05)(0.92,0.9)
  \psline(0.24,0.05)(1,1)
  \psline(0.96,0.05)(0.2,1)
  \psline[border=2pt]{->}(0.96,0.05)(0.28,0.9)
  \rput[bl]{0}(-0.05,0.1){$a$}
  \rput[br]{0}(1.25,0.1){$a$}
  \endpspicture
,
\end{equation}
for two anyons of definite charge $a$ by introducing an appropriate entanglement resource near these anyons and performing three consecutive forced measurement teleportations.

We assume these two anyons (partially) comprise the state
\begin{equation}
\rho \left(a,a,\ldots \right)=
 \pspicture[0.46666](-1,-1.5)(1.1,1.5)
  \small
  \psframe[linewidth=0.9pt,linecolor=black,border=0](-0.75,-0.25)(0.75,0.25)
  \rput[bl]{0}(-0.1,-0.15){$\rho$}
  \psset{linewidth=0.9pt,linecolor=black,arrowscale=1.5,arrowinset=0.15}
  \psline(0.6,0.25)(0.6,1)
  \psline(-0.6,0.25)(-0.6,1)
  \psline(0.6,-0.25)(0.6,-1)
  \psline(-0.6,-0.25)(-0.6,-1)
  \psline{->}(0.6,0.25)(0.6,0.8)
  \psline{->}(-0.6,0.25)(-0.6,0.8)
  \psline{-<}(0.6,-0.25)(0.6,-0.8)
  \psline{-<}(-0.6,-0.25)(-0.6,-0.8)
  \rput[bl](0.5,1.1){$a$}
  \rput[bl](-0.7,1.1){$a$}
  \rput[bl](0.5,-1.3){$a$}
  \rput[bl](-0.7,-1.3){$a$}
 \endpspicture
.
\end{equation}
Using projective topological charge measurements, we introduce the entanglement resource
\begin{equation}
\rho^{\text{ER}} = \left| \bar{a},a;0 \right\rangle \left\langle \bar{a},a ;0\right|
\end{equation}
and perform three forced projective measurements to get
\begin{eqnarray}
&&\breve{\Pi}_{M_{3}}^{\left(23\leftarrow 24 \right)} \circ \breve{\Pi}_{M_{2}}^{\left(24\leftarrow 21 \right)} \circ \breve{\Pi}_{M_{1}}^{\left(21 \leftarrow 23 \right)} \left[ \rho^{\text{ER} \left(23\right)} \otimes \rho^{\left( 14 \ldots \right) }\left(a,a,\ldots \right) \right] \notag \\
&&= \breve{\Pi}_{M_{3}}^{\left(23\leftarrow 24 \right)} \circ \breve{\Pi}_{M_{2}}^{\left(24\leftarrow 21 \right)} \circ \breve{\Pi}_{M_{1}}^{\left(21 \leftarrow 23 \right)} \left[
\pspicture[0.5](-1,-1.3)(1,1.1)
  \small
  \psframe[linewidth=0.9pt,linecolor=black,border=0](-0.75,-0.25)(0.75,0.25)
  \rput[bl]{0}(-0.1,-0.15){$\rho$}
  \psset{linewidth=0.5pt,linecolor=black,arrowscale=1.0,arrowinset=0.15}
  \psline(0.6,0.25)(0.6,0.75)
  \psline(0,0.5)(0.2,0.75)
  \psline(0,0.5)(-0.2,0.75)
  \psline(-0.6,0.25)(-0.6,0.75)
  \psline(0.6,-0.25)(0.6,-0.75)
  \psline(0,-0.5)(0.2,-0.75)
  \psline(0,-0.5)(-0.2,-0.75)
  \psline(-0.6,-0.25)(-0.6,-0.75)
  \psline{->}(0.6,0.25)(0.6,0.6875)
  \psline{->}(-0.6,0.25)(-0.6,0.6875)
  \psline{->}(0,0.5)(0.15,0.6875)
  \psline{->}(0,0.5)(-0.15,0.6875)
  \psline{-<}(0.6,-0.25)(0.6,-0.6875)
  \psline{-<}(-0.6,-0.25)(-0.6,-0.6875)
  \psline{-<}(0,-0.5)(0.15,-0.6875)
  \psline{-<}(0,-0.5)(-0.15,-0.6875)
  \rput[bl](0.5,0.85){$a$}
  \rput[bl](-0.7,0.85){$a$}
  \rput[bl](0.1,0.85){$a$}
  \rput[bl](-0.3,0.85){$\bar{a}$}
  \rput[bl](0.5,-1.05){$a$}
  \rput[bl](-0.7,-1.05){$a$}
  \rput[bl](0.1,-1.05){$a$}
  \rput[bl](-0.3,-1.05){$\bar{a}$}
 \endpspicture
\right]
= \pspicture[0.5](-1,-2.3)(1,2.1)
  \small
  \psframe[linewidth=0.9pt,linecolor=black,border=0](-0.75,-0.25)(0.75,0.25)
  \rput[bl]{0}(-0.1,-0.15){$\rho$}
  \psset{linewidth=0.5pt,linecolor=black,arrowscale=1.0,arrowinset=0.15}
  \psline(0.6,0.25)(-0.6,1.75)
  \psline(0,1.5)(0.2,1.75)
  \psline(0,1.5)(-0.2,1.75)
  \psline(-0.6,0.25)(0.6,1.75)
  \psline(0.6,-0.25)(-0.6,-1.75)
  \psline(0,-1.5)(0.2,-1.75)
  \psline(0,-1.5)(-0.2,-1.75)
  \psline(-0.6,-0.25)(0.6,-1.75)
  \psline{->}(0.6,0.25)(-0.55,1.6875)
  \psline{->}(-0.6,0.25)(0.55,1.6875)
  \psline{->}(0,1.5)(0.15,1.6875)
  \psline{->}(0,1.5)(-0.15,1.6875)
  \psline[border=1.5pt](-0.4,0.5)(0.4,1.5)
  \psline{-<}(0.6,-0.25)(-0.55,-1.6875)
  \psline{-<}(-0.6,-0.25)(0.55,-1.6875)
  \psline{-<}(0,-1.5)(0.15,-1.6875)
  \psline{-<}(0,-1.5)(-0.15,-1.6875)
  \psline[border=1.5pt](-0.4,-0.5)(0.4,-1.5)
  \rput[bl](0.5,1.85){$a$}
  \rput[bl](-0.7,1.85){$a$}
  \rput[bl](0.1,1.85){$a$}
  \rput[bl](-0.3,1.85){$\bar{a}$}
  \rput[bl](0.5,-2.05){$a$}
  \rput[bl](-0.7,-2.05){$a$}
  \rput[bl](0.1,-2.05){$a$}
  \rput[bl](-0.3,-2.05){$\bar{a}$}
 \endpspicture
\notag \\
&&= \rho^{\text{ER} \left(23\right)} \otimes R_{aa}^{\left(14\right)} \rho^{\left( 14 \ldots \right)}\left(a,a,\ldots \right) R_{aa}^{\dagger\left(14\right)}
,
\label{eq:projrhoR}
\end{eqnarray}
where the forced projective measurements used here are given by
\begin{eqnarray}
&& \breve{\Pi}_{M_{1}}^{\left(21 \leftarrow 23 \right)} = \Pi_{f_{n_{1}}=0}^{\left(21 \right)} \circ \Pi_{e_{n_{1}}}^{\left(23 \right)} \circ
\ldots \circ \Pi_{f_{1_{1}}}^{\left(21 \right)} \circ \Pi_{e_{1_{1}}=0}^{\left(23 \right)} \\
&& \breve{\Pi}_{M_{2}}^{\left(24 \leftarrow 21 \right)} = \Pi_{f_{n_{2}}=0}^{\left(24 \right)} \circ \Pi_{e_{n_{2}}}^{\left(21 \right)} \circ
\ldots \circ \Pi_{f_{1_{2}}}^{\left(24 \right)} \circ \Pi_{e_{1_{2}}=0}^{\left(21 \right)}\\
&& \breve{\Pi}_{M_{3}}^{\left(23 \leftarrow 24 \right)} = \Pi_{f_{n_{3}}=0}^{\left(23 \right)} \circ \Pi_{e_{n_{3}}}^{\left(24 \right)} \circ
\ldots \circ \Pi_{f_{1_{3}}}^{\left(23 \right)} \circ \Pi_{e_{1_{3}}=0}^{\left(24 \right)}
.
\end{eqnarray}
In these series, $e_{1_{1}}=0$ by the initialization assumption, $e_{1_{2}} = f_{n_{1}}$, and $e_{1_{3}} = f_{n_{2}}$, so the first measurement of each forced measurement series is actually already done by the previous forced measurement series and hence need not be repeated. Similarly, we have
\begin{eqnarray}
&&\breve{\Pi}_{M_{3}}^{\left(23\leftarrow 21 \right)} \circ \breve{\Pi}_{M_{2}}^{\left(21\leftarrow 24 \right)} \circ \breve{\Pi}_{M_{1}}^{\left(24 \leftarrow 23 \right)} \left[ \rho^{\text{ER} \left(23\right)} \otimes \rho^{\left( 14 \ldots \right) }\left(a,a,\ldots \right) \right] \notag \\
&& \qquad \qquad \qquad \qquad \qquad = \rho^{\text{ER} \left(23\right)} \otimes R_{aa}^{\dagger \left(14\right)} \rho^{\left( 14 \ldots \right)}\left(a,a,\ldots \right) R_{aa}^{\left(14\right)}.
\label{eq:projrhoR*}
\end{eqnarray}
using the forced projective measurements
\begin{eqnarray}
&& \breve{\Pi}_{M_{1}}^{\left(24 \leftarrow 23 \right)} = \Pi_{f_{n_{1}}=0}^{\left(24 \right)} \circ \Pi_{e_{n_{1}}}^{\left(23 \right)} \circ
\ldots \circ \Pi_{f_{1_{1}}}^{\left(24 \right)} \circ \Pi_{e_{1_{1}}=0}^{\left(23 \right)} \\
&& \breve{\Pi}_{M_{2}}^{\left(21\leftarrow 24 \right)} = \Pi_{f_{n_{2}}=0}^{\left(21 \right)} \circ \Pi_{e_{n_{2}}}^{\left(24 \right)} \circ
\ldots \circ \Pi_{f_{1_{2}}}^{\left(21 \right)} \circ \Pi_{e_{1_{2}}=0}^{\left(24 \right)}\\
&& \breve{\Pi}_{M_{3}}^{\left(23\leftarrow 21 \right)} = \Pi_{f_{n_{3}}=0}^{\left(23 \right)} \circ \Pi_{e_{n_{3}}}^{\left(21 \right)} \circ
\ldots \circ \Pi_{f_{1_{3}}}^{\left(23 \right)} \circ \Pi_{e_{1_{3}}=0}^{\left(21 \right)}
.
\end{eqnarray}
Eqs.~(\ref{eq:projrhoR},\ref{eq:projrhoR*}) respectively generate the counterclockwise and clockwise braiding transformations of anyons $1$ and $4$ on the state $\rho$. Thus, with the introduction of an appropriate entanglement resource $\rho^{\text{ER}}$ pair of anyons at positions $2$ and $3$, we can schematically write appropriate series of projective topological charge measurements as being equivalent to braiding transformations of anyons of charge $a$ at positions $1$ and $4$
\begin{eqnarray}
\breve{\Pi}_{M_{3}}^{\left(23\leftarrow 24 \right)} \circ \breve{\Pi}_{M_{2}}^{\left(24\leftarrow 21 \right)} \circ \breve{\Pi}_{M_{1}}^{\left(21 \leftarrow 23 \right)} &\cong& R_{aa}^{ \left(14\right)}
\label{eq:projR} \\
\breve{\Pi}_{M_{3}}^{\left(23\leftarrow 21 \right)} \circ \breve{\Pi}_{M_{2}}^{\left(21\leftarrow 24 \right)} \circ \breve{\Pi}_{M_{1}}^{\left(24 \leftarrow 23 \right)} &\cong & R_{aa}^{\dagger \left(14\right)} .
\label{eq:projR*}
\end{eqnarray}

\begin{figure}[t!]
\begin{center}
  \includegraphics[scale=.7]{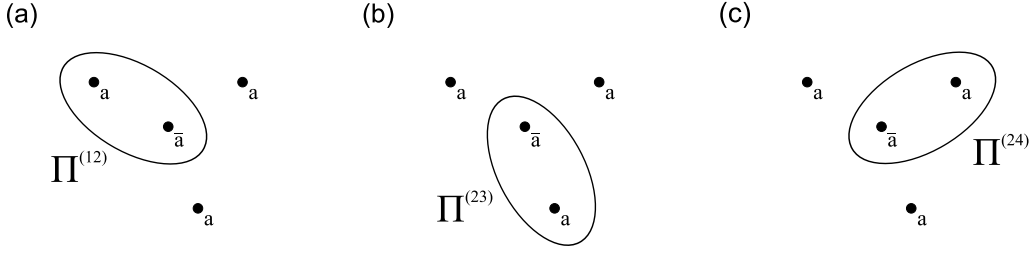}
  \caption{Projective topological charge measurements of pairs of anyons (a) $1$ and $2$, (b) $2$ and $3$, and (c) $2$ and $4$ are used to implement forced projective measurement anyonic state teleportation, which is used to produce braiding transformations as in Eqs.~(\ref{eq:projrhoR}\,--\,\ref{eq:projR*}). The ovals delineate the areas for which the contained collective topological charge is being measured. \newline}
  \label{fig:projmeasurements}
\end{center}
\end{figure}

When using interferometry measurements of topological charge, we restrict the state $\rho \left(a,a,\ldots \right) =\rho_{0}\left(a,a,\ldots \right)$ to have overall charge $0$ (because this restriction was needed to perform forced interferometry measurement anyonic teleportation in Section~\ref{sec:Interferometry_Teleportation}), and introduce the entanglement resource
\begin{equation}
\rho_{e}^{\text{ER}} = \frac{1}{d_{e}} \left| \bar{a},a;e \right\rangle \left\langle \bar{a},a ;e\right|
\end{equation}
where the particular value of $e$ is unimportant. To generate the braiding transformations, we perform three forced interferometry measurements
\begin{eqnarray}
&& \breve{\Delta}_{M_{3}}^{\left(23\leftarrow 24 \right)} \circ \breve{\Delta}_{M_{2}}^{\left(24\leftarrow 21 \right)} \circ \breve{\Delta}_{M_{1}}^{\left(21 \leftarrow 23 \right)} \left[ \rho_{e}^{\text{ER} \left(23\right)} \otimes \rho_{0}^{\left( 14 \ldots \right) } \left(a,a,\ldots \right) \right] \notag \\
&&= \breve{\Delta}_{M_{3}}^{\left(23\leftarrow 24 \right)} \circ \breve{\Delta}_{M_{2}}^{\left(24\leftarrow 21 \right)} \circ \breve{\Delta}_{M_{1}}^{\left(21 \leftarrow 23 \right)} \left[
\pspicture[0.5](-1,-1.3)(1,1.1)
  \small
  \psframe[linewidth=0.9pt,linecolor=black,border=0](-0.75,-0.25)(0.75,0.25)
  \rput[bl]{0}(-0.25,-0.15){$\rho_{0}$}
  \psset{linewidth=0.5pt,linecolor=black,arrowscale=1.0,arrowinset=0.15}
  \psline(0.6,0.25)(0.6,0.75)
  \psline(0,0.5)(0.2,0.75)
  \psline(0,0.5)(-0.2,0.75)
  \psline(-0.6,0.25)(-0.6,0.75)
  \psline(0.6,-0.25)(0.6,-0.75)
  \psline(0,-0.5)(0.2,-0.75)
  \psline(0,-0.5)(-0.2,-0.75)
  \psline(-0.6,-0.25)(-0.6,-0.75)
  \psline(0,0.5)(0.3,0.125)
  \psline(0,-0.5)(0.3,-0.125)
  \psline(0.3,-0.125)(0.3,0.125)
  \psline[border=1.5pt](0.1,0.375)(0.25,0.1875)
  \psline[border=1.5pt](0.1,-0.375)(0.25,-0.1875)
  \psline{->}(0.6,0.25)(0.6,0.6875)
  \psline{->}(-0.6,0.25)(-0.6,0.6875)
  \psline{->}(0,0.5)(0.15,0.6875)
  \psline{->}(0,0.5)(-0.15,0.6875)
  \psline{-<}(0.6,-0.25)(0.6,-0.6875)
  \psline{-<}(-0.6,-0.25)(-0.6,-0.6875)
  \psline{-<}(0,-0.5)(0.15,-0.6875)
  \psline{-<}(0,-0.5)(-0.15,-0.6875)
  \psline{->}(0.3,-0.125)(0.3,0.08)
  \rput[bl](0.5,0.85){$a$}
  \rput[bl](-0.7,0.85){$a$}
  \rput[bl](0.1,0.85){$a$}
  \rput[bl](-0.3,0.85){$\bar{a}$}
  \rput[bl](0.5,-1.05){$a$}
  \rput[bl](-0.7,-1.05){$a$}
  \rput[bl](0.1,-1.05){$a$}
  \rput[bl](-0.3,-1.05){$\bar{a}$}
  \scriptsize
    \rput[bl]{0}(0.35,-0.05){$e$}
 \endpspicture
\right]
= \pspicture[0.5](-1,-2.3)(1.3,2.1)
  \small
  \psframe[linewidth=0.9pt,linecolor=black,border=0](-0.75,-0.25)(0.75,0.25)
  \rput[bl]{0}(-0.15,-0.15){$\rho_{0}$}
  \psset{linewidth=0.5pt,linecolor=black,arrowscale=1.0,arrowinset=0.15}
  \psline(0.6,0.25)(-0.6,1.75)
  \psline(0,1.5)(0.2,1.75)
  \psline(0,1.5)(-0.2,1.75)
  \psline(-0.6,0.25)(0.6,1.75)
  \psline(0.6,-0.25)(-0.6,-1.75)
  \psline(0,-1.5)(0.2,-1.75)
  \psline(0,-1.5)(-0.2,-1.75)
  \psline(-0.6,-0.25)(0.6,-1.75)
  \psline(0,1.5)(1.05,0.1875)
  \psline(0,-1.5)(1.05,-0.1875)
  \psline(1.05,-0.1875)(1.05,0.1875)
  \psline{->}(0.6,0.25)(-0.55,1.6875)
  \psline{->}(-0.6,0.25)(0.55,1.6875)
  \psline{->}(0,1.5)(0.15,1.6875)
  \psline{->}(0,1.5)(-0.15,1.6875)
  \psline[border=1.5pt](-0.4,0.5)(0.1,1.125)
  \psline[border=1.5pt](0.1,1.375)(0.9,0.375)
  \psline{-<}(0.6,-0.25)(-0.55,-1.6875)
  \psline{-<}(-0.6,-0.25)(0.55,-1.6875)
  \psline{-<}(0,-1.5)(0.15,-1.6875)
  \psline{-<}(0,-1.5)(-0.15,-1.6875)
  \psline[border=1.5pt](-0.4,-0.5)(0.1,-1.125)
  \psline[border=1.5pt](0.1,-1.375)(0.9,-0.375)
  \psline{->}(1.05,-0.1875)(1.05,0.08)
  \rput[bl](0.5,1.85){$a$}
  \rput[bl](-0.7,1.85){$a$}
  \rput[bl](0.1,1.85){$a$}
  \rput[bl](-0.3,1.85){$\bar{a}$}
  \rput[bl](0.5,-2.05){$a$}
  \rput[bl](-0.7,-2.05){$a$}
  \rput[bl](0.1,-2.05){$a$}
  \rput[bl](-0.3,-2.05){$\bar{a}$}
  \scriptsize
    \rput[bl]{0}(1.15,-0.05){$e^{\prime}$}
 \endpspicture
\notag \\
&&= \rho_{e^{\prime}}^{\text{ER} \left(23\right)} \otimes R_{aa}^{\left(14\right)} \rho_{0}^{\left( 14 \ldots \right)}\left(a,a,\ldots \right) R_{aa}^{\dagger \left(14\right)} ,
\label{eq:intrhoR}
\end{eqnarray}
where $e^{\prime} = f_{n_{3}}$ is the collective charge (whose value is unimportant) of the entanglement resource pair given by the last $f$ charge measurement from the third forced measurement series $M_{3}$. The forced interferometry measurements used here are given by
\begin{eqnarray}
&& \breve{\Delta}_{M_{1}}^{\left(21 \leftarrow 23 \right)} = \Delta_{z_{n_{1}}=0}^{\left( 34 \ldots \right)} \circ \Delta_{f_{n_{1}}}^{\left(21 \right)} \circ \Delta_{e_{n_{1}}}^{\left(23 \right)} \circ \ldots \circ \Delta_{z_{1_{1}}}^{\left( 34 \ldots \right)} \circ \Delta_{f_{1_{1}}}^{\left(21 \right)} \circ \Delta_{e_{1_{1}}}^{\left(23 \right)} \\
&& \breve{\Delta}_{M_{2}}^{\left(24 \leftarrow 21 \right)} = \Delta_{z_{n_{2}}=0}^{\left( 13 \ldots \right)} \circ \Delta_{f_{n_{2}}}^{\left(24 \right)} \circ \Delta_{e_{n_{2}}}^{\left(21 \right)} \circ \ldots \circ \Delta_{z_{1_{2}}}^{\left( 13 \ldots \right)} \circ \Delta_{f_{1_{2}}}^{\left(24 \right)} \circ \Delta_{e_{1_{2}}}^{\left(21 \right)}\\
&& \breve{\Delta}_{M_{3}}^{\left(23 \leftarrow 24 \right)} = \Delta_{z_{n_{3}}=0}^{\left( 14 \ldots \right)} \circ \Delta_{f_{n_{3}}}^{\left(23 \right)} \circ \Delta_{e_{n_{3}}}^{\left(24 \right)} \circ \ldots \circ \Delta_{z_{1_{3}}}^{\left( 14 \ldots \right)} \circ \Delta_{f_{1_{3}}}^{\left(23 \right)} \circ \Delta_{e_{1_{3}}}^{\left(24 \right)}
,
\end{eqnarray}
where $e_{1_{2}} = f_{n_{1}}$ and $e_{1_{3}} = f_{n_{2}}$. Again, the first measurement of each forced measurement series is actually already done by the previous forced measurement series, so they need not be repeated. Similarly, we have
\begin{eqnarray}
&& \breve{\Delta}_{M_{3}}^{\left(23\leftarrow 21 \right)} \circ \breve{\Delta}_{M_{2}}^{\left(21\leftarrow 24 \right)} \circ \breve{\Delta}_{M_{1}}^{\left(24 \leftarrow 23 \right)} \left[ \rho_{e}^{\text{ER} \left(23\right)} \otimes \rho_{0}^{\left( 14 \ldots \right) } \left(a,a,\ldots \right) \right] \notag \\
&& \qquad \qquad \qquad \qquad \qquad = \rho_{e^{\prime}}^{\text{ER} \left(23\right)} \otimes R_{aa}^{\dagger \left(14\right)} \rho_{0}^{\left( 14 \ldots \right)}\left(a,a,\ldots \right) R_{aa}^{\left(14\right)} ,
\label{eq:intrhoR*}
\end{eqnarray}
using the forced interferometry measurements
\begin{eqnarray}
&& \breve{\Delta}_{M_{1}}^{\left(24 \leftarrow 23 \right)} = \Delta_{z_{n_{1}}=0}^{\left( 13 \ldots \right)} \circ \Delta_{f_{n_{1}}}^{\left(24 \right)} \circ \Delta_{e_{n_{1}}}^{\left(23 \right)} \circ \ldots \circ \Delta_{z_{1_{1}}}^{\left( 13 \ldots \right)} \circ \Delta_{f_{1_{1}}}^{\left(24 \right)} \circ \Delta_{e_{1_{1}}}^{\left(23 \right)} \\
&& \breve{\Delta}_{M_{2}}^{\left(21\leftarrow 24 \right)} = \Delta_{z_{n_{2}}=0}^{\left( 34 \ldots \right)} \circ \Delta_{f_{n_{2}}}^{\left(21 \right)} \circ \Delta_{e_{n_{2}}}^{\left(24 \right)} \circ \ldots \circ \Delta_{z_{1_{2}}}^{\left( 34 \ldots \right)} \circ \Delta_{f_{1_{2}}}^{\left(21 \right)} \circ \Delta_{e_{1_{2}}}^{\left(24 \right)}\\
&& \breve{\Delta}_{M_{3}}^{\left(23\leftarrow 21 \right)} = \Delta_{z_{n_{3}}=0}^{\left( 14 \ldots \right)} \circ \Delta_{f_{n_{3}}}^{\left(23 \right)} \circ \Delta_{e_{n_{3}}}^{\left(21 \right)} \circ \ldots \circ \Delta_{z_{1_{3}}}^{\left( 14 \ldots \right)} \circ \Delta_{f_{1_{3}}}^{\left(23 \right)} \circ \Delta_{e_{1_{3}}}^{\left(21 \right)}
.
\end{eqnarray}
Eqs.~(\ref{eq:intrhoR},\ref{eq:intrhoR*}) respectively generate the counterclockwise and clockwise braiding transformations of anyons $1$ and $4$ on the state $\rho_{0}$.  Thus, with the introduction of an appropriate entanglement resource $\rho^{\text{ER}}_{e}$ (with arbitrary $e$ allowed) pair of anyons at positions $2$ and $3$, we can schematically write appropriate series of interferometry topological charge measurements as being equivalent to braiding transformations of anyons of charge $a$ at positions $1$ and $4$
\begin{eqnarray}
\breve{\Delta}_{M_{3}}^{\left(23\leftarrow 24 \right)} \circ \breve{\Delta}_{M_{2}}^{\left(24\leftarrow 21 \right)} \circ \breve{\Delta}_{M_{1}}^{\left(21 \leftarrow 23 \right)} &\cong& R_{aa}^{ \left(14\right)}
\label{eq:intR} \\
\breve{\Delta}_{M_{3}}^{\left(23\leftarrow 21 \right)} \circ \breve{\Delta}_{M_{2}}^{\left(21\leftarrow 24 \right)} \circ \breve{\Delta}_{M_{1}}^{\left(24 \leftarrow 23 \right)} &\cong & R_{aa}^{\dagger \left(14\right)} .
\label{eq:intR*}
\end{eqnarray}

\begin{figure}[t!]
\begin{center}
  \includegraphics[scale=0.7]{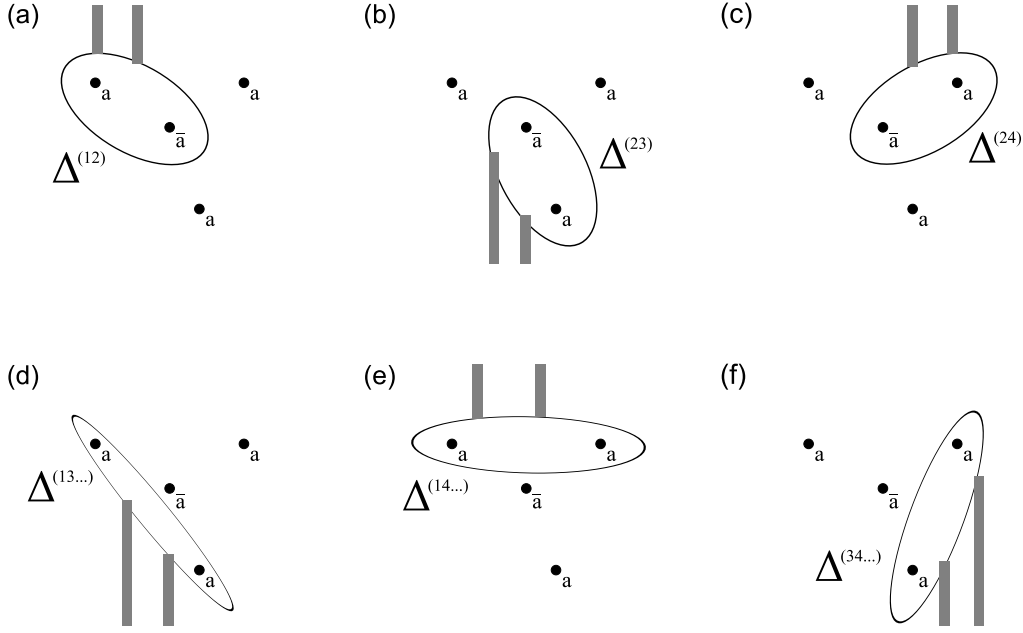}
  \caption{Interferometry topological charge measurements of pairs of anyons (a) $1$ and $2$; (b) $2$ and $3$; and (c) $2$ and $4$ are used together with measurements of multiple anyons (d) $1,3,\ldots$; (e) $1,4,\ldots$; and (f) $3,4,\ldots$; where ``$\ldots$'' represent the additional anyons (not shown) that also comprise the state $\rho_{0}$, to implement forced interferometry measurement anyonic state teleportation, which is used to produce braiding transformations as in Eqs.~(\ref{eq:intrhoR}\,--\,\ref{eq:intR*}). The ovals represent the interferometry loop of the probe anyons, delineating the areas for which the contained collective topological charge is being measured. The grey bars represent a ``safe'' choice of entry/exit paths of probe anyons. For a concrete example of this abstract depiction of interferometry measurements, see Fig.~\ref{fig:FQHint}. \newline}
  \label{fig:intmeasurements}
\end{center}
\end{figure}

An important point to emphasize is that, for both versions of measurement-generated braiding, the entanglement resource is fully replenished and returned to its original location at the end of these processes. This allows such measurement-generated braiding transformations to be employed repeatedly, without exhausting the entanglement resources. This is an advantage over the measurement-based approaches to conventional quantum computation, for which the entanglement resources are consumed by the computation processes.

In writing the particular measurements and corresponding diagrams in this section, there are implicit assumptions of the relative spatial configurations of the anyons and the manner in which the measurements are performed. It is important to understand that the precise positions and measurement geometries are not important, but rather it is the topologically invariant quantities that determine how the corresponding diagrams should be written. This can efficiently be encapsulated by the delineation of the spatial region (with respect to the anyon locations) for which the topological charge is being measured. The $2$-dimensional spatial configuration of the anyons and projective measurements used to generate the braiding transformation in Eq.~(\ref{eq:projR},\ref{eq:projR*}) are shown in Fig.~\ref{fig:projmeasurements}. There is, of course, also some freedom to make topologically different choices in the configurations and measurements used. This will generally change the details of the forced measurement procedures that must be used in a manner dependent upon the details of the configuration choice, however there are no conceptual differences between these procedures.

When using interferometry measurements, one must be careful not to allow the probe anyons to cause decoherence of important anyonic charge entanglement in the computational anyons' system. We recall from Section~\ref{sec:Interferometry_Measurement} that a stream of probe anyons passing between two regions will decohere the anyonic charge entanglement between anyons in the different regions. Thus, we need to make sure that the paths of probe anyons avoid partitioning space into regions that severe important charge entanglement lines in the computational anyons' system. We will therefore also indicate the probes' entry and exit paths (which will be chosen in to be ``safe'') when delineating the spatial regions of measurement. One also needs to be a bit careful to properly treat the measurements including ``\ldots'' anyons. The $2$-dimensional spatial configuration of the anyons and interferometry measurements used to generate the braiding transformation in Eq.~(\ref{eq:intR},\ref{eq:intR*}) are shown in Fig.~\ref{fig:intmeasurements}.

\section{Measurement-Only Topological Quantum Computation}
  \label{sec:MOTQC}

\begin{figure}[t!]
\begin{center}
  \includegraphics[scale=0.4]{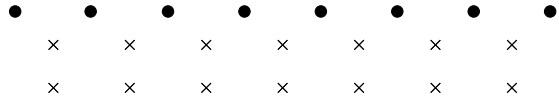}
  \caption{A quasi-linear array of stationary anyons used for measurement-only topological quantum computing. Entanglement resource pairs of anyons (denoted by X's) are situated between adjacent pairs of computational anyons (denoted by dots) to facilitate measurement generated braiding transformations used to implement computational gates on the topological qubits. \newline}
  \label{fig:quasiarray}
\end{center}
\end{figure}

Once we know how to generate braiding transformations using only topological charge measurements, it is clear that, given an appropriate array of anyons encoding qubits and providing necessary entanglement resources, we can perform topological quantum computation using topological charge measurements as the only computational primitive. We arrange our initialized computational anyons in a linear array and distribute maximally entangled pairs (more or less) between them, forming a quasi-one-dimensional array, as in Fig.~\ref{fig:quasiarray}. These anyons all remain stationary and computational gates on the topological qubits are implemented via measurements. Any quantum computation algorithm can be written in terms of a set of universal gates, using a quantum circuit model. Topological quantum computation in turn is based on the fact that computational gates, in particular those in the universal gate set one chooses to employ for a computation, can be generated (to arbitrary precision) from a series of braiding transformations of the computational anyons encoding the topological qubits. The relations in Eqs.~(\ref{eq:projR},\ref{eq:projR*},\ref{eq:intR},\ref{eq:intR*}) give the maps between braiding transformations and the topological charge measurements that will generate them. Combining these determines the series of measurements that should be performed to implement a particular quantum algorithm. It is important to remember that each forced measurement used in this implementation is a probabilistically determined adaptive series of measurement in the sense that a pre-specified pattern of measurements is repeated until a desired outcome is obtained. Since each repetition of the measurement pattern has a non-zero lower bound probability of resulting in the desired outcome, the number of repetitions needed to complete each forced measurement will be exponentially suppressed. This means that even though the time taken by each forced measurement is probabilistically determined, unacceptable delays (from the perspective of computational time scales) will generally not occur as a result of employing this procedure.

The implementation of measurement-only topological quantum computation described above is completely straightforward for the case of projective measurements, but there is a point one must be careful with when considering interferometry measurements that requires us to consider how qubits are encoded more carefully. There are many ways quantum information can be encoded in the non-local multi-dimensional Hilbert space of multiple non-Abelian anyons, and at some level the specific choice of how to encode is unimportant. However, different encodings have different properties, some of which may prove to be advantageous or detrimental, so in practice, the choice can be very important. For the sake of physically implementing a computational model, we require that there be a fixed upper bound (for arbitrary computation size) on the total number of anyons upon which any single measurement is performed. This is important because when we have a stationary array of computational and entanglement resource anyons, increasing the number of anyons in a measurement increases the perimeter of the measurement area, which in turn will generally decrease the fidelity and quantum coherence of the measurement\footnote{For example, the probe anyons in FQH interferometers are edge excitations, which have a finite coherence length determined by the system.}. For the case of projective MOTQC, this bound is automatically satisfied, as measurements are only performed on pairs of anyons. For interferometrical MOTQC (as we have described it), there are interferometry measurements involving ``$\ldots$'' anyons. A careless choice of encoding could require such measurements to include all the computational anyons, a number that scales with the number of qubits. Another point to keep in mind is that interferometry measurement generated teleportation and braiding (as we described it) required that the state upon which they act be restricted to have trivial overall topological charge $0$. At this point we limit our attention to TQC with anyon models that have only self-dual topological charges. This is not a severe limitation as it includes the most physically relevant examples of non-Abelian anyon models: Ising, Fib, and $\text{SU}(2)_{k}$ (see the Appendix for details). If the computational anyons have self-dual charge $a=\bar{a}$, we can use a more economical distribution of entanglement resource anyons, situating only one anyon from each maximally entangled pair between each adjacent pair of computational anyons, so that the second row of X's in Fig.~\ref{fig:quasiarray} is not needed, and the anyon playing the role of anyon $3$ in Eqs.~(\ref{eq:projrhoR},\ref{eq:projrhoR*},\ref{eq:intrhoR},\ref{eq:intrhoR*}) and Figs.~\ref{fig:projmeasurements},\ref{fig:intmeasurements} is instead situated either to the left of anyon $1$ or to the right of anyon $4$, depending on to which pair of computational anyons a braiding transformation is being applied\footnote{For TQC models in which the computational anyons do not all have the same anyonic charge, the same forced measurement anyonic teleportation generated braiding principles may be applied, but a greater number of entanglement resource anyons will be needed.}.

The standard choice of computational anyons for our example anyon models is
\begin{equation}
\begin{array}{ll}
a=\frac{1}{2}  \quad & \text{ for Ising}  \\
a=1  \quad & \text{ for Fib}  \\
a=\frac{1}{2} \quad & \text{ for SU$(2)_{k}$}
\end{array}
\end{equation}
which all obey the fusion rule
\begin{equation}
a \times a = 0 + 1
\end{equation}
(where $0=I$, $\frac{1}{2} = \sigma$, and $1=\psi$ for Ising; and $0=I$ and $1=\varepsilon$ for Fib). This naturally suggests the two fusion channels $0$ and $1$ as the basis in which to encode a topological qubit. For an encoding that grants us all of the necessary properties for implementing interferometrical MOTQC, we choose topological qubits to be encoded in the possible fusion channels of four charge $a$ non-Abelian anyons that have collective charge $0$, so the topological qubit basis states $\left| 0 \right\rangle$ and $\left| 1 \right\rangle$ are
\begin{equation}
\left| j \right\rangle= \left| a,a;c_{j} \right\rangle_{12} \left| a,a;c_{j} \right\rangle_{34} \left| c_{j},c_{j};0 \right\rangle_{\left(12\right)\left(34\right)} = \frac{1}{d_{a}}
 \pspicture[0.5](-1.5,0)(1.5,2.5)
  \small
  \psset{linewidth=0.9pt,linecolor=black,arrowscale=1.5,arrowinset=0.15}
  \psline(0.0,0.5)(1.2,2)
  \psline(0.0,0.5)(-1.2,2)
  \psline(0.8,1.5)(0.4,2)
  \psline(-0.8,1.5)(-0.4,2)
    \psline{->}(0.8,1.5)(0.5,1.875)
    \psline{->}(0.8,1.5)(1.1,1.875)
    \psline{->}(-0.8,1.5)(-0.5,1.875)
    \psline{->}(-0.8,1.5)(-1.1,1.875)
    \psline{->}(0,0.5)(-0.5,1.125)
  \rput[bl]{0}(-0.85,0.7){$c_{j}$}
  \rput[bl]{0}(-1.35,2.1){$a$}
  \rput[bl]{0}(-0.5,2.1){$a$}
  \rput[bl]{0}(0.3,2.1){$a$}
  \rput[bl]{0}(1.15,2.1){$a$}
 \endpspicture
\end{equation}
for $j=0,1$. Generally, there will be multiple fusion channels from which to choose $c_{j}$. For our favorite examples, there are exactly two fusion channels, so we use
\begin{equation}
c_{0}=0 \quad \text{and} \quad c_{1}=1 \quad \text{for Ising, Fib, and SU}(2)_{k}.
\end{equation}
This encoding is natural in the sense that it reproduces the standard qubit tensor product structure in the anyon model, i.e. there are no anyonic charge lines connecting the different topological qubits, and so no anyonic charge entanglement between them. Combining this property with the property that each $4$ anyon topological qubit has overall charge $0$ allows us to fix an upper bound on the number of anyons that must be included in any single topological charge measurement employed in interferometrical MOTQC. To see this, we recall from Section~\ref{sec:Interferometry_Teleportation} that the state upon which we perform forced interferometry measurements needed to have overall charge $0$ and the ``$\ldots$'' anyons were all anyons that originally had non-trivial anyonic charge entanglement with the anyon whose state is being teleported. Thus, as long as the topological qubits remain in the computational subspace of the encoding (i.e. with each qubit $4$-tuple of anyons having overall charge $0$), the ``$\ldots$'' only includes the other anyons of the $4$ encoding the qubit upon which an operation is being performed. Of course, this is generally only the case for single qubit gates. For multi-qubits gates in TQC, even though the topological qubits start and end in the computational subspace, there will generally be intermediate steps in the series of braiding transformations composing the gate during which there is non-trivial anyonic charge entanglement between the anyons of different topological qubits. For such $n$-qubit gates, the ``$\ldots$'' measurements could require measuring the collective topological charge of up to $4n$ anyon in one measurement. Fortunately, from the quantum circuit model, we know that computations may be performed with an upper bound on $n$. In fact, we can choose gate sets that only include single and $2$-qubit gates, setting the upper bound at $8$ anyons the must be included in any single topological charge measurement.

We summarize by stating that the concept of MOTQC is essentially to change how braiding transformations in TQC are implemented. Instead of physically transporting computational anyons around each other, one can instead perform a series of topological charge measurements. In particular, this means that the issue of computational universality of anyon models' braiding statistics is still present. A set of operations in computationally universal if they can densely populated the space of unitary transformations in the computational space. The braiding transformations of the Ising/SU$(2)_{2}$ anyons are the generators of the Clifford group, and hence are not computationally universal. To perform quantum computation with such anyons, one must find a way to supplement the braiding transformations with additional gates that makes the gate set universal. Two methods of doing this have been proposed so far: one involving non-topologically protected operations that are then ``distilled'' using topologically protected gates~\cite{Bravyi06}; the other involving dynamical topology change of the topological fluid in which the anyons exist~\cite{Freedman06a,FNW05b}. A nice feature of the Ising/SU$(2)_{2}$ anyons for the sake of implementing MOTQC is that the interferometry measurements that are used are actually projective measurement, so one can simply use the projective MOTQC protocol, instead of interferometry MOTQC. The reason for this is that one always measures pairs of $\sigma$ anyons, so the topological charge measurement outcomes $I$ and $\psi$ are always Abelian charges, which, as explained in Section~\ref{sec:Interferometry_Measurement}, means the interferometry measurement is projective. On the other hand, it has been shown~\cite{Freedman02a,Freedman02b} that the braiding transformations are computationally universal for the Fibonacci anyons and for SU$(2)_{k}$ when $k=3$ and $k \geq 5$.

\section{Implementation in Fractional Quantum Hall Systems}

Fractional quantum Hall systems are (currently) the most physically concrete candidates in which to implement topological quantum computing platforms, so we will address the anyonics for MOTQC devices in FQH systems in further detail. The braiding statistics of quasiparticles in the most physically relevant non-Abelian FQH states may be written in terms of an Ising, Fibonacci, or SU$(2)_{k}$ anyon model times an Abelian sector (see Appendix). For the purposes of TQC, one can simply focus on the non-Abelian sector of the anyon model describing quasiparticles in a FQH state, because the Abelian sector only contributes overall phases that are irrelevant for quantum computation. The topological charges of these anyon models are all self-dual, so we can employ the more economical distribution of entanglement resource anyon pairs needed for MOTQC.

An anyon with definite topological charge $a$ in a FQH system can be realized by either a single quasiparticle excitation that carries that topological charge, or by a cluster of quasiparticles that are always treated collectively and have (one way or another) been projected into a state with definite collective topological charge $a$. One example of how to controllably localize anyons in FQH systems is through the use of anti-dots. By placing a gate on top of the Hall bar that allows the Hall fluid beneath it to be selectively depleted through application of voltage, an anti-dot can be created that localizes a controllable number of quasiparticle excitations. Quasiparticle excitations localized on a single anti-dot are essentially fused together, and so are treated collectively and have definite collective charge (as a result of superselection). It should be clear that the use of antidots are not easily amenable to physically transporting anyons (i.e. entire antidots) around each other, which demonstrates why MOTQC is a desirable technique.

The initialization step for implementing MOTQC in FQH systems involves forming a quasi-linear array, formed by two rows of anyons. One row contains the computational anyons, and has $4n$ anyons in order to encode $n$ topological qubits, as explained in Section~\ref{sec:MOTQC}. The second row contains $4n$ entanglement resource anyons. To initialize this array of anyons for computation, the anyons in each row are divided into $2n$ adjacent pairs which are put into the collective topological charge $0$ state. This can be carried out by several techniques, one of which is to simply measure the topological charge of each pair and, if it has the wrong value, throw it away and use a new pair. This initializes each qubit in the $\left| 0 \right\rangle$ state, and each entanglement resource pair to have charge $0$.

\subsection{Topological charge measurement}

As topological charge measurements are the keystone of MOTQC, it is important to address how they will be carried out in FQH implementations. In order to perform topological charge measurements in a manner that is nondemolitional and leaves the measured anyons stationary, the only choice for FQH systems is to use double point-contact interferometers. Such FQH interferometers have been studied in detail~\cite{Chamon97,Fradkin98,DasSarma05,Stern06a,Bonderson06a,Bonderson06b,Chung06,Fendley06a,Fendley07a,Fidkowski07c,Ardonne07a,Bonderson07b,Bonderson07c,Bishara08}.
Furthermore, they have been successfully implemented experimentally for Abelian FQH systems~\cite{Camino05a,Camino07a}, and recently for non-Abelian FQH systems~\cite{Willett08}.

The concept of a double-point contact interferometer is to opportunistically use of the edge current as a natural supply of probe anyons. By deforming the Hall fluid edges, for example using gates placed on the Hall bar, one can form a point contact -- a constriction in the Hall fluid where two separate edge regions are brought into close enough proximity that edge quasiparticle excitations can tunnel through the Hall fluid from one edge to the other. Creating two such point contacts around a region of Hall fluid using the same edges establishes an interferometry loop around this region that allows the different tunneling paths to give rise to interference effects that can be experimentally observed in the resulting tunneling current (related to the Hall conductance) from one edge to the other. The interference will depend on the total magnetic flux as well as the total collective topological charge enclosed in the interference loop, and running such a double-point contact interferometer will thus measure the collective topological charge contained inside the interference loop. Because of the chirality of FQH systems, these double-point contact interferometers are of the Fabry--P\'{e}rot type, allowing multiple passes of probe anyons around the interferometry loop. To suppress such terms higher order terms, we restrict to point contacts that are in the weak tunneling limit. Restricting to this limit also has the important effect of ensuring that quasiparticle tunneling across the point contacts will be dominated by fundamental quasiholes, which are the most RG relevant contributions because they are the excitations with smallest conformal scaling dimension.

\begin{figure}[t!]
\begin{center}
  \includegraphics[scale=0.7]{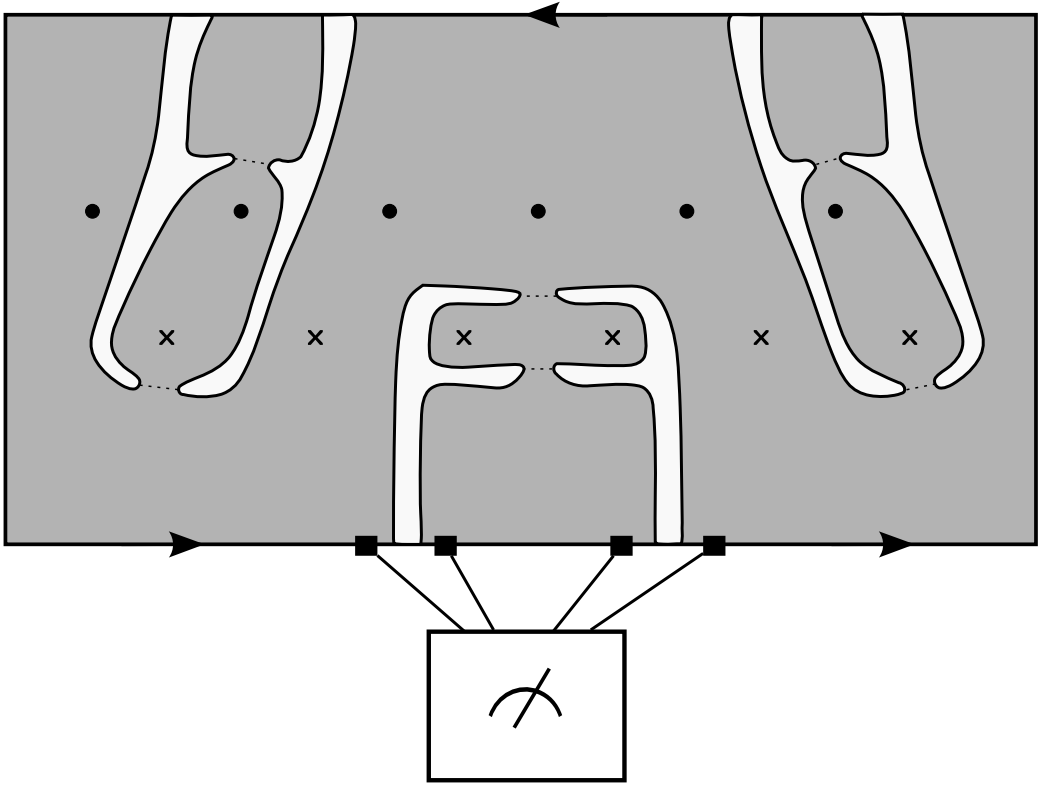}
  \caption{For measurement-only topological quantum computation in fractional quantum Hall systems, we set up a quasi-linear array of stationary anyons, and use double point-contact interferometers to perform interferometrical topological charge measurements. Here we show a section of an array in a Hall bar with three interferometers measuring the topological charges of different pairs of anyons. The bulk of the FQH fluid is the grey area, and its edge chirality is indicated by the arrows. The computational anyons are denoted here by dots, and the entanglement resource anyons are denoted by Xs. The FQH edge is deformed into the bulk by depleting the Hall fluid (light grey areas) in order to construct interferometers enveloping anyons to be measured. Once a measurement is completed, the edge protrusions are retracted, destructing the no longer needed interferometer. Tunneling across the point contacts is indicated by dashed lines. The ``arms'' reaching in from the side carry the incoming and outgoing current of probe anyons along their edges, and thus represent the entry/exit paths of probe anyons depicted abstractly in Fig.~\ref{fig:intmeasurements}. Topological charge measurement outcomes are distinguished by the observed values of current that tunnels an interferometer. For one of the interferometers, we show where leads are attached (black squares) to the Hall bar to measure the tunneling current across the interferometer. \emph{Note: figure not drawn to scale.} \newline}
  \label{fig:FQHint}
\end{center}
\end{figure}

Once such an interferometer has collapsed superpositions of collective topological charge inside it to be $a_{\text{in}}$, the tunneling current (in the weak tunneling limit) will be proportional to the tunneling probability
\begin{equation}
\label{eq:2PCpint}
p^{\shortleftarrow}_{a_{\text{in}}} \simeq \left| t_{1} \right|^{2} + \left| t_{2} \right|^{2}+ 2 \left| t_{1} t_{2} \right| \text{Re} \left\{ M_{a_{\text{in}} B} e^{i \beta} \right\}
,
\end{equation}
where $t_{1}$ and $t_{2}$ are the tunneling amplitudes of the two point contacts, $\beta$ is due to the background contribution, and $M_{a_{\text{in}} B}$ is the monodromy scalar component for which the probe $B$ is the topological charge of the fundamental quasihole. It can be seen from the analysis in Refs.~\cite{Bonderson07b,Bonderson07c} that deviation from the weak tunneling limit ($\left| t_{j} \right| \ll 1$) will amount to a uniform suppression factor $Q \in \left[ 0,1 \right] $ of the interference term. In fact, most factors that decrease the interferometer's coherence can be lumped into this $Q$ factor, such as noise, background and path fluctuations, edge--bulk tunneling, thermal effects, and finite coherence lengths of edge excitations. Fortunately, the effects encoded in the $Q$ factor do not introduce errors, but, rather, they only reduce the visibility of the quantum interference in the experiment, i.e. $\Delta p$ is multiplied by $Q$. Of course, this means the number of probes needed for a desired level of confidence in a measurement is multiplied by $Q^{-2}$, so the duration of measurements will need to be increased, but a simple multiplicative constant to such time scales such as this will cause no major trouble.

The key challenge for the purposes of MOTQC is to have such interferometers where we need them only during specified time intervals. For this, we envision an array of gates on the Hall bar that allow us to deform the edge in a controlled fashion used to construct and destruct arm-like protrusions of the edge reaching into the bulk of the Hall fluid that establish interferometers where necessary (see Fig.~\ref{fig:FQHint}). Creating and removing these interferometer arms must be done carefully so as not to disturb the rest of the system, particularly the topological qubits.

\subsection{Calibration}

We notice from Eq.~\ref{eq:2PCpint} that the tunneling probability distinguishes topological charge measurement outcomes $a_{\text{in}}$ by the interference term proportional to
\begin{equation}
\text{Re} \left\{ M_{a_{\text{in}} B} e^{i \beta} \right\} = \left| M_{a_{\text{in}} B} \right| \cos \left( \theta_{a_{\text{in}}B} + \beta \right)
,
\end{equation}
where $\beta$ is an experimental variable, and $\theta_{a_{\text{in}}B} = \arg \left\{ M_{a_{\text{in}} B} \right\}$. (This is still true when there is a $Q$ factor.) Trying to distinguish topological charges $a$ and $a^{\prime}$ for which the monodromy matrix elements have the same magnitudes $\left| M_{a B} \right| = \left| M_{a^{\prime} B} \right|$ can get a bit tricky. Since these monodromies and their resulting tunneling probabilities and currents are only distinguished by their relative phase, in order to correctly identify the topological charge measurement outcome, one must initially calibrate the interferometers, associating fixed values of tunneling current with the different measurement outcome possibilities (tuning $\beta$ to a preferable value). Additionally, one must have the ability to reproduce tuned values of $\beta$ with reliable precision. This may be quite difficult to achieve.

When trying to distinguish topological charges $a$ and $a^{\prime}$ for which the monodromy matrix elements have the different magnitudes $\left| M_{a B} \right| \neq \left| M_{a^{\prime} B} \right|$, there is a more robust method of identifying the measurement outcome. By varying $\beta$, one can observe the magnitude of the interference term, and in doing so distinguish between $a$ and $a^{\prime}$. For this, the interferometers' calibration only involves associating interference magnitudes with the different measurement outcome possibilities.

Unfortunately, when using Ising anyons for TQC, we have to deal with the first situation, because we need to distinguish between the Ising charges $I$ and $\psi$ using $\sigma$ probes, which have $M_{I \sigma} = 1$ and $M_{\psi \sigma} = -1$. As a minor concession, these monodromies are at least also maximally distinguishable ($\Delta M = 2$), giving them the most amount of leeway one can hope for when precision tuning of $\beta$ is required. When using Fibonacci anyons for TQC, we fortunately have the benefit of the second situation, because we need to distinguish between the Fib charges $I$ and $\varepsilon$ using $\varepsilon$ probes, which have $M_{I \varepsilon} = 1$ and $M_{\varepsilon \varepsilon} = -\phi^{-2} \approx -0.38$. When using SU$(2)_{k}$ anyons for TQC, we need to distinguish between the charges $0$ and $1$ using charge $\frac{1}{2}$ probes, which have $M_{0 \frac{1}{2}} = 1$ and $M_{1 \frac{1}{2}} = 1 - 4 \sin^{2}\left( \frac{\pi}{k+2} \right)$. For $k=2$, this is exactly the same as the Ising anyons. For $k \geq 3$, these monodromies have different magnitudes, so we have the benefit of the second situation described above. However, we also point out that as $k$ increases, $M_{1 \frac{1}{2}} \rightarrow 1$ and $\Delta M \rightarrow 0$, so the distinguishability also decreases, eventually to the point of making either method incapable of distinguishing the charges in practice. Thus, $k=3$, which is essentially the same as the Fibonacci anyons, is the optimal case from the perspective of measuring topological charge.

\subsection{Time scales}

It is difficult to make a sound estimate of the time $\tau_{r}$ it takes to repattern the edges of the FQH fluid in order to construct/destruct interferometers as desired. We can, however, make a na\"{i}ve estimate assuming that the bulk Hall fluid can be displaced a distance $L_{\text{int}}$ corresponding to the size of the interferometer, at a rate determined by the velocity $v_{e}$ of electrons at the edge of the Hall fluid. This gives an estimated interferometer construction/destruction time scale of
\begin{equation}
\tau_{r} \sim v_{e} L_{\text{int}}
.
\end{equation}
The electron edge velocity in experiments is found to be about $v \sim 10^{3}$~m/s, and we can approximate the interferometer length as $L_{\text{int}} \sim 1$~$\mu$m, giving the time estimate of $\tau_{r} \sim 1$~ns. It is quite possible this is an overly optimistic estimate and that other rate determining factors will necessitate longer durations, but more accurate time scales for such processes are, in any case, best obtained empirically.

We can estimate the measurement duration $\tau_{m}$ one needs to run a properly tuned double point-contact interferometer (in the weak tunneling limit) to achieve a desired confidence level $1-\alpha$ for the interferometry topological charge measurement by using Eq.~(\ref{eq:N2pointest}) to give
\begin{equation}
\tau_{m} \sim \frac{e^{\ast} t^{2} N}{I_{\text{t}} } \gtrsim \frac{ 8 e^{\ast} \left[ \text{erfc}^{-1} \left( \alpha \right) \right]^{2} }{ I_{\text{t}} \left( \Delta M \right)^{2}}
,
\end{equation}
where $e^{\ast}$ is the electric charge of the tunneling quasiparticles (i.e. the probe anyons of the interferometer), $I_{\text{t}}$ is the tunneling current through a single point contact with tunneling amplitude $t$, and $N$ is the estimated number of probes needed. For example, to achieve $\alpha = 10^{-4}$ in the $\nu =5/2$ MR state, where the probes are fundamental quasiholes (which have $e/4$ electric charge and $\sigma$ Ising topological charge) and the measurements are distinguishing between $I$ and $\psi$ Ising topological charges, a tunneling current $I_{\text{t}} \sim 1$~nA typical of point contacts in experiments~\cite{Miller07a,Dolev08,Radu08,Willett08} conducted at $\nu =5/2$ gives the necessary measurement duration estimate $\tau_{m} \sim 1$~ns. This analysis neglected a possible suppression $Q$ factor, which, as discussed above, will increase the measurement time needed by a factor of $Q^{-2}$, but it is difficult to quantify all the contributions to $Q$, so it is also best obtained empirically.

Combining these two time estimates with the average number of attempts $\left\langle n \right\rangle \sim d_{a}^{2}$ needed in each forced measurement (which is equal to $2$ for the MR state) and the number of topological charge measurements in each attempt ($2$ or $3$), we estimate the time it takes to implement a single measurement generated braiding transformation $R_{aa}$ as
\begin{equation}
\tau_{R} \sim 3 d_{a}^{2} \left( 2 \tau_{r} + \tau_{m} \right)
.
\end{equation}
For the $\nu =5/2$ MR state, this gives a time estimate for implementing a braiding transformation on the order of $\tau_{R} \sim 10$~ns. The number of braiding transformations needed will depend on the computation being performed and the anyon model of the non-Abelian FQH state being used.

\subsection{Error sources}

Our measurement-only scheme is strongly dependent on the accuracy
of measurements. Suppose, for instance, that we measure the total
topological charge of a pair of $\sigma$ quasiparticles in the $\nu=5/2$ MR state.
Let us further suppose that we perform the measurement for long enough
that the state is projected onto $I$ or $\psi$ topological charge with the desired accuracy.
There is still the possibility that if the state of the pair is $I$
then we may misidentify it as $\psi$, which would be a fatal error.
How could such a calamity occur? Our measurement technique is
interferometry, so we must be able to distinguish the two possible
conductances which can occur, depending on the collective topological charge
of the pair of $\sigma$s. In other words, we must be able to distinguish between
two possible values of the tunneling current through the interferometer.
However, the current will invariably be noisy (if not as a result of thermal
noise, then at least as a result of quantum noise). Thus, it is important
for us to make the noise as small as possible.

Let us suppose that we are at low enough temperatures that
we can ignore all sources of noise apart from quantum noise.
(We can estimate the temperature which must be maintained in
order to avoid Johnson-Nyquist noise.) At low frequencies,
there will be `shot noise,' which can be understood as follows.
During a measurement time $\tau_{m}$, $N_{t}$ quasiparticles will tunnel across the interferometer,
leading to a current $I_{t}=e^{\ast} N_{t}/\tau_{m}$. If these are independent events
with a Poisson distribution, as can be shown to be the case
by direct computation, then the fluctuations in the current will
be $\Delta I_{t} = e^{\ast} \sqrt{2N_{t}}/\tau_{m} = \sqrt{2 e^{\ast} I_{t}/\tau_{m}}$. This is often
written as $\Delta I_{t} = \sqrt{2 e^{\ast} I_{t} \Delta \omega}$, where
$\omega$ is the frequency bandwidth of the measurement.
Consequently, we can minimize the fluctuations in the current
by performing the measurement over as long a time as possible.
For the sake of concreteness, let us suppose that we have configured
our interferometer so that the conductance through the interferometer
is $g>0$ if the two quasiparticles inside the interferometer fuse to $\psi$
and the conductance is zero (perfect destructive interference) is the
two quasiparticles fuse to $I$. Suppose the two quasiparticles
fuse to $\psi$. If an average current $I_{t}$ would flow through
the interferometer in a measurement of infinite duration then,
for a Poisson distribution, there is a probability $e^{-\tau_{m} I_{t} /e^{\ast}}$ that zero current would
flow through the interferometer during a measurement time $\tau_{m}$.
This probability can be made smaller by making either $I_{t}$ or $\tau_{m}$
larger. However, we cannot make $I_{t}$ too large as $I_{t} = gV$ and we need
$ e^{\ast} V \ll \Delta$, where $\Delta$ is the energy gap. Hence, the probability
of an erroneous measurement is greater than $e^{-\tau_{m} g \Delta / \left(e^{\ast}\right)^2}$.
By making $\tau_{m}$ large, we can make this error probability small. In
principle, we are free to make $\tau_{m}$ as large as we like, so long
as it remains much smaller than the decoherence time of the qubit,
$\tau_{m} \ll \tau_{\text{dec}}$. As we discuss below, we expect
$\tau_{\text{dec}} \propto e^{\Delta/2T}$ for temperature $T$. Thus, by taking
\begin{equation}
1/\Delta \ll \tau_{m} \ll \tau_{\text{dec}}
,
\end{equation}
we can make the probability of an erroneous measurement very small. For the $\nu = 5/2$ state, the currently largest measured values of the gap $\Delta_{5/2} \sim 500$~mK, together with experimentally accessed temperatures of about $T \sim 10$~mK, indicate that desirable measurement times for avoiding errors should be $0.1$~ns~$\ll \tau_{m} \ll 10^{3}$~s.

Stray excitations are another source of error. In any real device, there
will be local potential wells where quasiparticles get trapped. As long as
they do not move, they will not be an overly serious problem, as their effect can be neutralized in a number of ways. In the Ising case,
each computational $\sigma$ anyon can actually be a collection of an odd number of
quasiparticles, some of which may be on an antidot, while the others are stray excitations
localized in the bulk. In order for this identification to be correct, it
is important that, when we perform measurements of the combined topological charge of
two $\sigma$ anyons, we always include the stray localized excitations that we are associating with
a computational anyon. Thus, when we group some stray excitations
with our anti-dots, it makes sense to include in our grouping those stray excitations which
are enclosed by the arms of the possible interferometers
and to exclude those which are not. If there are an odd number of quasiparticles on an anti-dot and an odd number of stray localized
excitations that is grouped with an anti-dot in this way, then we should either
modify the interferometer arms so that they enclose an even number of strays, somehow remove one of the strays, or instead localize an even number of quasiparticles on the anti-dot. The same strategy works for Fibonacci anyons, except that in this case we can have either an even or an odd number of
quasiparticles that define a computational anyon, so long as the total topological charge of the group (anti-dot + strays)
is $\varepsilon$. If it is $I$, then we proceed with the same methods as in the Ising case to change the number of quasiparticles comprising the computational anyon. For general non-Abelian anyons, essentially the same strategy applies, but there may be more fusion channels to worry about.

If the strays excitations move around, this is a more serious problem.
Motion within a grouping is not a problem since this cannot change
the topological charge of the group, and hence does not effect the encoded qubits. However, the motion of a stray excitation
out of a group or encircling parts of two different groups will cause an error.
Let us suppose that the quasiparticles diffuse with a diffusion constant
$D$ which is exponentially small with Arrhenius-type dependence $D \sim D_{0} e^{-\Delta / 2T}$. If the groups
have a linear scale $L$ which is roughly the distance between anti-dots,
then a stray quasiparticle must move a distance of approximately $L$ in order to cause
an error. This will take a time $\tau \sim {L^2}/D$. The number of such quasiparticles
is approximately $n_{s} L^2$, where $n_{s}$ is the density of stray localized quasiparticles. Hence,
the error rate
\begin{equation}
\label{eq:Gamma}
\Gamma \sim n_{s} {L^2} / \tau \sim n_{s} D \sim n_{s} D_{0} e^{-\Delta/2T}
.
\end{equation}
By making the temperature sufficiently small, we can keep this error rate low. For an empirical estimate, we use the relation
\begin{equation}
\sigma_{xx} = e^{2} \frac{ \partial n_{s}}{\partial \mu} D \approx \frac{e^{2} n_{s} D}{\Delta}
\end{equation}
with the experimental data for the $\nu=5/2$ state from Ref.~\cite{Choi07}, which found $R_{xx} \simeq R_{0} e^{-\Delta / 2T}$ with $R_{0} \simeq 170$~$\Omega$ and $\Delta \simeq 544$~mK (from measurements in the temperature range $T \simeq 40-200$~mK). Extrapolating down to $T \sim 10$~mK (which can be reached experimentally, but makes $R_{xx}$ too impractically small to measure) gives $\Gamma \sim 10^{-3}$~s$^{-1}$, i.e a decoherence time on the order of $\tau_{\text{dec}} \sim 10^{3}$~s. If necessary, this decoherence time can be greatly increased further with only a modest temperature decrease, since it has $\tau_{\text{dec}} \propto e^{\Delta/2T}$ dependence on temperature.

\appendix
\section{Examples of Anyon Models}

In this Appendix, we give detailed descriptions of the Ising, Fibonacci, and SU$(2)_{k}$ anyons models, and explain where they occur in non-Abelian fractional quantum Hall states.

\subsection{Ising}
  \label{sec:Ising}

The Ising anyon model is derived from the CFT that
describes the Ising model at criticality~\cite{Moore89b}. It is related to $\text{SU}(2)_{2}$ as its CFT can be obtained using the coset construction $\text{SU}(2)_{2}/\text{U}(1)$. It has anyonic
charges $\mathcal{C}=\left\{I,\sigma,\psi \right\}$ (which respectively
correspond to vacuum, spin, and Majorana fermions in the CFT, and are sometimes denoted $0$, $\frac{1}{2}$, and $1$, because of the relation with $\text{SU}(2)_{2}$). The anyon model
is described by (listing only the non-trivial $F$-symbols and $R$-symbols, i.e. those not listed are equal to one if their vertices are permitted by fusion, and equal to zero if they are not permitted):%
\begin{equation*}
\begin{tabular}{|l|l|}
\hline
\multicolumn{2}{|l|}{$\mathcal{C}=\left\{I,\sigma,\psi \right\}, \quad I\times a=a,\quad \sigma \times \sigma%
=I+\psi,\quad \sigma \times \psi=\sigma,\quad \psi \times \psi=I$} \\ \hline
\multicolumn{2}{|l|}{$\left[ F_{\sigma}^{\sigma \sigma \sigma}\right] _{ef}=
\left[ F_{\sigma \sigma}^{\sigma \sigma}\right] _{ef}=
\left[
\begin{array}{rr}
\frac{1}{\sqrt{2}} & \frac{1}{\sqrt{2}} \\
\frac{1}{\sqrt{2}} & \frac{-1}{\sqrt{2}}%
\end{array}\right] _{ef}^{\phantom{T}}$} \\
\multicolumn{2}{|l|}{$\left[ F_{\psi}^{\sigma \psi \sigma}\right] _{\sigma \sigma}=%
\left[ F_{\sigma}^{\psi \sigma \psi}\right] _{\sigma \sigma_{\phantom{j}}}\!\!=
\left[ F_{\psi \sigma}^{\sigma \psi}\right] _{\sigma \sigma}=
\left[ F_{\sigma \psi}^{\psi \sigma}\right] _{\sigma \sigma}=-1 $} \\ \hline
\multicolumn{2}{|l|}{
$R_{I}^{\sigma \sigma}=e^{-i\frac{\pi }{8}},\quad R_{\psi}^{\sigma \sigma}=e^{i\frac{3\pi }{8}},
\quad R_{\sigma}^{\sigma \psi}=R_{\sigma}^{\psi \sigma}=e^{-i\frac{\pi }{2}},\quad R_{I}^{\psi \psi}=-1$} \\ \hline
$S=\frac{1}{2}\left[
\begin{array}{rrr}
1 & \sqrt{2} & 1 \\
\sqrt{2} & 0 & -\sqrt{2} \\
1 & -\sqrt{2} & 1%
\end{array}%
\right]^{\phantom{T}}_{\phantom{j}} $ & $M=\left[
\begin{array}{rrr}
1 & 1 & 1 \\
1 & 0 & -1 \\
1 & -1 & 1%
\end{array}%
\right] $ \\ \hline
$d_{I}=d_{\psi}=1,\quad d_{\sigma_{\phantom{j}}}\!\!=\sqrt{2}, \quad \mathcal{D}=2$ & $\theta _{I}=1,\quad \theta
_{\sigma}=e^{i\frac{\pi }{8}},\quad \theta _{\psi}=-1$ \\ \hline
\end{tabular}%
\end{equation*}%
where $e,f\in \left\{ I,\psi\right\} $.

Probes anyons of definite charge $b=\sigma$ have $M_{ab}= 1,0,-1$ for $I,\sigma,\psi$ respectively, and so have trivial monodromy only with the vacuum charge $I$. Distinguishing between $I$ and $\psi$, these probes have $\Delta M = 2$.

\subsection{Fib}
  \label{sec:Fib}

The Fibonacci (Fib) anyon model (also known as $\text{SO}(3)_{3}$, since it may be obtained from the $\text{SU}\left( 2\right) _{3}$ anyon model by restricting to integer spins $j=0,1$; as a Chern-Simons or WZW theory, this is more properly equated with $\left( \text{G}_{2}\right)_{1}$, since $\text{SO}\left( 3\right) _{k}$ is only allowed for $k=0~{\rm mod}~4$) is known to be universal for TQC~\cite{Freedman02b}. It has two charges $\mathcal{C}=\left\{I,\varepsilon \right\} $ (sometimes denoted $0$ and $1$, respectively, because of the relation with $\text{SU}\left( 2\right) _{3}$) and is described by (listing only the non-trivial $F$-symbols and $R$-symbols):%
\begin{equation*}
\begin{tabular}{|l|l|}
\hline
\multicolumn{2}{|l|}{$\mathcal{C}=\left\{I,\varepsilon \right\}, \quad I\times I=I,\quad I\times \varepsilon=\varepsilon,\quad \varepsilon\times \varepsilon=I+\varepsilon$} \\
\hline
\multicolumn{2}{|l|}{$\left[ F_{\varepsilon}^{\varepsilon \varepsilon \varepsilon}\right] _{ef}=\left[ F_{\varepsilon \varepsilon}^{\varepsilon \varepsilon}\right] _{ef}
=\left[
\begin{array}{cc}
\phi ^{-1} & \phi ^{-1/2} \\
\phi ^{-1/2} & -\phi ^{-1}%
\end{array}%
\right] _{ef_{\phantom{j}}}^{\phantom{T}}$} \\ \hline
\multicolumn{2}{|l|}{$R_{I}^{\varepsilon \varepsilon}=e^{-i4\pi /5},\quad R_{\varepsilon}^{\varepsilon \varepsilon}=e^{i3\pi /5}$} \\ \hline
$S=\frac{1}{\sqrt{\phi +2}}\left[
\begin{array}{rr}
1 & \phi \\
\phi & -1%
\end{array}%
\right]^{\phantom{T}}_{\phantom{j}} $ & $M=\left[
\begin{array}{cc}
1 & 1 \\
1 & -\phi ^{-2}%
\end{array}%
\right] $ \\ \hline
$d_{I}=1,\quad d_{\varepsilon}=\phi,\quad \mathcal{D}=\sqrt{\phi+2}$ & $\theta _{I}=1,\quad \theta _{\varepsilon}=e^{i\frac{%
4\pi }{5}}$ \\ \hline
\end{tabular}%
\end{equation*}%
where $\phi =\frac{1+\sqrt{5}}{2}$ is the Golden ratio. We denote the anyon
model given by this with the complex conjugate values of the $R$-symbols and topological
spins as $\overline{\text{Fib}}$.

Probes anyons of definite charge $b=\varepsilon$ have $M_{ab}= 1,-\phi^{2}$ for $I,\varepsilon$ respectively, and so have trivial monodromy only with the vacuum charge $I$. Distinguishing between $I$ and $\varepsilon$, these probes have $\Delta M = 1+ \phi^{-2} \approx 1.38$.

\subsection{$\text{SU}\left(2\right)_{k}$}
  \label{sec:SU(2)_k}
The SU$\left( 2\right) _{k}$ anyon models (for $k$ an integer) are ``$q$-deformed'' versions of the usual SU$\left( 2\right) $ for
$q=e^{i\frac{2\pi }{k+2}}$, which, roughly speaking, means integers $n$ are replaced by $\left[ n\right]_{q}\equiv \frac{q^{n/2}-q^{-n/2}}{q^{1/2}-q^{-1/2}}$. These describe SU$\left(2\right) _{k}$ Chern-Simons theories~\cite{Witten89} and
WZW CFTs~\cite{Wess71,Witten83}, and give rise to the Jones polynomials of knot theory~\cite{Jones85}. Their braiding statistics are known to be universal for TQC~\cite{Freedman02a} all $k$, except $k=1$, $2$, and $4$. They are described by:%
\begin{equation*}
\begin{tabular}{|l|l|}
\hline
\multicolumn{2}{|l|}{$\mathcal{C}=\left\{ 0,\frac{1}{2},\ldots ,\frac{k}{2}\right\}, \quad j_{1}\times j_{2}=\sum\limits_{j=\left|
j_{1}-j_{2}\right| }^{\min \left\{ j_{1}+j_{2},k-j_{1}-j_{2}\right\} }j$} \\
\hline
\multicolumn{2}{|l|}{$\left[ F_{j}^{j_{1},j_{2},j_{3}}\right]
_{j_{12},j_{23}}=\left( -1\right) ^{j_{1}+j_{2}+j_{3}+j}\sqrt{\left[
2j_{12}+1\right] _{q}\left[ 2j_{23}+1\right] _{q}}\left\{
\begin{array}{ccc}
j_{1} & j_{2} & j_{12} \\
j_{3} & j & j_{23}%
\end{array}%
\right\} _{q}^{\phantom{T}},$} \\
\multicolumn{2}{|l|}{$\left\{
\begin{array}{ccc}
j_{1} & j_{2} & j_{12} \\
j_{3} & j & j_{23}%
\end{array}%
\right\} _{q}=\Delta \left( j_{1},j_{2},j_{12}\right) \Delta \left(
j_{12},j_{3},j\right) \Delta \left( j_{2},j_{3},j_{23}\right) \Delta \left(
j_{1},j_{23},j\right) $} \\
\multicolumn{2}{|l|}{$\quad \quad \quad \quad \quad \quad \quad \quad \times
\sum\limits_{z}\left\{ \frac{\left( -1\right) ^{z}\left[ z+1\right] _{q}!}{%
\left[ z-j_{1}-j_{2}-j_{12}\right] _{q}!\left[ z-j_{12}-j_{3}-j\right] _{q}!%
\left[ z-j_{2}-j_{3}-j_{23}\right] _{q}!\left[ z-j_{1}-j_{23}-j\right] _{q}!}%
\right. $} \\
\multicolumn{2}{|l|}{$\quad \quad \quad \quad \quad \quad \quad \quad \quad
\quad \quad \times \left. \frac{1}{\left[ j_{1}+j_{2}+j_{3}+j-z\right] _{q}!%
\left[ j_{1}+j_{12}+j_{3}+j_{23}-z\right] _{q}!\left[ j_{2}+j_{12}+j+j_{23}-z%
\right] _{q}!}\right\}_{\phantom{g}} ,$} \\
\multicolumn{2}{|l|}{$\Delta \left( j_{1},j_{2},j_{3}\right) =\sqrt{\frac{%
\left[ -j_{1}+j_{2}+j_{3}\right] _{q}!\left[ j_{1}-j_{2}+j_{3}\right] _{q}!%
\left[ j_{1}+j_{2}-j_{3}\right] _{q}!}{\left[ j_{1}+j_{2}+j_{3}+1\right]
_{q}!}}^{\phantom{T}}_{\phantom{g}},\quad \quad \quad \left[ n\right] _{q}! \equiv \prod\limits_{m=1}^{n}%
\left[ m\right] _{q}$} \\ \hline
\multicolumn{2}{|l|}{$R_{j}^{j_{1},j_{2}}=\left( -1\right)
^{j-j_{1}-j_{2}}q^{\frac{1}{2}\left( j\left( j+1\right) -j_{1}\left(
j_{1}+1\right) -j_{2}\left( j_{2}+1\right) \right) }$} \\ \hline
$S_{j_{1},j_{2}}=\sqrt{\frac{2}{k+2}}\sin \left( \frac{\left(
2j_{1}+1\right) \left( 2j_{2}+1\right) \pi }{k+2}\right) $ & $%
M_{j_{1},j_{2}}=\frac{\sin \left( \frac{\left( 2j_{1}+1\right) \left(
2j_{2}+1\right) \pi }{k+2}\right)^{\phantom{T}} \sin \left( \frac{\pi }{k+2}\right) }{\sin
\left( \frac{\left( 2j_{1}+1\right) \pi }{k+2}\right) \sin \left( \frac{%
\left( 2j_{2}+1\right) \pi }{k+2}\right)_{\phantom{g}} }$ \\ \hline
$d_{j}= \left[ 2j+1 \right]_{q} =\frac{\sin \left( \frac{\left( 2j+1\right) \pi }{k+2}\right)^{\phantom{T}} }{\sin
\left( \frac{\pi }{k+2}\right)_{\phantom{g}} }, \quad \mathcal{D}=\frac{\sqrt{ \frac{k+2}{2}}}{\sin \left( \frac{\pi}{k+2} \right)_{\phantom{g}}}$ & $\theta _{j}=e^{i2\pi \frac{j\left(
j+1\right) }{k+2}}$ \\ \hline
\end{tabular}%
\end{equation*}
where $\left\{ \quad \right\} _{q}$ is a ``$q$-deformed'' version of the usual SU$\left( 2\right) $ $6j$-symbols.

Probes anyons of definite charge $b=\frac{1}{2}$ have $M_{jb}= \frac{ \cos\left(\frac{\left(2j+1 \right) \pi} {k+2} \right) }{\cos\left(\frac{\pi} {k+2} \right)}$, and so monodromy with each other charge is distinguishable, and trivial only with the vacuum charge $0$. Distinguishing between $0$ and $1$, these probes have $\Delta M = 4 \sin^{2} \left( \frac{\pi}{k+2} \right)$.

\subsection{Non-Abelian Fractional Quantum Hall States}

Following Ref.~\cite{Moore91}, it is natural to construct FQH wavefunctions using CFT correlators. Though the CFT is needed to generate explicit wavefunctions and to describe the details of the edge physics~\cite{Wen92b}, one can determine an anyon model describing the fusion and braiding statistics of the quasiparticles of a FQH state directly from its associated CFT. In FQH systems, the anyons $1,\ldots,m$ are bulk quasiparticles, and the probe anyons are edge excitations. Nonetheless, edge excitations have well-defined topological properties, and for the purposes of this paper, the anyon model contains all the pertinent information.

\subsubsection{Moore--Read and Bonderson--Slingerland}

The MR state~\cite{Moore91} for $\nu = 5/2, 7/2 $ is described by (a spectrum restriction of) the product of the Ising CFT with an Abelian $\text{U}\left( 1 \right)$. Specifically, the is
\begin{equation}
\text{MR} = \left. \text{Ising} \times \text{U}\left( 1 \right)_{2} \right|_{\mathcal{C}}
\end{equation}
where the spectrum restriction is such that $I$ and $\psi$ Ising charges are paired with integer $\text{U}\left( 1 \right)$ fluxes, while $\sigma$ Ising charges are paired with half-integer $\text{U}\left( 1 \right)$ fluxes. The fundamental quasihole of the MR state has electric charge $e/4$ and carrying the Ising topological charge $\sigma$. The particle-hole conjugate of MR simply has non-Abelian statistics that are complex conjugated.

The BS states~\cite{Bonderson07d} are obtained from MR by applying a hierarchical (or alternatively a composite fermion) construction to the $\text{U}\left( 1 \right)$ sector. They may be written as
\begin{equation}
\text{BS}_{K} = \left. \text{Ising} \times \text{U}\left( 1 \right)_{K} \right|_{\mathcal{C}}
\end{equation}
where the $K$-matrix is determined by the details of the hierarchical construction over MR. This Ising based candidate states for all other second Landau level FQH filling fractions (i.e. including those observed at $\nu = 7/3$, $12/5$, $8/3$, and $14/5$). The quasiparticle excitation spectra of the BS states include excitations that carry Ising topological charge $\sigma$, but these do not always also carry the minimal electric charge.

\subsubsection{$k=3$ Read--Rezayi and NASS}

The particle-hole conjugate of the $k=3$, $M=1$ RR state~\cite{Read99} is a candidate for $\nu=12/5$, which is constructed from  the $\mathbb{Z}_{3}$-Parafermion (Pf$_{3}$) CFT and an Abelian $\text{U}\left( 1 \right)$. The braiding statistics of this state is described by the direct product of anyon models
\begin{equation}
\overline{\text{RR}}_{k=3,M=1} = \overline{ \text{Pf}_{3} \times \text{U}\left(1\right) } = \overline{\text{Fib}} \times \mathbb{Z}_{10}^{\left(3\right)}
,
\end{equation}
where the overline indicates complex conjugation and $\mathbb{Z}_{10}^{\left(3\right)}$ is an Abelian anyon model (using the notation of Ref.~\cite{Bonderson07b,Bonderson07c}). The fundamental quasiholes of this state have electric charge $e/5$ and Fib topological charge $\varepsilon$.

The $k=2$, $M=1$ NASS state~\cite{Ardonne99}, based on $\text{SU}\left(3\right)_{k}$-parafermions, is a candidate for $\nu = 4/7$. Its braiding statistics is described by
\begin{equation}
\text{NASS}_{k=2,M=1} = \overline{\text{Fib}} \times \text{D}^{\prime} \left( \mathbb {Z}_{2} \right) \times \text{U}\left( 1 \right)
,
\end{equation}
where $\text{D}^{\prime} \left( \mathbb {Z}_{2} \right)$ is an Abelian theory similar to $\text{D} \left( \mathbb {Z}_{2} \right)$, the quantum double of $\mathbb {Z}_{2}$ (a.k.a. the toric code). Its data is listed in~\cite{Bonderson07b} and also as $\nu=8$ in Table~2 of~\cite{Kitaev06a}. The fundamental quasiholes of this state carrying Fib topological charge $\varepsilon$, and electric charge of either $e/7$ or $2e/7$.

As these theories are the direct product of a Fibonacci theory with Abelian sectors, the braiding statistics of quasiparticle excitations carrying the non-trivial Fibonacci charge are computationally universal.

\subsubsection{NAF states}

Wen's NAF states~\cite{Wen91a,Blok92} have non-Abelian statistics based on $\text{SU}\left(N\right)_{k}$. We will not describe these states in detail, but simply state that the fundamental quasiholes of the $\text{SU}\left(2\right)_{k}$ states have carry the non-Abelian topological charge $\frac{1}{2}$.


\begin{thebibliography}{10}
\expandafter\ifx\csname url\endcsname\relax
  \def\url#1{\texttt{#1}}\fi
\expandafter\ifx\csname urlprefix\endcsname\relax\def\urlprefix{URL }\fi

\bibitem{vonNeumann55}
J.~von Neumann, Mathematical Foundations of Quantum Mechanics, Princeton
  University Press, Princeton, 1955.

\bibitem{Kitaev03}
A.~Y. Kitaev, Fault-tolerant quantum computation by anyons, Ann. Phys. 303
  (2003) 2, quant-ph/9707021.

\bibitem{Freedman03b}
M.~H. Freedman, A.~Kitaev, M.~J. Larsen, Z.~Wang, Topological quantum
  computation, Bull. Amer. Math. Soc. (N.S.) 40 (2003) 31--38,
  quant-ph/0101025.

\bibitem{Bonderson08a}
P.~Bonderson, M.~Freedman, C.~Nayak, Measurement-only topological quantum
  computation, Phys. Rev. Lett. 101 (2008) 010501, arXiv:0802.0279.

\bibitem{Bonderson07a}
P.~Bonderson, K.~Shtengel, J.~K. Slingerland, Decoherence of anyonic charge in
  interferometry measurements, Phys. Rev. Lett. 98 (2007) 070401,
  quant-ph/0608119.

\bibitem{Bonderson07b}
P.~H. Bonderson, Non-{A}belian {A}nyons and {I}nterferometry, Ph.D. thesis
  (2007).

\bibitem{Bonderson07c}
P.~Bonderson, K.~Shtengel, J.~K. Slingerland, Interferometry of non-Abelian
  anyons, Ann. Phys. 323 (2008) 2709, arXiv:0707.4206.

\bibitem{Moore91}
G.~Moore, N.~Read, Nonabelions in the fractional quantum {H}all effect, Nucl.
  Phys. B 360 (1991) 362--396.

\bibitem{Read99}
N.~Read, E.~Rezayi, Beyond paired quantum {H}all states: Parafermions and
  incompressible states in the first excited {L}andau level, Phys. Rev. B 59
  (1999) 8084--–8092, cond-mat/9809384.

\bibitem{Bonderson07d}
P.~Bonderson, J.~K. Slingerland, Fractional quantum Hall hierarchy and the
  second Landau level, Phys. Rev. B 78 (2008) 125323, arXiv:0711.3204.

\bibitem{Willett08}
R.~L. Willett, M.~J. Manfra, L.~N. Pfeiffer, K.~W. West, Interferometric
  measurement of filling factor 5/2 quasiparticle charge (2008),
  arXiv:0807.0221.

\bibitem{Turaev94}
V.~G. Turaev, Quantum Invariants of Knots and 3-Manifolds, Walter de Gruyter,
  Berlin, New York, 1994.

\bibitem{Bakalov01}
B.~Bakalov, A.~Kirillov, Lectures on Tensor Categories and Modular Functors,
  Vol.~21 of University Lecture Series, American Mathematical Society, 2001.

\bibitem{Preskill-lectures}
J.~Preskill, Topological quantum computation, lecture notes (2004).
\newline http://www.theory.caltech.edu/$\sim$preskill/ph219/topological.ps

\bibitem{Kitaev06a}
A.~Kitaev, Anyons in an exactly solved model and beyond, Ann. Phys. 321 (2006)
  2--111, cond-mat/0506438.

\bibitem{Camino05a}
F.~E. Camino, W.~Zhou, V.~J. Goldman, Realization of a {L}aughlin quasiparticle
  interferometer: Observation of fractional statistics, Phys. Rev. B 72 (2005)
  075342, cond-mat/0502406.

\bibitem{Camino07a}
F.~E. Camino, W.~Zhou, V.~J. Goldman, $e/3$ {L}aughlin quasiparticle
  primary-filling $\nu=1/3$ interferometer, Phys. Rev. Lett. 98 (2007) 076805,
  cond-mat/0610751.

\bibitem{Ji03}
Y.~Ji, Y.~Chung, D.~Sprinzak, M.~Heiblum, D.~Mahalu, H.~Shtrikman, An
  electronic {M}ach--{Z}ehnder interferometer, Nature 422 (2003) 415--418,
  cond-mat/0303553.

\bibitem{Einstein35}
A.~Einstein, B.~Podolsky, N.~Rosen, Can quantum-mechanical description of
  physical reality be considered complete?, Phys. Rev. 47 (1935) 777.

\bibitem{Bell64}
J.~S. Bell, On the {E}instein-{P}oldolsky-{R}osen paradox, Physics 1 (1964)
  195.

\bibitem{Bennett93}
C.~H. Bennett, G.~Brassard, C.~Crepeau, R.~Jozsa, A.~Peres, W.~K. Wootters,
  Teleporting an unknown quantum state via dual classical and
  {E}instein-{P}odolsky-{R}osen channels, Phys. Rev. Lett. 70 (1993) 1895.

\bibitem{Bouwmeester97}
D.~Bouwmeester, J.-W. Pan, K.~Mattle, M.~Eibl, H.~Weinfurter, A.~Zeilinger,
  Experimental quantum teleportation, Nature 390 (1997) 575--579.

\bibitem{Gottesman99}
D.~Gottesman, I.~Chuang, Demostrating the viability of universal quantum
  computation using teleportation and single-qubit operations, Nature 402
  (1999) 390, quant-ph/9908010.

\bibitem{Raussendorf01}
R.~Raussendorf, H.~J. Briegel, A one-way quantum computer, Phys. Rev. Lett. 86
  (2001) 5188 -- 5191, quant-ph/0010033.

\bibitem{Nielsen03}
M.~A. Nielsen, Universal quantum computation using only projective measurement,
  quantum memory, and preparation of the 0 state, Phys. Lett. A. 308 (2003)
  96--100, quant-ph/0108020.

\bibitem{Aliferis04}
P.~Aliferis, D.~W. Leung, Computation by measurements: a unifying picture,
  Phys. Rev. A 70 (2004) 062314, quant-ph/0404082.

\bibitem{Bravyi06}
S.~Bravyi, Universal quantum computation with the $\nu=5/2$ fractional quantum
  {H}all state, Phys. Rev. A 73 (2006) 042313, quant-ph/0511178.

\bibitem{Freedman06a}
M.~Freedman, C.~Nayak, K.~Walker, Towards universal topological quantum
  computation in the $\nu=5/2$ fractional quantum {H}all state, Phys. Rev. B 73
  (2006) 245307, cond-mat/0512066.

\bibitem{FNW05b}
M.~Freedman, C.~Nayak, K.~Walker, Tilted interferometry realizes universal
  quantum computation in the {I}sing {TQFT} without overpasses (2005),
  cond-mat/0512072.

\bibitem{Freedman02a}
M.~H. Freedman, M.~J. Larsen, Z.~Wang, A modular functor which is universal for
  quantum computation, Commun. Math. Phys. 227 (2002) 605--622,
  quant-ph/0001108.

\bibitem{Freedman02b}
M.~H. Freedman, M.~J. Larsen, Z.~Wang, The two-eigenvalue problem and density
  of {J}ones representation of braid groups, Commun. Math. Phys. 228 (2002)
  177--199, math/0103200.

\bibitem{Chamon97}
C.~de~C.~Chamon, D.~E. Freed, S.~A. Kivelson, S.~L. Sondhi, X.~G. Wen, Two
  point-contact interferometer for quantum {H}all systems, Phys. Rev. B 55
  (1997) 2331--43, cond-mat/9607195.

\bibitem{Fradkin98}
E.~Fradkin, C.~Nayak, A.~Tsvelik, F.~Wilczek, A {C}hern-{S}imons effective
  field theory for the {P}faffian quantum {H}all state, Nucl. Phys. B 516
  (1998) 704--18, cond-mat/9711087.

\bibitem{DasSarma05}
S.~Das~Sarma, M.~Freedman, C.~Nayak, Topologically protected qubits from a
  possible non-{A}belian fractional quantum {H}all state, Phys. Rev. Lett. 94
  (2005) 166802, cond-mat/0412343.

\bibitem{Stern06a}
A.~Stern, B.~I. Halperin, Proposed experiments to probe the non-abelian
  $\nu=5/2$ quantum {H}all state, Phys. Rev. Lett. 96 (2006) 016802,
  cond-mat/0508447.

\bibitem{Bonderson06a}
P.~Bonderson, A.~Kitaev, K.~Shtengel, Detecting non-{A}belian statistics in the
  $\nu=5/2$ fractional quantum {H}all state, Phys. Rev. Lett. 96 (2006) 016803,
  cond-mat/0508616.

\bibitem{Bonderson06b}
P.~Bonderson, K.~Shtengel, J.~K. Slingerland, Probing non-{A}belian statistics
  with quasiparticle interferometry, Phys. Rev. Lett. 97 (2006) 016401,
  cond-mat/0601242.

\bibitem{Chung06}
S.~B. Chung, M.~Stone, Proposal for reading out anyon qubits in non-{A}belian
  $\nu = 12/5$ quantum {H}all state, Phys. Rev. B 73 (2006) 245311,
  cond-mat/0601594.

\bibitem{Fendley06a}
P.~Fendley, M.~P.~A. Fisher, C.~Nayak, Dynamical disentanglement across a point
  contact in a non-{A}belian quantum {H}all state, Phys. Rev. Lett. 97 (2006)
  036801, cond-mat/0604064.

\bibitem{Fendley07a}
P.~Fendley, M.~P.~A. Fisher, C.~Nayak, Edge states and tunneling of
  non-{A}belian quasiparticles in the $\nu=5/2$ quantum {H}all state and p+ip
  superconductors, Phys. Rev. B 75 (2007) 045317, cond-mat/0607431.

\bibitem{Fidkowski07c}
L.~Fidkowski, Double point contact in the k=3 {R}ead--{R}ezayi state (2007),
  arXiv:0704.3291.

\bibitem{Ardonne07a}
E.~Ardonne, E.-A. Kim, Hearing non-abelian statistics from a {M}oore--{R}ead
  double point contact interferometer, J. Stat. Mech. (2008)
  L04001arXiv:0705.2902.

\bibitem{Bishara08}
W.~Bishara, C.~Nayak, Edge states and interferometers in the {P}faffian and
  anti-{P}faffian states, Phys. Rev. B 77 (2008) 165302, arXiv:0708.2704.

\bibitem{Miller07a}
J.~B. Miller, I.~P. Radu, D.~M. Zumbuhl, E.~M. Levenson-Falk, M.~A. Kastner,
  C.~M. Marcus, L.~N. Pfeiffer, K.~W. West, Fractional quantum {H}all effect in
  a quantum point contact at filling fraction 5/2, Nature Physics 3 (2007) 561, cond-mat/0703161.

\bibitem{Dolev08}
M.~Dolev, M.~Heiblum, V.~Umansky, A.~Stern, D.~Mahalu, Observation of a quarter
  of an electron charge at the $\nu = 5/2$ quantum hall state, Nature 452
  (2008) 829--834, ar{X}iv:0802.0930.

\bibitem{Radu08}
I.~P. Radu, J.~B. Miller, C.~M. Marcus, M.~A. Kastner, L.~N. Pfeiffer, K.~W.
  West, Quasiparticle tunneling in the fractional quantum hall state at $\nu =
  5/2$, Science 320 (2008) 899, ar{X}iv:0803.3530.

\bibitem{Choi07}
H.~C. Choi, W.~Kang, S.~D. Sarma, L.~N. Pfeiffer, K.~W. West, Activation gaps of fractional
  quantum hall effect in the second landau level, Phys. Rev. B 77 (2008) 081301, ar{X}iv:0707.0236.

\bibitem{Moore89b}
G.~Moore, N.~Seiberg, Classical and quantum conformal field theory, Commun.
  Math. Phys. 123 (1989) 177--254.

\bibitem{Witten89}
E.~Witten, Quantum field theory and the {J}ones polynomial, Comm. Math. Phys.
  121 (1989) 351--399.

\bibitem{Wess71}
J.~Wess, B.~Zumino, Consequences of anomalous {W}ard identities, Phys. Lett. B
  37 (1971) 95.

\bibitem{Witten83}
E.~Witten, Global aspects of current algebra, Nucl. Phys. B 223 (1983)
  422--432.

\bibitem{Jones85}
V.~F.~R. Jones, A polynomial invariant for knots via von {N}eumann algebras,
  Bull. Am. Math. Soc. 12 (1985) 103--111.

\bibitem{Wen92b}
X.~G. Wen, Theory of the edge states in fractional quantum {H}all effects,
  Intl. J. Mod. Phys. B 6 (1992) 1711--62.

\bibitem{Ardonne99}
E.~Ardonne, K.~Schoutens, A new class of non-{A}belian spin-singlet quantum
  {H}all states, Phys. Rev. Lett. 82~(25) (1999) 5096--5099, cond-mat/9811352.

\bibitem{Wen91a}
X.~G. Wen, Non-{A}belian statistics in the fractional quantum {H}all states,
  Phys. Rev. Lett. 66 (1991) 802--5.

\bibitem{Blok92}
B.~Blok, X.~G. Wen, Many-body systems with non-{A}belian statistics, Nucl.
  Phys. B 374 (1992) 615.

\end{thebibliography}

\end{document}